\documentclass{pasa}%

\usepackage{graphicx}
\usepackage{epsfig}
\usepackage{epstopdf}
\usepackage{subfigure}
\usepackage{amssymb}
\usepackage[figuresright]{rotating}  
\usepackage{textcomp}
\usepackage[usenames,dvipsnames]{color}
\usepackage{float,lscape}
\usepackage[flushleft]{threeparttable} 
\newcommand{\farcm}{\mbox{\ensuremath{.\mkern-4mu^\prime}}}

\newcommand{\nsources}{1,863 } 
\newcommand{\arcdeg}{\ensuremath{^{\circ}}}
\makeatletter

\title[Defining the GLEAM 4-Jy Sample]{The GLEAM 4-Jy (G4Jy) Sample: \newline I. Definition and the catalogue}

\author[Sarah V. White et al.]{Sarah V. White$^{1, 2}$\thanks{sarahwhite.astro@gmail.com}, Thomas M.O. Franzen$^{1, 3}$, Chris J. Riseley$^{4,5,6}$, O. Ivy Wong$^{7}$, \newline Anna D. Kapi{\'n}ska$^{7,8}$, Natasha Hurley--Walker$^{1}$, Joseph R. Callingham$^{3}$, Kshitij Thorat$^{2,9}$, \newline Chen Wu$^{7}$, Paul Hancock$^{1}$, Richard W. Hunstead$^{10}$, Nick Seymour$^{1}$, Jesse Swan$^{11}$, Randall Wayth$^{1}$, John Morgan$^{1}$, Rajan Chhetri$^{1}$, Carole Jackson$^{3}$, Stuart Weston$^{12}$, Martin Bell$^{13}$, Bi-Qing For$^{7}$, B.\,M. Gaensler$^{14}$, Melanie Johnston--Hollitt$^{1}$, Andr{\'e} Offringa$^{3}$, and Lister Staveley--Smith$^{7}$ 
\affil{$^{1}$International Centre for Radio Astronomy Research (ICRAR), Curtin University,  Bentley, WA 6102, Australia}%
\affil{$^{2}$Department of Physics and Electronics, Rhodes University, PO Box 94, Grahamstown, 6140, South Africa}%
\affil{$^{3}$ASTRON: the Netherlands Institute for Radio Astronomy, Oude Hoogeveensedijk 4, 7991 PD, Dwingeloo, The Netherlands}%
\affil{$^{4}$CSIRO Astronomy and Space Science, PO Box 1130, Bentley, WA 6102, Australia}%
\affil{$^{5}$Dipartimento di Fisica e Astronomia, Universit\`a degli Studi di Bologna, via P. Gobetti 93/2, 40129 Bologna, Italy}%
\affil{$^{6}$INAF -- Istituto di Radioastronomia, via P. Gobetti 101, 40129 Bologna, Italy}%
\affil{$^{7}$ICRAR, University of Western Australia (M468), 35 Stirling Highway, Crawley, WA 6009, Australia}%
\affil{$^{8}$National Radio Astronomy Observatory (NRAO), 1003 Lopezville Rd, Socorro NM 87801, USA}%
\affil{$^{9}$South African Radio Astronomy Observatory (SARAO), 2 Fir Street, Observatory, Cape Town, 7925, South Africa}%
\affil{$^{10}$Sydney Institute for Astronomy (SIfA), School of Physics, University of Sydney, NSW 2006, Australia}%
\affil{$^{11}$School of Physical Sciences, University of Tasmania, Private Bag 37, Hobart, Tasmania, 7001 Australia}%
\affil{$^{12}$Institute for Radio Astronomy and Space Research (IRASR), Auckland University of Technology, Auckland 1010, New Zealand}%
\affil{$^{13}$University of Technology Sydney, 15 Broadway, Ultimo NSW 2007, Australia}%
\affil{$^{14}$Dunlap Institute for Astronomy and Astrophysics, University of Toronto, Toronto, ON M5S 3H4, Canada}%
}%

\jid{PASA}
\doi{10.1017/pasa.\the\year.xxx}
\jyear{\the\year}

\usepackage{aas_macros}
\usepackage{hyperref} 
\hypersetup{colorlinks,citecolor=blue,linkcolor=blue,urlcolor=blue}

\hypersetup{draft}

\begin{document}

\begin{frontmatter}
\maketitle

\begin{abstract}
The Murchison Widefield Array (MWA) has observed the entire southern sky (Declination, $\delta< 30^{\circ}$) at low radio-frequencies, over the range 72--231\,MHz. These observations constitute the GaLactic and Extragalactic All-sky MWA (GLEAM) Survey, and we use the extragalactic catalogue (Galactic latitude, $|b| >10^{\circ}$) to define the GLEAM 4-Jy (G4Jy) Sample. This is a complete sample of the `brightest' radio-sources ($S_{\mathrm{151\,MHz}}>4$\,Jy), the majority of which are active galactic nuclei with powerful radio-jets. Crucially, low-frequency observations allow the selection of such sources in an orientation-independent way (i.e. minimising the bias caused by Doppler boosting, inherent in high-frequency surveys). We then use higher-resolution radio images, and information at other wavelengths, to morphologically classify the brightest components in GLEAM. We also conduct cross-checks against the literature, and perform internal matching, in order to improve sample completeness (which is estimated to be $>95.5$\%). This results in a catalogue of 1,863 sources, making the G4Jy Sample over 10 times larger than that of the revised Third Cambridge Catalogue of Radio Sources (3CRR; $S_{\mathrm{178\,MHz}}>10.9$\,Jy). Of these G4Jy sources, 78 are resolved by the MWA (Phase-I) synthesised beam ($\sim$2$'$ at 200\,MHz), and we label 67\% of the sample as `single', 26\% as `double', 4\% as `triple', and 3\% as having `complex' morphology at $\sim$1\,GHz (45$''$ resolution). We characterise the spectral behaviour of these objects in the radio, and find that the median spectral-index is $\alpha=-0.740 \pm 0.012$ between 151\,MHz and 843\,MHz, and $\alpha=-0.786 \pm 0.006$ between 151\,MHz and 1400\,MHz (assuming a power-law description, $S_{\nu} \propto \nu^{\alpha}$), compared to $\alpha=-0.829 \pm 0.006$ within the GLEAM band. Alongside this, our value-added catalogue provides mid-infrared source associations (subject to $6''$ resolution at 3.4\,$\mu$m) for the radio emission, as identified through visual inspection and thorough checks against the literature. As such, the G4Jy Sample can be used as a reliable training set for cross-identification via machine-learning algorithms. We also estimate the angular size of the sources, based on their associated components at $\sim$1\,GHz, and perform a flux-density comparison for 67 G4Jy sources that overlap with 3CRR. Analysis of multi-wavelength data, and spectral curvature between 72\,MHz and 20\,GHz, will be presented in subsequent papers, and details for accessing all G4Jy overlays are provided at https://github.com/svw26/G4Jy.
\end{abstract}

\begin{keywords}
catalogues -- galaxies: active -- galaxies: evolution -- radio continuum: galaxies \newline

(Received 15 October 2019; revised 23 March 2020; accepted 23 March 2020)
\end{keywords}

\end{frontmatter}

\section{INTRODUCTION}
\label{sec:intro}

There are two key processes that influence how a galaxy evolves: star formation and black-hole accretion. The former involves the collapse of molecular gas to form stars, resulting in the build-up of stellar mass. However, such growth may be halted (typically in low-mass galaxies) if the power of supernovae is enough to expel gas from the system \citep{Efstathiou2000}, or if gas is stripped away during interaction with another galaxy \citep{Mihos1991} or within a cluster \citep{Kenney2014}. Meanwhile, material may be accreting onto the galaxy's central, supermassive black-hole. As it does so, a large amount of energy is released over a wide wavelength range [see reviews by \citet{Urry1995}, \citet{Wilkes1999} and \citet{Netzer2015}], and the galaxy is described as having an {\it active} galactic nucleus (AGN). AGN activity has been shown to affect the host galaxy, through both the {\it suppression} and {\it promotion} of star formation (referred to as `negative' and `positive' feedback, respectively). In the case of star formation being suppressed, the halo of the galaxy is heated by thermal energy from the accretion disc of the AGN, thereby preventing gas from cooling sufficiently to collapse to form stars \citep{Croton2006,Teyssier2011}. In addition, some AGN have radio jets associated with them, which may impact upon a molecular cloud, triggering its collapse and subsequent star formation \citep[e.g.][]{Davies2006, Ishibashi2012}.

A great strength of radio observations is that they are unaffected by dust obscuration, allowing both star formation and black-hole accretion to be detected out to higher redshift than is possible at other wavelengths (e.g. \citealt{Collier2014}). This includes finding high-redshift (proto-)clusters, by exploiting the tendency of `radio-loud' AGN to reside in dense environments \citep{Wylezalek2013}. The added advantage of {\it low-frequency} radio data is that they allow us to select a radio-source sample in an orientation-independent way. This is because the low-frequency emission of powerful AGN is dominated by the radio lobes, which are not subject to relativistic beaming \citep[also known as `Doppler boosting';][]{Rees1966,Blandford1979}. The same cannot be said for the radio core, hotspots and jets that dominate the emission of sources at high radio-frequencies. As a result of this beaming effect (which may push the observed radio-brightness above the flux-density limit), radio sources selected at high frequencies tend to be biased towards AGN that have their jet axis close to the line-of-sight.

In addition, low-frequency measurements allow us to probe older radio-emission, thereby revealing a population of galaxies that had an AGN in the past but show no signs of recent activity (as verified at higher radio-frequencies, e.g. \citealt{HurleyWalker2015}). The ability to constrain the radio spectrum over a broad frequency range also exposes `restarted radio-galaxies', which can be used to investigate episodic jet activity \citep{Blundell2011,Walg2014}. This provides an idea of the timescale over which AGN activity may promote or suppress star formation in the host galaxy. Furthermore, we can use low-frequency data to uncover poorly-studied processes in galaxies, such as the energetics within radio lobes. Doing so allows the internal pressure and magnetic-field strengths of the lobes to be determined \citep[e.g.][]{Harwood2016}. Extended frequency coverage also highlights sources with a turnover in their radio spectrum, showing that the canonical, power-law description ($S_{\nu} \propto \nu^{\alpha}$, with spectral index, $\alpha$) is too simplistic for many sources \citep[e.g.][]{Callingham2017}. The spectral curvature in the radio indicates that either ionised gas is present (leading to free-free absorption) or that synchrotron self-absorption is taking place \citep{Lacki2013}.

The revised Third Cambridge Catalogue of Radio Sources \citep[3CRR;][]{Laing1983} is currently the best-studied low-frequency radio-source sample, complete with optical data. This has enabled seminal pieces of work, such as the correlation between radio-jet power and optical line luminosity, found by \citet{Rawlings1991}. This correlation suggests that extragalactic radio-sources have a common central-engine mechanism driving their emission. In addition, \citet{Barthel1989} used the 3CRR sample to show that a unification model, based on orientation of the AGN, can explain the observed properties of quasars and radio galaxies. Another example of ground-breaking work using 3CRR is that of \citet{Heckman1986}, whose follow-up campaign concluded that a significant fraction of very powerful radio-sources may be driven by galaxy interactions and mergers.

However, the flux-density limit of 3CRR (10.9\,Jy at 178\,MHz\footnote{\label{3CRRlimit}This flux-density limit follows \citet{Laing1980} in using the flux-density scale of \citet{Roger1973}, hereafter the `RBC scale'. Corrected 178-MHz flux-densities are available at https://3crr.extragalactic.info and through VizieR (http://vizier.u-strasbg.fr). However, at the time of writing, the latter retains the outdated description for the flux-density column: `Flux at 178\,MHz (KPW scale)'. The `KPW scale' is that of \citet{Kellermann1969}, where 10.0\,Jy corresponds to 10.9\,Jy on the RBC scale \citep{Laing1980}.}) restricts the detection of radio-loud galaxies to 173 sources. As such, there is not a sufficient number of objects for studying their cosmological evolution in detail, in terms of age or environmental density \citep{Wang2008}. This is a far-reaching problem, as it is thought that such sources have a significant impact in proto-clusters, through powerful jets preventing gas from cooling and falling onto proto-galaxies \citep{Rawlings2004}. This is supported by X-ray observations of clusters showing `cavities' that have been carved out by radio jets \citep[e.g.][]{Fabian2000}, and hydrodynamical simulations that demonstrate the effect of buoyant `bubbles' -- inflated by the AGN -- on the intracluster medium \citep[e.g.][]{Sijacki2006}. Also, the relatively small number of high-excitation radio-galaxies (HERGs; \citealt{Best2012}) in the 3CRR sample means that how their active lifetime and jet power differs from that of low-excitation radio-galaxies (LERGs) can not be tested reliably \citep{Turner2015}. As a result, whether these properties are connected to the underlying accretion mode -- thought to be different for HERGs and LERGs -- requires further investigation.

AGN of similar radio flux-density have been identified in the Molonglo Southern 4-Jy (MS4) Sample \citep{Burgess2006a}, which consists of 228 sources detected above 4\,Jy at 408\,MHz. The brightest of these (137 sources) form a subset that is the southern equivalent of the 3CRR sample, known as `SMS4'. \citet{Burgess2006a} find that this subset has a greater proportion of sources larger than 5$'$, when compared to 3CRR, which they suggest may be due to 3CRR missing sources with low surface-brightness. However, the 178-MHz flux-densities for the SMS4 radio-sources are derived through either extrapolation from, or interpolation of, measurements at other frequencies (namely 80, 86, 160, 408, and 843\,MHz, where available). This, therefore, complicates the comparison with 3CRR, as some of the sources may have a spectral turnover at low radio-frequencies.

For this work we use observations at low radio-frequencies, obtained via the Murchison Widefield Array \citep[MWA;][]{Tingay2013}. This telescope is situated in a protected radio-quiet zone, which means that there is little radio-frequency interference, leading to very good spectral coverage. With 50 of the 128 antenna-tiles located less than 100\,m from the centre of the instrument (in the original Phase-I configuration), the MWA is also sensitive to large-scale, diffuse radio emission. All-sky data have been taken through the GaLactic and Extragalactic All-sky MWA \citep[GLEAM;][]{Wayth2015} survey, and we use the `brightest' detections in the extragalactic catalogue \citep{HurleyWalker2017} to construct the GLEAM 4-Jy (G4Jy) Sample \citep{Jackson2015, White2018}. Our sample contains \nsources sources and is over 10 times larger than 3CRR, due to its lower flux-density limit and larger survey area. Like 3CRR, the majority of these sources are galaxies with an active black-hole at the centre, and many have radio jets associated with them. By using this larger sample to study radio-bright active galaxies, we can gain a better understanding of their connection with their environment, investigate their fuelling mechanism, and more-closely analyse how these radio sources evolve over cosmic time. Furthermore, being the brightest radio-sources in the southern sky makes them excellent candidates for detailed studies using the Square Kilometre Array (SKA) and its precursor/pathfinder telescopes. 

However, in order to study the brightest GLEAM sources in detail, we first need to ensure that associated radio emission is collected together correctly. The necessity of this is clear for individual sources that have multiple radio detections in the GLEAM catalogue. In addition, we attempt to identify the galaxy that hosts the radio emission, so that the G4Jy Sample can be cross-matched more easily with catalogues at other wavelengths. For this we employ visual inspection, which is the most-reliable method for cross-identifying complex, extended radio-sources \citep[e.g.][]{Williams2019}.

\subsection{Paper outline}

In this paper we describe how we construct the G4Jy Sample, which consists of radio sources that are brighter than 4\,Jy at 151\,MHz. This involves using multi-wavelength data to collapse a list of GLEAM {\it components} into a list of GLEAM {\it sources}. Doing so is particularly important for ensuring that GLEAM flux-densities incorporate all of the radio emission associated with extended sources. The resulting G4Jy catalogue includes positions for the likely host-galaxy, to enable simpler cross-matching with other datasets. We also provide flux densities and angular sizes at $\sim1$\,GHz, and calculate multiple sets of spectral indices.

The data used for this work are summarised in Section~\ref{sec:observations}, and Section~\ref{sec:sampleselection} clarifies our initial sample selection. In Section~\ref{sec:centroids} we explain how we derive brightness-weighted centroids, and our visual inspection is detailed in Section~\ref{sec:visualinspection}. Contents of the G4Jy catalogue are outlined in Section~\ref{sec:finalcatalogue}, with column descriptions and an excerpt of the catalogue provided in Appendix~E. Sample completeness is discussed in Section~\ref{sec:completeness}, and initial analysis is described in Section~\ref{sec:discussion}. We then summarise our work in Section~\ref{sec:finalsummary}, and refer the reader to the accompanying paper (Paper~II; \citealt{White2020b}), where we demonstrate the wide variety of bright radio-sources in the G4Jy Sample and document additional literature checks.

Unless otherwise specified, we use integrated flux-densities (as opposed to peak surface-brightnesses) throughout this paper. In addition, we use a $\Lambda$CDM cosmology, with $H_{0} = 70$\,km\,s$^{-1}$\,Mpc$^{-1}$, $\Omega_{m}=0.3$, $\Omega_{\Lambda}=0.7$. Source names that are based on B1950 co-ordinates are indicated via the prefix `B', whilst all other position-derived names refer to J2000 co-ordinates. The sign convention that we use for a spectral index, $\alpha$, is as defined by $S_{\nu} \propto \nu^{\alpha}$.

\section{Data} \label{sec:observations}


The GLEAM Survey allows us to study the entire southern sky at frequencies below 300\,MHz. These MWA observations provide wide spectral coverage but, in order to assess the morphology of the radio sources, we require the better spatial-resolution that is afforded by other radio surveys. As such, we use data at 843\,MHz, 1.4, 4.8, 8.6, and 20\,GHz, which are described below, but also draw on the literature for further information (see Paper~II). In addition, we collate mid-infrared and optical data for the G4Jy Sample. The former allows us to identify the likely host-galaxy, including cases where the AGN is obscured by dust \citep[e.g.][]{Lacy2004}. Meanwhile, optical spectra enable redshifts to be determined, and provide information about the sources' star-forming and/or AGN properties \citep[e.g.][]{Baldwin1981,Kewley2001,Sadler2002}. Optical identifications for the G4Jy Sample will be presented in Paper~III by White et al. (in preparation).

\subsection{Radio data} 

\subsubsection{GLEAM catalogue and images (72--231\,MHz)}
\label{sec:GLEAMdata}

We use the extragalactic catalogue (EGC) of the GLEAM Survey \citep{HurleyWalker2017}, created using MWA observations of the southern sky (Declination, $\delta< 30^{\circ}$; Galactic latitude, $|b| >10^{\circ}$) at low radio-frequencies (72--231\,MHz). The resolution of the GLEAM Survey is declination dependent and, at the central frequency of 154\,MHz, is approximated by $2.5 \times 2.2$\,arcmin$^{2}/ \cos(\delta + 26.7^{\circ})$ \citep{Wayth2015}. This corresponds to a typical synthesised beam of $\sim$2$'$ at 200\,MHz. 20 flux densities are measured across the 72--231\,MHz band via priorised fitting, at positions determined from the `wide-band image'. This image was created by combining the data collected between 170 and 231\,MHz, in order to achieve greater signal-to-noise alongside the best possible resolution. The source-finding algorithm, \textsc{Aegean} v1.9.6 \citep{Hancock2012,Hancock2018}\footnote{https://github.com/PaulHancock/Aegean}, was performed over this image, and all Gaussian components detected above 5\,$\sigma$ ($S_{\mathrm{200\,MHz}}\gtrsim$ 50\,mJy) were retained for the catalogue. As such, the catalogue contains 307,455 GLEAM components. In addition, we use cutouts from the wide-band image for the visual inspection described in Section~\ref{sec:visualinspection}. 

\subsubsection{TGSS ADR1 catalogue and images (150\,MHz)}
\label{sec:TGSSdata}
The Giant Metrewave Radio Telescope (GMRT; \citealt{Swarup1991}) previously surveyed the sky above Dec.~$=-55^{\circ}$ at 150\,MHz, creating the TIFR GMRT Sky Survey (TGSS). However, due to poor data-quality at low elevations, only observations at Dec.~$>-53^{\circ}$ were retained for the first alternative data release (ADR1; \citealt{Intema2017}), which we use for this work. In addition, \citet{Intema2017} note that there is incomplete coverage at $6.5<$~R.A./h~$<9.5$, $25<$~Dec./$^{\circ}<39$, so we do not use TGSS data over this region. With a resolution of $25 \times 25$\,arcsec$^{2}$ [or $25 \times 25$\,arcsec$^{2}/ \cos(\delta - 19^{\circ})$ for Dec.~$<19^{\circ}$], this survey provides useful spatial information at low frequencies, complementing the broad frequency-range and surface-brightness sensitivity of the MWA. The typical rms is $<$\,5\,mJy\,beam$^{-1}$ (a 7-$\sigma$ threshold being used for the associated catalogue), and the astrometric accuracy is $<2''$ in R.A. and Dec. For this work, we note the flux-density-scale correction found by \citet{HurleyWalker2017b} to obtain consistency between TGSS and GLEAM.

\subsubsection{SUMSS catalogue and images (843\,MHz)}
\label{sec:SUMSSdata}

For GLEAM components at Dec.~$<-39.5^{\circ}$, we use images and flux densities from Version 2.1\footnote{Dated 2012-Feb-16 and obtained via VizieR (http://vizier.u-strasbg.fr/viz-bin/VizieR?-source=VIII\%2F81B).} of the Sydney University Molonglo Sky Survey (SUMSS) catalogue \citep{Mauch2003,Murphy2007}. This survey was conducted at a frequency of 843\,MHz using the Molonglo Observatory Synthesis Telescope \citep{Mills1981,Robertson1991}, and reaches a $\sim$5-$\sigma$ sensitivity limit of between 6\,mJy\,beam$^{-1}$ (Dec.~$\leq-50^{\circ}$) and 10\,mJy\,beam$^{-1}$ ($-50^{\circ} < $~Dec.~$\leq -30^{\circ}$). The resolution of these data is $45 \times 45\ \mathrm{cosec} |\delta|$\,arcsec$^{2}$, and the {\it largest} positional error ($\sqrt{(\Delta \alpha)^{2} + (\Delta \delta)^{2}}$, where $\alpha = $ Right Ascension, R.A.) is $\sim30''$. However, the positional error is more typically 1--2$''$ for sources brighter than 200\,mJy at 843\,MHz.

\subsubsection{NVSS catalogue and images (1.4\,GHz)}
\label{sec:NVSSdata}

The Very Large Array (VLA; \citealt{Thompson1980}) surveyed the northern sky at 1.4\,GHz, down to a declination of $-40^{\circ}$. The resulting NRAO (National Radio Astronomy Observatory) VLA Sky Survey \citep[NVSS;][]{Condon1998} has a 5-$\sigma$ limit in peak source-brightness of $\sim$2.5\,mJy\,beam$^{-1}$, and a resolution of 45$''$. We use images and flux densities from the NVSS catalogue for GLEAM components at Dec.~$\geq-39.5^{\circ}$, which corresponds to 77\% of the G4Jy sources. The NVSS components associated with these sources are brighter than 15\,mJy\,beam$^{-1}$, and so have a positional accuracy of $\lesssim$1$''$. 

\subsubsection{The AT20G catalogue (20\,GHz)}
\label{sec:AT20Gdata}

The Australia Telescope 20-GHz (AT20G) Survey \citep{Murphy2010} is a blind survey over the southern sky (Dec.~$<0^{\circ}$, $|b| >1.5^{\circ}$) at 20\,GHz, down to a flux-density limit of 40\,mJy (8\,$\sigma$) and with a positional error of $\sim1''$. The survey was conducted using the Australia Telescope Compact Array (ATCA; \citealt{Frater1992}), and for the majority of sources below Dec.~= $-15^{\circ}$, includes near-simultaneous observations at 4.8 and 8.6\,GHz (which will be used for a future paper by White et al.). As noted by \citet{Murphy2010}, the shortest baseline being 30.6\,m limits the sensitivity of the instrument to extended emission, and so biases AT20G detections towards AGN cores and hotspots. In addition, observations at high radio-frequencies (i.e. 20\,GHz) are strongly affected by weather conditions. As such, the blind-scan component of the AT20G catalogue has varying completeness, ranging from 92\% at 50\,mJy\,beam$^{-1}$ to 98\% at 70\,mJy\,beam$^{-1}$ \citep{Hancock2011}. 



\subsection{Mid-infrared data: AllWISE catalogue and images}
\label{sec:WISEdata}

The {\it Wide-field Infrared Survey Explorer} ({\it WISE}; \citealt{Wright2010}) has imaged the entire sky in the mid-infrared, at 3.4, 4.6, 12, and 22\,$\mu$m. These observing bands are referred to as W1, W2, W3, and W4, and correspond to resolutions of 6.1, 6.4, 6.5, and 12.0$''$, respectively. We use the AllWISE data release \citep{Cutri2013}, which involved combining data from the cryogenic and post-cryogenic phases of the survey. The result is improved sensitivity (0.054, 0.071, 0.73, and 5.0\,mJy, respectively, at 5\,$\sigma$) and astrometric accuracy ($\ll$1$''$) with respect to the {\it WISE} All-Sky data release \citep{Cutri2012}. 


\subsection{Optical data: The 6dFGS catalogue} \label{sec:6dFGSopticaldata}

The 6-degree Field Galaxy Survey (6dFGS; \citealt{Jones2004}) used the UK Schmidt Telescope (UKST; \citealt{Tritton1978}) to obtain optical spectroscopy over the southern hemisphere (Dec.~$<0^{\circ}$, $|b| >10^{\circ}$). We use the final data release (DR3; \citealt{Jones2009}), which presents redshifts for all southern sources brighter than $K=12.65$ in the 2MASS (Two Micron All Sky Survey) Extended
Source Catalogue (XSC; \citealt{Jarrett2000}). The resulting median redshift is 0.053.



\section{Initial sample definition} \label{sec:sampleselection}

Our starting point for defining the G4Jy Sample is to select all components in the GLEAM extragalactic catalogue \citep{HurleyWalker2017} that have an integrated flux-density greater than 4\,Jy at 151\,MHz ($S_{\mathrm{151\,MHz}}>4$\,Jy). This flux-density limit is chosen in order to construct a sample that is over 10 times larger than the 3CRR sample, from which we can create a radio-galaxy sub-sample\footnote{Whilst we expect the vast majority of extragalactic radio-emission above this high flux-density threshold to be due to AGN, we note that there are a few other types of radio source within the G4Jy Sample. These are described in section~4 of Paper~II.} that allows AGN properties to be investigated more robustly (e.g. as a function of redshift and/or environment). The resulting list of 1,879 GLEAM components is then `collapsed' into a {\it source} list, where we define a source as being the object from which the radio emission originates. This is done through visual inspection (as detailed in Section~\ref{sec:visualinspection}), and is necessary as some radio sources have multiple GLEAM components. For example, a single AGN may have three entries in the GLEAM catalogue: two components marking radio lobes (where jets are interacting with the surrounding environment), and another due to an accreting `core' (associated with the central supermassive black-hole). Their individual flux densities can then be summed together to calculate the source's total flux density, at each of the 20 frequencies that span the GLEAM band.  

Additional GLEAM components enter the sample by association (Section~\ref{sec:morphology}), and we also search for sources that are brighter than 4\,Jy but have been missed from this initial selection (Section~\ref{sec:completeness}). Given how the GLEAM source counts vary with flux density \citep{Franzen2019}, and that visual inspection and cross-checks are very time-consuming, it is currently infeasible to extend this work to a flux-density limit lower than $S_{\mathrm{151\,MHz}}=4$\,Jy. Meanwhile, concerning very bright radio-sources, the following sub-section lists those that are known to be absent from the GLEAM catalogue in the first instance.

\subsection{Masked sources and the Orion Nebula}
\label{sec:AteamOrion}

For readers unfamiliar with the GLEAM Survey, we clarify that the very brightest sources at Dec.~$< 30^{\circ}$ and $|b| >10^{\circ}$ (belonging to a group of radio sources colloquially referred to as the `A-team') are masked for the GLEAM extragalactic catalogue (EGC), and so do not appear in the G4Jy Sample. The sources in question are listed in Table~\ref{GLEAMexclusions} and, due to the difficulty in calibrating and imaging them at low frequencies, will be presented in a separate paper (White et al., in preparation). Also masked for the EGC are the Large and Small Magellanic Clouds, for which multi-frequency, integrated flux-densities (e.g. $S_{\mathrm{150\,MHz}} = 1450$\,Jy and $S_{\mathrm{150\,MHz}} = 258$\,Jy, respectively) are presented by \citet{For2018}. Details of the masked regions are provided in table 3 of \citet{HurleyWalker2017}, with $<474$\,deg$^2$ of sky coverage being flagged due to the aforementioned sources. 

In addition, Table~\ref{GLEAMexclusions} includes the Orion Nebula (or `Orion~A'). Although its 151-MHz flux-density is well above the 4-Jy threshold (Appendix~\ref{sec:orion}), it was excluded by {\sc Aegean} source-fitting during the creation of the GLEAM catalogue. This happens when an object's integrated flux-density is more than 10 times its peak flux-density, in which case the object is considered to be highly resolved, and so is removed from the catalogue. This criterion is specified because {\sc Aegean} is optimised for fitting point sources, and so would not provide reliable measurements for diffuse radio-emission. 


\begin{table}
\centering 
\caption{A list of the brightest sources in the southern sky (Dec.~$< 30^{\circ}$, $|b| >10^{\circ}$) that currently do not appear in the G4Jy Sample. Below, we use `Cen A' as shorthand for `Centaurus A'. The flux densities ($S_{\mathrm{151\,MHz}}$) and spectral indices ($\alpha$) shown are approximate values \citep{HurleyWalker2017}, based on measurements (spanning 60--1400\,MHz) from the NASA/IPAC Extragalactic Database (NED)\protect\footnotemark. The exception is for *Orion~A (the Orion Nebula), where these values are determined via the method described in Appendix~\ref{sec:orion}. Note that its spectral index is valid only very locally at 151\,MHz, due to the high degree of spectral curvature.}
\begin{tabular}{@{}lcccc@{}}
 \hline
Source & R.A.  & Dec. & $S_{\mathrm{151\,MHz}}$ & $\alpha$ \\
 &  (h:m:s) &  (d:m:s) & /\,Jy &  \\
 \hline
Cen A & 13:25:28 & $-$43:01:09 & 1577 & $-$0.50 \\
Taurus A & 05:34:32 & +22:00:52& 1425 & $-$0.22 \\
Hercules A & 16:51:08 & +04:59:33 & 509 & $-$1.07 \\
Hydra A & 09:18:06 & $-$12:05:44 & 367 & $-$0.96 \\
Pictor A & 05:19:50 & $-$45:46:44 & 515 & $-$0.99 \\
Virgo A & 12:30:49 & +12:23:28 & 1096 & $-$0.86 \\
*Orion A & 05:35:17 & $-$05:23:23 & 67 & $+1.1$\\
\hline
\label{GLEAMexclusions}
\end{tabular}
\end{table}

\footnotetext{http://ned.ipac.caltech.edu/}

\section{Brightness-weighted centroids}
\label{sec:centroids}

The typical resolution of the MWA beam is $\sim$2$'$, and so 1,785 of the final \nsources sources (Section~\ref{sec:finalcatalogue}) consist of a single component in GLEAM. For the remainder, the low-frequency radio emission is so extended that it is detected as multiple GLEAM components. In order to determine which components are associated with the same `parent' source, we exploit the better-resolution data afforded by the longer baselines of GMRT and higher-frequency radio surveys. Since SUMSS and NVSS offer comparable sensitivity to extended emission as GLEAM, we only consider these two datasets for this section, but supplement this with information from TGSS ADR1 in Section~\ref{sec:visualinspection}.

Firstly, we automatically cross-correlate the 1,879 GLEAM components (Section~\ref{sec:sampleselection}) with SUMSS data at Dec~$< -39.5^{\circ}$, and with NVSS data at Dec.~$\geq -39.5^{\circ}$. This is done by using all pixels in the SUMSS/NVSS image that are within the 3-$\sigma$ contour level, enclosing the GLEAM position being considered (and where $\sigma$ is the local rms noise in SUMSS/NVSS), to set the `integration area'. We then deem all catalogued SUMSS/NVSS components lying within the integration area as being associated with the GLEAM component in question. The flux densities and positions of the associated SUMSS/NVSS components are then used to calculate the brightness-weighted centroid (of the SUMSS/NVSS emission) for each GLEAM component. Based on symmetry arguments regarding the radio emission, this position therefore estimates the location of the host galaxy (i.e. the `parent' source). This is useful for when we try to identify the mid-infrared position that corresponds to the G4Jy radio-source (Section~\ref{sec:identifyhost}). For the G4Jy sources where this is not possible(/relevant), the centroid position then becomes the best reference position for cross-matching against other datasets.

When calculating the centroid's positional errors in R.A. and Dec. ($\sigma_{\alpha}$ and $\sigma_{\delta}$, respectively), we take a conservative approach by assuming that the positional errors of the individual SUMSS/NVSS components are correlated. If the centroid position is obtained using NVSS, we typically find that $\sigma_{\alpha} \approx 0.5''$ and $\sigma_{\delta} \approx 0.6''$. If the centroid position is instead obtained using SUMSS, then typically $\sigma_{\alpha} \approx 1.5''$ and $\sigma_{\delta} \approx 1.7''$. In addition, we sum the SUMSS/NVSS flux-densities to obtain the total, integrated flux-density at 843\,MHz/1.4\,GHz. For the error on the total flux-density we assume that the component flux-density errors are uncorrelated, and so sum them in quadrature.

Using this technique, SUMSS/NVSS counterparts for a GLEAM component may be missed if there is no extended emission linking them in SUMSS/NVSS. (That is, the SUMSS/NVSS components are well separated and may wrongly be assumed to be unrelated.) This can be the case for very extended radio-sources. Conversely, unrelated point-sources lying within the integration area of a GLEAM component will be misclassified as associated emission at 843\,MHz/1.4\,GHz. In order to identify and correct these errors, we visually inspect the centroid positions for each of the 1,879 GLEAM components, using overlays detailed in the next section.

\section{Visual inspection}
\label{sec:visualinspection}

Considering the bright radio flux-densities involved ($S_{\mathrm{151\,MHz}}>4$\,Jy), it is expected that AGN dominate this sample, with many having a radio morphology that is multi-component. This poses a problem for combining radio catalogues with data at other wavelengths, where sources tend to be single-component and (subject to the flux-density limit) have a higher spatial density across the sky. As a result -- and particularly for complex sources -- a simple, nearest-neighbour cross-match will lead to incorrect association of multi-wavelength emission.

To aid the construction of multi-wavelength spectral energy distributions (SEDs) for the G4Jy Sample, we use several datasets (Section~\ref{sec:observations}) for visual inspection of the selected GLEAM components. Doing so allows us to classify the morphology of the sources in question, and also enables us to identify the most-likely host galaxy for the radio emission. This is especially important for cases where calculation of the centroid position (Section~\ref{sec:centroids}) has been affected by: (a) unrelated sources being blended by the NVSS/SUMSS beam (i.e. confusion); (b) unrelated -- but distinct -- sources in NVSS/SUMSS being incorrectly treated as `associated', due to $>$3-$\sigma$ emission between them; (c) the absence of extended $>$3-$\sigma$ emission linking NVSS/SUMSS components that {\it should} be associated; or (d) the radio emission not being axisymmetric (e.g. a wide-angle tail [WAT] radio-galaxy, see section~4.7 of Paper~II). 

By limiting this work to the brightest GLEAM components (where we have good signal-to-noise ratios), ionospheric effects and confusion noise will have little impact on our definition of the G4Jy Sample. (This is because these bright sources dominate the signal during calibration of the radio data.) In addition, the time-consuming nature of visual inspection means that we cannot justify consideration of a larger sample to a lower flux-density limit (see Section~\ref{sec:sampleselection}). To this end, automated algorithms for morphology classification \citep[e.g. {\sc ClaRAN};][]{Wu2019} and cross-identification will need to be developed. Until such prototype tools become proven technology, visual classification remains the most reliable method for sources with complicated morphology. In which case, an approach akin to the Radio Galaxy Zoo project \citep{Banfield2015} may be needed.

\subsection{Creating the overlays}

\begin{figure*}
\centering
\includegraphics[scale=2.0]{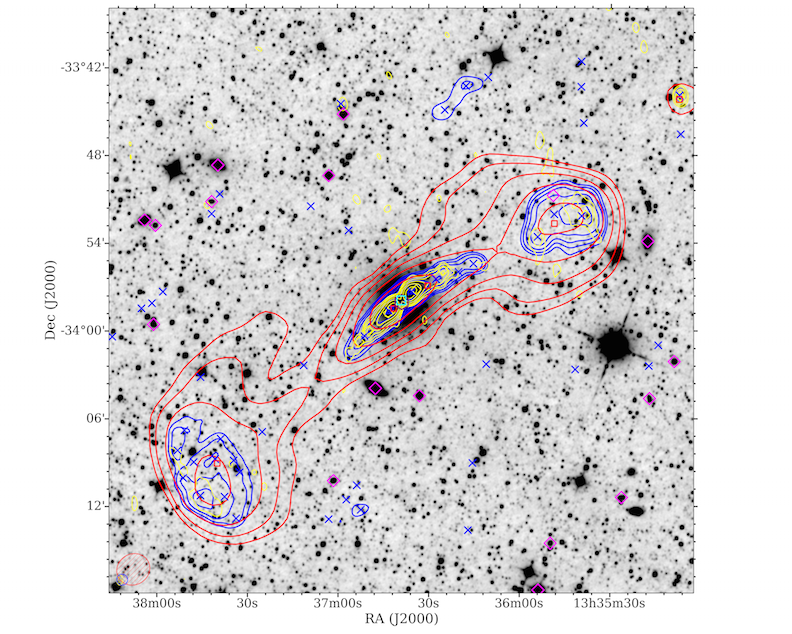}
 \caption{An overlay, centred at R.A. = 13:36:39, Dec. = $-$33:57:57 (J2000), for an extended radio-galaxy in the G4Jy Sample (G4Jy~1080, also known as IC~4296, at $z=0.012$). Radio contours from TGSS (150\,MHz; yellow), GLEAM (170--231\,MHz; red) and NVSS (1.4\,GHz; blue) are overlaid on a mid-infrared image from AllWISE ($3.4\,\mu$m; inverted greyscale). For each set of contours, the lowest contour is at the 3\,$\sigma$ level (where $\sigma$ is the local rms), with the number of $\sigma$ doubling with each subsequent contour (i.e. 3, 6, 12\,$\sigma$, etc.). Also plotted, in the bottom left-hand corner, are ellipses to indicate the beam sizes for TGSS (yellow with `+' hatching), GLEAM (red with `/' hatching), and NVSS (blue with `\textbackslash' hatching). This source is an unusual example, in that its GLEAM-component positions (red squares) needed to be re-fitted using \textsc{Aegean} \citep{Hancock2012,Hancock2018} -- see Appendix~D.1. Also plotted are catalogue positions from TGSS (yellow diamonds) and NVSS (blue crosses). The brightness-weighted centroid position, calculated using the NVSS components, is indicated by a purple hexagon. The cyan square represents an AT20G detection, marking the core of the radio galaxy. Magenta diamonds represent optical positions for sources in 6dFGS, and so we see above that G4Jy~1080 is not in this survey.}
\label{tiefighterOverlay}
\end{figure*}

We use the APLpy Python module \citep{Robitaille2012} to overlay radio contours from GLEAM, TGSS, and NVSS (or SUMSS, for Dec.~$<-39.5^{\circ}$) onto mid-infrared (W1) images from {\it WISE} (e.g. Figure~\ref{tiefighterOverlay}). GLEAM images are obtained via the online GLEAM Postage Stamp Service\footnote{http://mwa-web.icrar.org/gleam\_postage/q/form}, whilst TGSS, NVSS, SUMSS and {\it WISE} images are downloaded from the SkyView Virtual Observatory\footnote{https://skyview.gsfc.nasa.gov/current/cgi/query.pl}. For all images, orthographic (i.e. sine) projection is used, with GLEAM images having a pixel scale of 28\,arcsec\,pixel$^{-1}$. {\it WISE} images are at 1.375\,arcsec\,pixel$^{-1}$, TGSS images are downloaded at 5\,arcsec\,pixel$^{-1}$, and a scale of 10\,arcsec\,pixel$^{-1}$ is set for the NVSS and SUMSS images. For each set of radio contours, the lowest contour-level that we plot is 3\,$\sigma$ (where $\sigma$ is the local rms).

The reason behind using mid-infrared images as the greyscale `base' for our overlays is that this allows us to identify even the most-dust-obscured host galaxies. This would not be possible if optical images were used instead. Furthermore, mid-infrared emission includes contributions from evolved stellar populations, and avoids the bias of optical surveys towards actively star-forming galaxies. Of the four possible {\it WISE} bands, W1 is chosen for the imaging as this offers the best sensitivity and resolution.

Originally our overlays were chosen to be 20$'$ across, but first inspection of the sample revealed that a number of sources extended far beyond this size. Following a few iterations, we decided to create two sets of overlays: one set consisting of images 1\arcdeg\ across (in order to encompass all of the relevant emission for the largest sources, and so more-accurately classify the morphology -- Section~\ref{sec:morphology}) and another set using images 10$'$ across (acting as `close-ups' for identifying the likely host-galaxy -- Section~\ref{sec:identifyhost}). For the 1\arcdeg\ overlays, the GLEAM component's R.A. and Dec. specifies the centre of the image. As for the 10$'$ overlays, these are centred on the brightness-weighted centroid positions described in Section~\ref{sec:centroids}\footnote{One exception is for G4Jy~517. Its `close-up' overlay needs to be 20$'$ across in order to show both the host galaxy (confirmed by the literature) and the position of the centroid, as these are $10\farcm9$ apart (section~4.8 of Paper~II). The centring for this overlay is on a position 4$'$ north of the centroid for this source.}.

A problem faced when downloading images that are 1\arcdeg\ across is that this size increases the likelihood of running into artefacts associated with poor image-processing, or the source being too close to the edge of a mosaic/tile (resulting in a truncated image). Such was the case for the NVSS images of three components: GLEAM~J045610$-$215922, GLEAM~J122039$-$374017, and GLEAM~J154030$-$051436. This was remedied by obtaining multiple images from the NVSS Postage Stamp Server\footnote{http://www.cv.nrao.edu/nvss/postage.shtml}, offset in R.A. and Dec., and stitching them together using {\sc SWarp} \citep{Bertin2002}.

In addition to overlaying radio contours on the mid-infrared images, we plot positions from the GLEAM, TGSS, NVSS/SUMSS, AT20G and 6dFGS catalogues. Although AT20G is incomplete, detections from this survey indicate the presence of AGN cores, or hotspots in the radio lobes \citep{Massardi2011}. Meanwhile, 6dFGS positions help to identify host galaxies that are nearby/bright enough to have a spectrum from this all-sky -- albeit shallow -- optical survey. We also mark the centroid positions, described in the previous section, and use the errors in this position ($\sigma_{\alpha}$ and $\sigma_{\delta}$) to draw an error ellipse. However, in most overlays this ellipse is so small that it appears as a dot. Each of these datasets feature in the overlay presented in Figure~\ref{tiefighterOverlay}.

Both sets of overlays (1\arcdeg\ and 10$'$ across), and the images from which they are made, are available online\footnote{Please see https://github.com/svw26/G4Jy for details of how to download the overlays and/or cutouts.}. As the overlays are created {\it per GLEAM component}, radio sources that are multi-component will appear multiple times.

\subsection{Morphological classification}
\label{sec:morphology}

As part of visually inspecting the GLEAM components, we provide a classification based on the morphology of the source in NVSS/SUMSS (and/or TGSS, where available). This classification is one of the following four categories:

\begin{itemize}
\item `single' -- the source has a simple (typically compact) morphology in TGSS and NVSS/SUMSS;
\item `double' -- the source has two lobes evident in TGSS or NVSS/SUMSS, but there is no distinct detection of a core; {\it or} it has an elongated structure that is suggestive of lobes, but is accompanied by a single, catalogued detection;
\item `triple' -- the source has two lobes evident in TGSS or NVSS/SUMSS, and there is a distinct detection of a core in the same survey;
\item `complex' -- the source has a complicated morphology that does not clearly belong to any of the above categories.
\end{itemize} 

When determining the morphology, we take into account extra information provided by the underlying distribution of mid-infrared sources (i.e. potential host galaxies) and the positions of AT20G detections. This helps to resolve ambiguities, particularly in cases where (for example) two nearby NVSS detections may be interpreted as either a `double' radio source or two unrelated sources. For a `double', we expect the host galaxy to lie about half-way between the two NVSS components, as indicated by a mid-infrared source at the centroid position. If instead there is mid-infrared emission coincident with one (or both) of the radio components, then they are likely to be unrelated.  However, it can still be difficult to distinguish between a source with two radio lobes and two unrelated radio sources that are close to one another. In these situations we consult notes by \citet{Jones1992} on the observed structure of southern, extragalactic sources, and also consider the criterion defined by \citet{Magliocchetti1998}. This is where two radio components are likely to be associated if their flux densities are within a factor 4 of each other. 

In addition to the morphology, for each G4Jy source we record:

\renewcommand{\theenumi}{\roman{enumi}}%
\begin{enumerate}
\item the number of NVSS/SUMSS detections associated with the radio source. The integrated flux-densities for these detections are summed together to determine the total radio emission at $\sim$1\,GHz.  
\item the number of GLEAM components associated with the radio source. The integrated flux-densities for these components are then summed together to determine the total radio emission, in each of GLEAM's 20 sub-bands.
\item whether multiple sources (as judged visually\footnote{We recognise that this is subjective, and heavily influenced by the resolution of the available data. Therefore, after the first {\it two} passes of visual inspection, we compare the findings of four assessors (SVW, TMOF, OIW, ADK) and debate any disagreements until a conclusion is reached. This is revised, where necessary, following additional passes of visual inspection and checks against the literature (see Paper~II).}) contribute to the GLEAM component(s) under inspection. This acts as a `confusion flag', indicating cases where the MWA beam has blended unrelated sources together.
\end{enumerate} 

Regarding (i), we check whether these detections match those used for the calculation of the centroid position (Section~\ref{sec:centroids}). In cases where there is disagreement, the centroid positions are re-calculated following manual intervention. We refer to this as `re-centroiding', and direct the reader to Section~\ref{sec:recentroiding} for further details. As for the confusion flag (iii), our criteria are that: (1) unrelated sources are detected above 6\,$\sigma$ in NVSS/SUMSS, and (2) the positions of the unrelated sources' peak emission (at $\sim$1\,GHz) are within the 3-$\sigma$ GLEAM contour for the G4Jy source.

We emphasise that steps (i) and (ii) above are especially important for extended sources (typically larger than 3$'$ across), as otherwise their total radio emission may be severely underestimated. Meanwhile, step (iii) highlights cases where {\it multiple} sources contribute towards a particular detection in GLEAM. Since we are typically interested in only one of these contributing sources, the measured GLEAM flux-densities will overestimate the low-frequency radio emission, and therefore must be treated with caution. In light of this, we exploit the better resolution of TGSS to judge whether the GLEAM detection crosses the $S_{\mathrm{151\,MHz}}>4$\,Jy threshold as a consequence of confusion. However, we do not rely solely on the TGSS flux-densities for this assessment, as \citet{HurleyWalker2017b} found there to be significant variation in the flux-density scale over the TGSS survey-area. Hence, we consider what {\it fraction} that each blended source contributes to the total emission (corresponding to the GLEAM component) at 150\,MHz. If {\it none} of the blended sources has a TGSS (150\,MHz) integrated flux-density that corresponds to $S_{\mathrm{151\,MHz}}>4$\,Jy, we remove the $S_{\mathrm{151\,MHz}}>4$\,Jy detection from the GLEAM-component list. Hence the removal of the following components: GLEAM~J093918+015948, GLEAM~J101051$-$020137, GLEAM~J201707$-$310305, GLEAM~J202336$-$191144, and GLEAM~J222751$-$303344 (Appendix~\ref{sec:removedcomponents}). 

Meanwhile, the identification of extended low-frequency emission results in 84 components being added to the GLEAM-component list by association. These are GLEAM components that {\it individually} have $S_{\mathrm{151\,MHz}}<4$\,Jy but where visual inspection indicates that the emission should be combined with one or more other components for a particular radio source (resulting in a {\it summed} $S_{\mathrm{151\,MHz}}$ that is $>4$\,Jy; see also Section~\ref{sec:completeness}). We create individual overlays for these new components, and inspect them in the same way to ensure consistency. For a list of all the sources that are multi-component in GLEAM, see Table~\ref{multiGLEAMsources} in Appendix~\ref{app:multiGLEAMsources}. The overlays for these sources are presented in Figures~\ref{tiefighterOverlay}, 3--9,  Appendices~\ref{app:multiGLEAMsources} and D.3, and Paper~II (figures 3--4, 6, 8, 12, 16--17, 19, 21, 23). 

\subsubsection{Artefacts in the TGSS catalogue}
\label{sec:TGSSartefacts}

\begin{figure*}
\centering
\subfigure[G4Jy~679]{
	\includegraphics[width=0.42\linewidth]{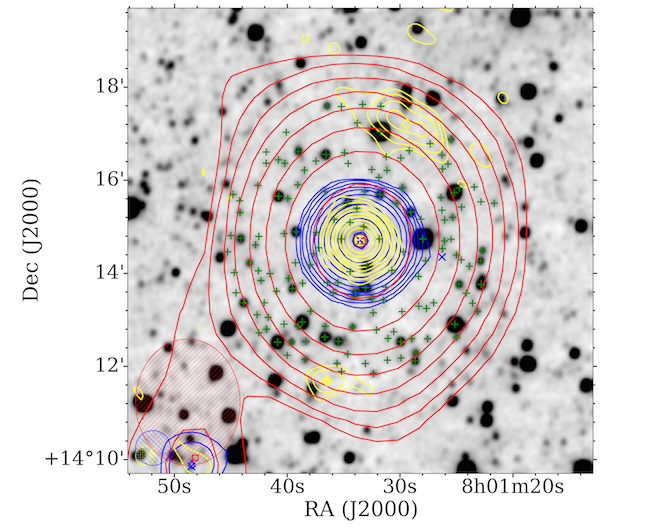} 
	} 
\subfigure[G4Jy~938]{
	\includegraphics[width=0.42\linewidth]{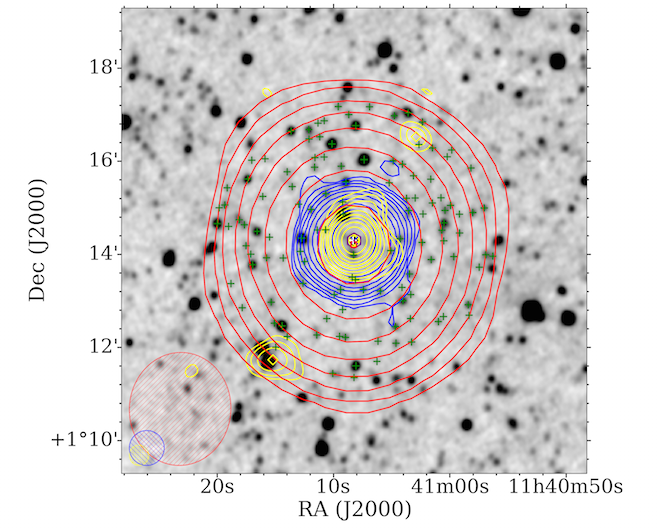} 
	}  \\[-0.35cm]
\subfigure[G4Jy~1005]{
	\includegraphics[width=0.42\linewidth]{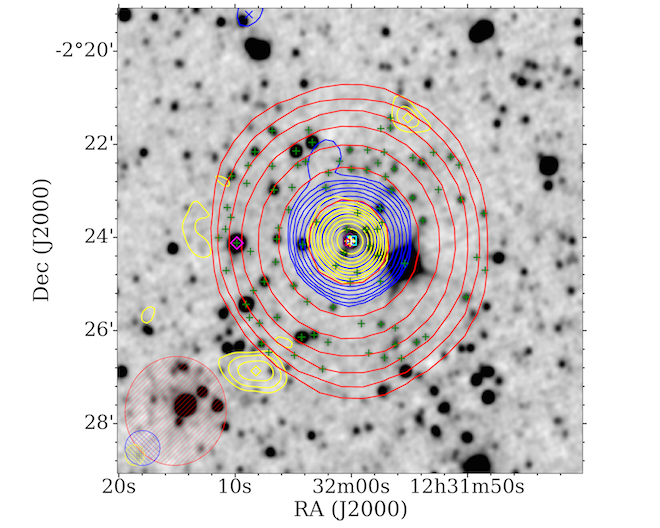} 
	} 
\subfigure[G4Jy~1085]{
	\includegraphics[width=0.42\linewidth]{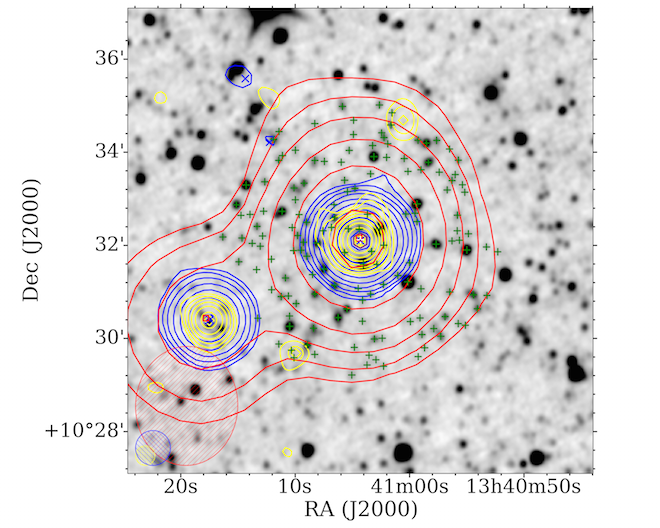} 
	}  \\[-0.35cm]
\subfigure[G4Jy~1209]{
	\includegraphics[width=0.42\linewidth]{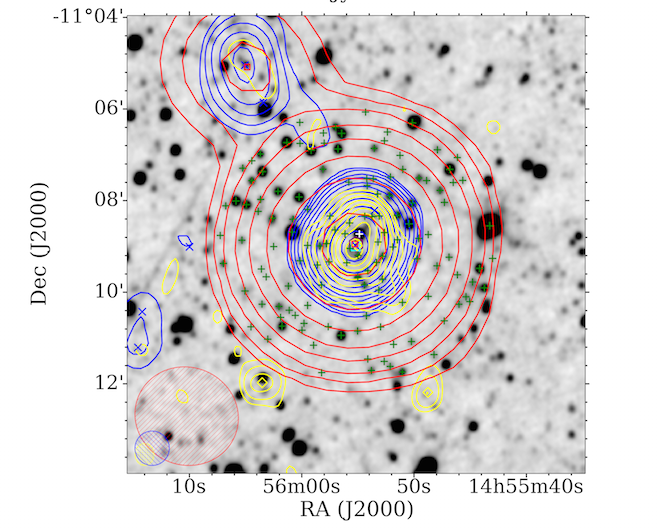} 
	} 
\subfigure[G4Jy~1239]{
	\includegraphics[width=0.42\linewidth]{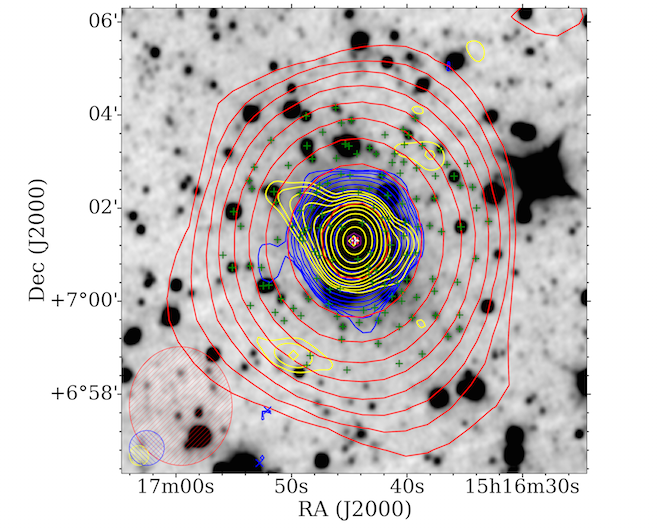} 
	} 
\caption{Examples of sources that have TGSS artefacts (Section~\ref{sec:TGSSartefacts}), with contours, symbols, and beams as described for Figure~\ref{tiefighterOverlay}. In addition, AllWISE positions (green plus signs) within 3$'$ of the centroid position (purple hexagon) are plotted, with the host galaxy highlighted in white. \label{artefactexamples}}
\end{figure*}

Through our visual inspection, we notice that several bright radio-sources (such as those in Figure~\ref{artefactexamples}) have low-level TGSS contours at a certain position-angle ($149.0\pm5.4$\arcdeg\ and/or $330.4\pm7.1$\arcdeg) and distance from the source ($161.9\pm13.3''$)\footnote{We quote median values, where the error is the median absolute deviation.}. Recognising that these are likely sidelobe artefacts, we take care not to misinterpret the morphology of the source in question (which would lead to an incorrect morphology classification). We find that these artefacts are exhibited by 63 sources (listed in Table~\ref{TGSSartefactstable}), which we use to characterise the position angle and distance quoted above. However, there may be other cases where an artefact coincides with a nearby, unrelated source, making them more difficult to identify. Unfortunately, for the 63 sources considered (which have $S_{\mathrm{151\,MHz}}$ ranging from 4.0\,Jy to 55.9\,Jy), the majority of the artefacts appear as detections in the TGSS catalogue, as indicated by yellow-diamond markers in the overlays.

\begin{table*}
\centering 
\caption{63 G4Jy sources identified as most likely having artefacts in the TGSS catalogue (Section~\ref{sec:TGSSartefacts}). }
\begin{tabular}{@{}rcrc@{}}
 \hline
Source & Corresponding GLEAM component & Source & Corresponding GLEAM component   \\
 \hline
   G4Jy 28  &          GLEAM J001619$-$143009 & G4Jy 971  &          GLEAM J120232$-$024005  \\   
   G4Jy 36  &          GLEAM J002021$-$023305 & G4Jy 978  &          GLEAM J121256+203237 \\   
   G4Jy 109  &          GLEAM J010010$-$174841 & G4Jy 1005  &          GLEAM J123200$-$022406 \\   
   G4Jy 114  &          GLEAM J010244$-$273124 & G4Jy 1014  &          GLEAM J124219$-$044616 \\   
   G4Jy 116  &          GLEAM J010249+255219  & G4Jy 1019  &          GLEAM J124357+162250 \\   
   G4Jy 138  &          GLEAM J011815$-$255148 & G4Jy 1023  &          GLEAM J124823$-$195915 \\   
   G4Jy 162  &          GLEAM J013027$-$260956 & G4Jy 1054  &          GLEAM J130949$-$001238 \\   
   G4Jy 169  &          GLEAM J013212$-$065232 & G4Jy 1064  &          GLEAM J132025+064439 \\   
   G4Jy 406  &          GLEAM J040107+003636 & G4Jy 1083  &          GLEAM J133808$-$062709 \\   
   G4Jy 414  &          GLEAM J040724+034049 & G4Jy 1085  &          GLEAM J134104+103209 \\   
   G4Jy 469  &          GLEAM J043106+011252 & G4Jy 1086  &          GLEAM J134243+050431 \\   
   G4Jy 594  &          GLEAM J060657$-$492928 & G4Jy 1156  &          GLEAM J142409+185249  \\   
   G4Jy 642  &          GLEAM J070554$-$424849 & G4Jy 1167  &          GLEAM J142740+283327  \\   
   G4Jy 674  &          GLEAM J074528+120930 & G4Jy 1170  &          GLEAM J142831$-$012402 \\   
   G4Jy 679  &          GLEAM J080133+141441 & G4Jy 1209  &          GLEAM J145555$-$110856 \\   
   G4Jy 706  &          GLEAM J082717$-$202619 & G4Jy 1239  &          GLEAM J151644+070118 \\   
   G4Jy 717  &          GLEAM J083710$-$195152 & G4Jy 1255  &          GLEAM J152357+105545 \\   
   G4Jy 748  &          GLEAM J090225$-$051639 & G4Jy 1267  &          GLEAM J153315+133221  \\   
   G4Jy 763  &          GLEAM J091829+223234 & G4Jy 1335  &          GLEAM J162514+265034  \\   
   G4Jy 768  &          GLEAM J092212$-$142845 & G4Jy 1338  &          GLEAM J162732+211224 \\   
   G4Jy 802  &          GLEAM J095338+251623 & G4Jy 1371  &          GLEAM J165258+001908 \\   
   G4Jy 819  &          GLEAM J100557$-$414849 & G4Jy 1400  &          GLEAM J172004$-$142601 \\   
   G4Jy 837  &          GLEAM J102003$-$425130 & G4Jy 1456  &          GLEAM J180139+135121 \\   
   G4Jy 863  &          GLEAM J103848$-$043115 & G4Jy 1482  &          GLEAM J182248+293131 \\   
   G4Jy 881  &          GLEAM J105517+020541 & G4Jy 1521  &          GLEAM J191739$-$453025 \\   
   G4Jy 884  &          GLEAM J105817+195203 & G4Jy 1564  &          GLEAM J194019$-$032719 \\   
   G4Jy 890  &          GLEAM J110231$-$094122 & G4Jy 1626  &          GLEAM J202807$-$152116 \\   
   G4Jy 911  &          GLEAM J111917$-$052714 & G4Jy 1639  &          GLEAM J203534$-$174522 \\   
   G4Jy 918  &          GLEAM J112610$-$191154 & G4Jy 1657  &          GLEAM J205108$-$143439 \\   
   G4Jy 924  &          GLEAM J113259+102341 & G4Jy 1658  &          GLEAM J205125+165251 \\   
   G4Jy 938  &          GLEAM J114108+011412 & G4Jy 1731  &          GLEAM J215104+121944 \\   
   G4Jy 959  &          GLEAM J114956+124719 & &  \\ 
\hline
\label{TGSSartefactstable}
\end{tabular}
\end{table*}

\subsection{Re-fitting with {\sc Aegean} }
\label{sec:refitting}

Also connected to our visual inspection, we identify radio sources that require re-fitting using {\sc Aegean}. This may be due to source-fitting not taking into account all of the relevant emission, or the original GLEAM components appearing to have inappropriate positions (given the morphology of the radio emission). Full details regarding such sources are provided in Appendix~D, where we also explain how we correct for the re-fitting process either under- or over-estimating the integrated flux-densities.

We describe the re-fitting as `unconstrained' when it corresponds to {\sc Aegean} being re-run, in its usual mode for source-fitting and characterisation, over a larger region of the sky than previously. A `re-fitted flag' of `1' is used in the G4Jy catalogue to denote GLEAM components that have been re-fitted this way. For one source the re-fitting is unconstrained but requires additional work. We use a re-fitting flag of `2' for this scenario. In the case of `priorised re-fitting', we constrain {\sc Aegean} to use pre-determined positions for the GLEAM components. The components resulting from this type of re-fitting are assigned a re-fitting flag of `3'. The total number of G4Jy sources that required re-fitting is eight, corresponding to 15 GLEAM components. The remaining 1,945 GLEAM components, that are not re-fitted, retain the default flag of `$0$'. 

However, we caution that {\sc Aegean} may still struggle to characterise the flux density for particularly-extended radio-sources. This is because -- like the source-fitting program, {\sc vsad} \citep{Condon1998}, used for both the NVSS and SUMSS catalogues -- it fits radio components using elliptical Gaussians, and so is optimised for point sources.

\subsection{Re-centroiding after manual intervention}
\label{sec:recentroiding}

Following visual inspection, we find that a total of 54~sources require their brightness-weighted centroid position (Section~\ref{sec:centroids}) to be corrected. In the majority of cases, the error was due to incorrect association of unrelated sources, and so we specify exactly which NVSS/SUMSS components should be used when re-calculating the centroid position (and integrated flux-density at $\sim$1\,GHz). Such manual intervention is also needed for extended sources with {\it well-separated} NVSS/SUMSS components, as illustrated by G4Jy~1080 in Figure~\ref{tiefighterOverlay}. (Re-centroiding would usually be unnecessary for sources that are multi-component in GLEAM but have their NVSS/SUMSS components enveloped by a single 3-$\sigma$ NVSS/SUMSS contour.) The G4Jy sources, with centroids updated for these two reasons, are assigned a `centroid flag' of `1'.

\begin{figure*}
\centering
\subfigure[G4Jy~1173]{
	\includegraphics[scale=1.1]{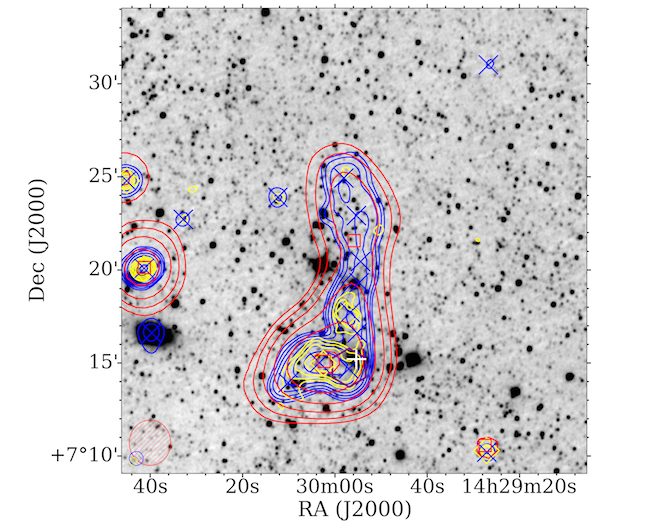}
	}  
\subfigure[G4Jy~1282]{
	\includegraphics[scale=1.1]{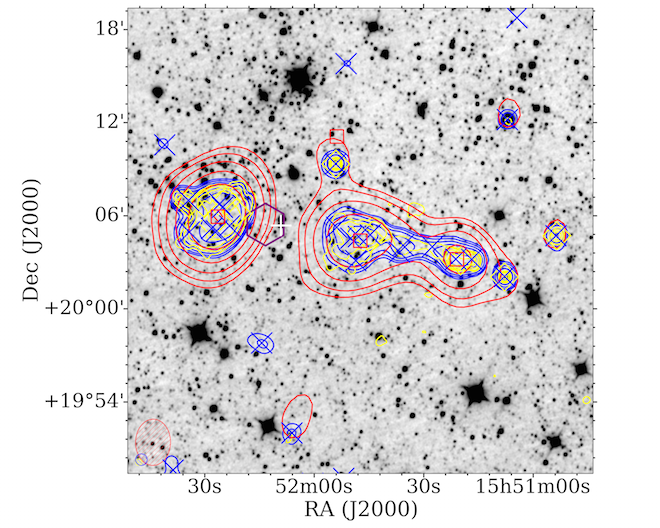}
	}  
\caption{(a) An overlay for the source G4Jy~1173, centred on the component GLEAM~J142955+072134. (b) An overlay for the source G4Jy~1282, centred on the component GLEAM~J155147+200424. Radio contours from TGSS (150\,MHz; yellow), GLEAM (170--231\,MHz; red), and NVSS (1.4\,GHz; blue), are overlaid on a mid-infrared image from {\it WISE} (3.4\,$\mu$m; inverted greyscale). For each set of contours, the lowest contour is at the 3\,$\sigma$ level (where $\sigma$ is the local rms), with the number of $\sigma$ doubling with each subsequent contour (i.e. 3, 6, 12\,$\sigma$, etc.). As
discussed in Section~\ref{sec:recentroiding}, manual re-centroiding was required for both sources shown here, due to their complex morphology. Updated centroid-positions (Section~\ref{sec:recentroiding}) are indicated by purple hexagons, and also plotted are catalogue positions from TGSS (yellow diamonds), GLEAM (red squares), and NVSS (blue crosses).}
\label{recentroidedexamples}
\end{figure*}

In addition, we note G4Jy sources with non-axisymmetric, or very-extended, emission. Regarding the former, their morphology may be indicative of radio jets interacting with an inhomogeneous environment. Alternatively, the morphology could be the result of the galaxy's radio jets being `bent backwards' as it falls into a cluster (see Paper~II). In these cases (e.g. Figure~\ref{recentroidedexamples}a) we use only the NVSS/SUMSS components that are closest to the core of the radio galaxy, as the centroid would otherwise be influenced by the geometry of the outermost regions. For extended `doubles' showing evidence of multiple knots of radio emission, we also use only the innermost NVSS/SUMSS components when re-calculating the centroid position. To reflect these two cases, we specify `2' as the centroid flag. This is applicable for seven sources, with the updated centroid-position acting as a better guide for identifying the host galaxy, as described in the next section. 

Another example of association, leading to re-centroiding, involves the intriguing morphology of GLEAM~J155147+200424 and GLEAM~J155226+200556. Both of these components have $S_{\mathrm{151\,MHz}}>4$\,Jy, and the larger overlays created for them suggest that they are part of a single object ({\bf G4Jy~1282}; Figure~\ref{recentroidedexamples}b). Indeed, this source appears as 3C~326 in the 3CRR sample \citep{Laing1983} and has been classified as an `FR II' radio galaxy. [Such a classification is used for `edge-brightened' radio galaxies, where the brightest radio-emission is located in the lobes, far from the AGN  \citep{Fanaroff1974}. Other sources, where the radio luminosity decreases with distance from the AGN, are labelled `FR I'.] Based on this morphological interpretation, the component GLEAM~J155120+200312 is added to the G4Jy Sample by association. Consequently, the NVSS components for G4Jy~1282 are re-determined manually, and used for the updated centroid position. 

Although \citet{Mauch2003} invested effort into removing image artefacts from the SUMSS catalogue, we note that some still remain amongst the cutouts for the G4Jy Sample. As a result, erroneous components were being used in the centroid calculation for some sources. We rectify this by updating the centroid position, using only reliable SUMSS components (as identified via visual inspection). The affected sources are also given a centroid flag of `1'. The remaining 1,802 sources, which did {\it not} have their centroid position updated for any reason, retain the default centroid flag of `0'. 

\subsection{Identifying the likely host-galaxy}
\label{sec:identifyhost}

Through {\sc topcat} software \citep{Taylor2005}, we obtain a subset of the AllWISE catalogue, where all objects are within 3$'$ of a centroid position belonging to the G4Jy Sample (this radius being the maximum value allowed by the `CDS Upload X-Match' facility of {\sc topcat}). We add these AllWISE positions (green plus signs, `+') to the overlays that are 10$'$ across, and initially use a white `+' to highlight the AllWISE source that is closest to the centroid position (at the centre of the overlay). We then inspect these overlays to determine whether the highlighted mid-infrared source is the likely host-galaxy for the G4Jy source in question. In doing so we also consider the errors in the centroid position (represented by an ellipse), having noted that the errors in the AllWISE positions are negligible by comparison. For 1,388 (i.e. 75\%) of the \nsources G4Jy sources, we find that the appropriate mid-infrared source has been highlighted (e.g. see Figures~\ref{artefactexamples}a--d, \ref{artefactexamples}f, and \ref{JMsources1}a). 

Conversely, 475 G4Jy sources require additional attention. For these radio sources, the nearest AllWISE source does {\it not} appear to be the host galaxy for the radio emission (or there is ambiguity), and so they are set aside for re-inspection. This is done via interactive Multi-Catalogue Visual Cross-Matching (MCVCM) software\footnote{https://github.com/kasekun/MCVCM -- This software creates overlays from the input images, and allows the user to {\it click} on catalogue positions, which are also plotted. A cross-identification tag/string (referring to the two catalogues being cross-matched) is output to a text file, as well as a flag for indicating (for example) the user's certainty as to the selection.} (Swan et al., in preparation), which allows us to {\it manually} select the most-likely host galaxy. The corresponding 10$'$ overlay is then updated, so that the white `+' highlights this selected source (e.g. see Figures~\ref{artefactexamples}e and \ref{JMsources1}b--f). The result, across the 10$'$ overlays for the full sample, is that this symbol indicates the AllWISE host-galaxy identification for the G4Jy source.

Having inspected each G4Jy source, we assign a `host flag' that corresponds to one of the following four categories: 
\begin{itemize}
\item{`i' -- a host galaxy has been identified in the AllWISE catalogue, with the position and mid-infrared magnitudes (W1, W2, W3, W4) recorded as part of the G4Jy catalogue (Section~\ref{sec:finalcatalogue}),}
\item{`u' -- it is unclear which AllWISE source is the most-likely host galaxy, due to the complexity of the radio morphology and/or the spatial distribution of mid-infrared sources (leading to ambiguity),}
\item{`m' -- identification of the host galaxy is limited by the mid-infrared data, with the relevant source either being too faint to be detected in AllWISE, or affected by bright mid-infrared emission nearby,}
\item{`n' -- no AllWISE source should be specified, given the type of radio emission involved.} 
\end{itemize}

Manual identification of the host galaxy was usually required for the multi-component radio-sources, where the geometry of the NVSS/SUMSS radio emission meant that the centroid position was more subject to error. In 37\% of such cases, the G4Jy source had a `core' indicated by a detection in 6dFGS and/or AT20G. G4Jy sources with a host galaxy in 6dFGS are noted for later analysis (Franzen et al., in preparation; White et al., in preparation), whilst those with AT20G information are explored further in a separate paper on broadband radio spectra (White et al., in preparation).

As mentioned previously, differing spatial scales of radio emission, and the fact that a single source may have multiple radio components, makes it particularly difficult to cross-match radio catalogues with data at other wavelengths (where sources typically have a singular morphology). This is complicated further by the greater density of sources seen at shorter wavelengths, leading to ambiguity when trying to identify the corresponding galaxy. Therefore, even after careful re-inspection and investigation, we cannot always determine which mid-infrared source is the `correct' host -- hence our use of the `u' flag for 129 G4Jy sources.

In some cases we find that the radio position is robust -- as suggested by the coincidence of detections from multiple radio surveys -- but the likely host-galaxy is too faint in the mid-infrared to appear in the AllWISE catalogue. This could be due to the radio source being at very high redshift, with confirmation of this requiring follow-up observations, such as optical/near-infrared spectroscopy (as discussed further in section~4.10 of Paper~II). For these situations (i.e. 126 G4Jy sources) we use the `m' flag. However, the reader should note that this label is also used for G4Jy sources that have a {\it bright} mid-infrared host that is absent from the AllWISE catalogue, due to its photometry being affected by (for example) source confusion or a diffraction spike from a nearby star.  
 
Our final host-galaxy flag, `n', is used for 2 G4Jy sources for which it is inappropriate to select a single AllWISE source, as there is no `host galaxy' to identify. Such is the case for extended radio emission associated with a nebula and a cluster relic (both of which are presented in Paper~II).

\subsubsection{Consulting the literature}

The fact that radio sources can exhibit complex and/or asymmetric morphology, coupled with the limited resolution provided by TGSS/NVSS/SUMSS (25--45$''$), prompts us to consult the literature as part of our host-galaxy identification. For details regarding individual G4Jy sources, we refer the reader to the accompanying paper, Paper~II \citep{White2020b}. Here we summarise our methods and considerations:

\begin{enumerate}
\item{We use a mixture of radio and (candidate) mid-infrared positions to search the NED and SIMBAD\footnote{http://simbad.u-strasbg.fr/simbad/} databases for existing cross-identifications. E.g., PKS~B0503$-$290 and ESO~422-G028 appear as separate entries in NED, despite referring to the same source (G4Jy~517; section~4.8 of Paper~II). The only NED cross-identification that is common to both entries is `MSH 05$-$202'.}
\item{However, we do not `blindly' use identifications from databases, but instead inspect the original images or supporting, follow-up observations ourselves (if they are published/accessible). This allows us to corroborate (or disregard) the identification, which often involves converting between B1950 and J2000 co-ordinates. E.g., 4.9-GHz radio contours \citep{Massaro2012} lead us to question the identification for G4Jy~700 (3C~198), which dates back to \citet{Wyndham1966}. See section~5.2 of Paper~II for details.}
\item{We bear in mind that many historical identifications were obtained by overlaying radio contours onto optical images, in which case they are biased against dust-obscured sources. Using our overlays (of radio contours on {\it mid-infrared} images), we consider whether there are plausible alternatives to the existing identification. If this is the case, we search for additional evidence in order to hopefully resolve the ambiguity. E.g., ATCA observations in the literature confirm our host-galaxy identification for G4Jy~1525 (B1910$-$800; see Section~\ref{sec:checkagainstJM}), which is in disagreement with \citet{Jones1992}.}
\item{For some sources, we are able to find higher-resolution ($<$25--45$''$) radio images that are presented `directly' in the literature, or are available online (e.g. cutouts from FIRST\footnote{https://third.ucllnl.org/cgi-bin/firstcutout}; Faint Images of the Radio Sky at Twenty-Centimeters; \citealt{White1997}). We look for evidence of the innermost part of any radio jets (if applicable) and, ideally, the radio-core position. E.g., FIRST reveals `triple' morphology for G4Jy~367 (3C~89), allowing us to determine the correct host amongst clustered mid-infrared sources. We find this higher-resolution radio survey useful for another 20 G4Jy sources, all of which are noted individually in Paper~II.} 
\item{Spectral-index maps are particularly valuable for our visual checks, as we expect the radio core to be easily distinguished via its flat-spectrum emission. E.g., the map provided by \citet{Safouris2009}, between 1378 and 2368\,MHz, confirms the radio-core position for G4Jy~347 (B0319$-$453; section~4.8 of Paper~II), and that it is not coincident with the `obvious', SUMSS-detected, mid-infrared source lying roughly midway between the two lobes.}
\item{Evidence of X-ray emission at the position of the host galaxy may also enable us to confirm whether or not the identification is correct. E.g., \citet{Massaro2012} find no detection of the putative host in the X-ray observation for G4Jy~700 (3C~198), throwing the existing identification into further doubt.}
\item{For cases where the host galaxy appears to be blended, faint, or affected by artefacts in the mid-infrared, we examine optical images that are at higher resolution and may be of greater depth. E.g., a SuperCOSMOS image \citep{Hambly2001} suggests that two AllWISE candidates for G4Jy~1079 (section~4.8 of Paper~II) are likely a result of the host's extended structure in the mid-infrared.}
\item{Although the result is that we have fewer mid-infrared identifications in the first version of the G4Jy catalogue, our stance is to err on the side of caution until sufficient data become available.} 
\end{enumerate}

\subsubsection{Excluding possible stars}
\label{sec:stars}


Having identified a host galaxy in the AllWISE catalogue (`host flag' = `i') for the majority of the G4Jy sources, we subsequently check that we have not mistakenly selected a mid-infrared source that is a foreground star. We do this by first applying the following {\it WISE}-colour criteria, for separating stars from galaxies: $[3.4]<10.5$\,mag, $[4.6]-[12]<1.5$\,mag and $[3.4]-[4.6] < 0.4$\,mag \citep{Jarrett2011}. This identifies 16 G4Jy sources for which the AllWISE source is a possible star, but re-inspection confirms that either the host galaxy is unambiguous or is supported by a high-resolution radio-image. If we replace the $[3.4]-[4.6]$ criterion with one that employs the W4 band, i.e. $[12] - [22]<1.2$\,mag \citep{Jarrett2011}, we select {\it zero} AllWISE host-galaxies for re-inspection. Hence, we are satisfied that none of the mid-infrared sources in the G4Jy catalogue (Section~\ref{sec:finalcatalogue}) are stars.

For some sources where we are {\it uncertain} as to the host-galaxy identification, this may be due to obscuration by stars. This is particularly problematic for G4Jy sources at low Galactic latitude, and is borne in mind during our visual inspection and checks against the literature.

For the interested reader, note that the distribution of G4Jy sources in {\it WISE} colour-colour space will be presented in Paper~III, along with other multi-wavelength analysis (White et al., in preparation).

\section{The GLEAM 4-Jy catalogue}
\label{sec:finalcatalogue}

This section summarises information in the G4Jy catalogue that supplements 307 columns from the parent, GLEAM extragalactic catalogue \citep[EGC;][]{HurleyWalker2017}. For a full list of the 76 new columns that we provide, and first-row entries as examples, see Table~\ref{tab:catalogueexample} in Appendix~E.


\subsection{Naming of the G4Jy sources}

Having identified which GLEAM components are associated with each other (Section~\ref{sec:visualinspection}), and which additional GLEAM components are to be included in the G4Jy Sample (Section~\ref{sec:completeness}), we sort the catalogue in order of increasing R.A. The `ncmp\_GLEAM' column is added to indicate the number of GLEAM {\it components} that correspond to each {\it source}. We then use simple numbering as our naming scheme: `G4Jy~1', `G4Jy~2', `G4Jy~3', etc. This both allows a short-hand way of referring to sources, and avoids `hard-coding' a co-ordinate position that may later be refined. Similarly, we use `A', `B', `C', etc. to label individual GLEAM components belonging to multi-component sources. For example, GLEAM~J000456+124810 is the eastern radio-lobe of G4Jy~7 and can be referred to as `G4Jy~7B'.

\subsection{Morphology}

The morphology of the source (Section~\ref{sec:morphology}) is determined through visual inspection, and is based on NVSS/SUMSS contours, or TGSS contours where coverage allows. Although literature checks uncover radio images of higher resolution for some sources (see Paper~II for details), we do not change the morphology label as we wish these to be consistent across the entire sample. Furthermore, we note that some `doubles' may actually be `core-jet' sources \citep[e.g.][]{Kellermann1981,Pearson1988}, where the radio-jet emission is one-sided. However, we do not have sufficient resolution to confirm these, and so we apply Occam's razor and leave the morphology label as `double'. We expect many of these morphology labels -- the `singles' especially -- to be updated as better-resolution ($<25$--$45''$) radio-images come to light.

\subsection{Information at $\sim$1\,GHz}

The `Freq' column indicates whether NVSS (1400\,MHz) or SUMSS (843\,MHz) has been considered for the source in question. Alongside this, we provide the number of associated NVSS/SUMSS components (`ncmp\_NVSSorSUMSS'), the summed flux-density across these components (`S\_NVSSorSUMSS'), and the brightness-weighted centroid position (`centroid\_RAJ2000', `centroid\_DEJ2000') based on these components. The `centroid flag' column indicates whether the centroid position is from the original, automated calculation (centroid\_flag = `0'; see Section~\ref{sec:centroids}), or has been updated following manual intervention (centroid\_flag = `1' or `2'; see Section~\ref{sec:recentroiding}). The `confusion flag' is based upon visual inspection (Section~\ref{sec:morphology}), with G4Jy sources potentially having their GLEAM flux-densities affected by unrelated radio-sources (confusion\_flag = `1'; e.g. G4Jy~935 in Figure~\ref{JMsources1}d) or not (confusion\_flag = `0'; e.g. G4Jy~1628 in Figure~\ref{JMsources1}f).

\subsubsection{Angular sizes}
\label{sec:angsizesforcatalogue}

We provide an estimate of the angular size at $\sim$1\,GHz (`angular\_size') but warn the user that these values are only to give an {\it indication} of the extent of the radio emission. This is because the apparent size is affected by resolution (leading to over-estimation) and projection (leading to under-estimation). Investigating the orientation of the G4Jy sources is beyond the scope of this work, but would need to be borne in mind when using the angular sizes to estimate true, physical sizes. In addition, the angular-size distribution is complicated by sources with bent-tail morphology (see Figure~\ref{recentroidedexamples}a and section~4.7 of Paper~II).

For G4Jy sources that have a single component in NVSS/SUMSS, we adopt (where possible) the deconvolved major-axis measurement from the respective catalogue (i.e. the MajAxis value from NVSS, or the major\_axis\_arcsec\_afterdeconvolution value from SUMSS). For single-NVSS-component sources, we inherit the limit associated with the MajAxis value and place this in our `angular\_size\_limit' column. Meanwhile, the SUMSS catalogue does not provide a deconvolved major-axis measurement for unresolved sources. For such cases, we instead set the angular size equal to the major\_axis\_arcsec value -- this being the original, fitted value dictated by the survey's spatial resolution -- and accompany this with angular\_size\_limit = `$<$'. The inequality therefore indicates which of our angular-size estimates should be interpreted as upper limits. For the remaining angular sizes presented in the G4Jy catalogue, the angular\_size\_limit column is left blank. 

For G4Jy sources that are multi-component at $\sim$1\,GHz, we use the largest angular-separation between associated NVSS/SUMSS components as our angular-size estimate (see Section~\ref{sec:sizeanalysis} for the full-sample distribution). However, again, this value should be taken as a guide, because the fitted NVSS/SUMSS positions may not fully describe the spatial extent of the GLEAM emission. Users of the G4Jy catalogue may instead wish to consider the semi-major-axis measurements output by {\sc Aegean} (e.g. the a\_wide and a\_151 columns), but then the low spatial-resolution of GLEAM becomes an issue, as these angular sizes are not deconvolved. The reasons why we do not use TGSS positions to calculate angular sizes are that this catalogue: (i) is biased towards compact emission, (ii) does not provide coverage for all G4Jy sources, and (iii) contains artefacts around numerous bright radio-sources (Section~\ref{sec:TGSSartefacts}).

\subsection{Mid-infrared data for the host galaxies}

Alongside our visual inspection (Section~\ref{sec:visualinspection}) and extensive checks against the literature (sections~4 and 5 of Paper~II), we use the `host\_flag' to indicate whether or not we are able to identify the host galaxy of the radio emission in the mid-infrared (Section~\ref{sec:identifyhost}). For G4Jy sources that {\it are} identified (host\_flag = `i'), we provide the AllWISE name, position, and mid-infrared magnitudes (and errors) from the AllWISE catalogue. This information is being used for collating additional multi-wavelength data for the G4Jy Sample and subsequent analysis (White et al., in preparation).


\subsection{Total, integrated GLEAM flux-densities}
\label{sec:totalGLEAMfluxdensities}

For each of the 20 sub-band measurements provided by the GLEAM Survey, we calculate the {\it total}, integrated flux-density, summed over all of the GLEAM components associated with a particular G4Jy source. The errors in these total flux-densities are determined by adding in quadrature (per sub-band) the integrated flux-density errors for the individual GLEAM components. If the G4Jy source is single-component in GLEAM, these `total' columns are simply a repeat of the integrated flux-densities (and errors) for that single GLEAM component. Note that it is the total, integrated flux-density at 151\,MHz (`total\_int\_flux\_151') that must exceed 4\,Jy for a radio source to be listed in the G4Jy Sample.

Furthermore, we remind the reader that some of the individual, integrated flux-densities provided in the G4Jy catalogue do not appear in the EGC. They are instead the result of re-fitting (and re-scaling, in some cases), as described in Section~\ref{sec:refitting} and Appendix~D. We use the `refitted\_flag' to indicate for which GLEAM components this applies.

\subsection{Four sets of spectral indices}
\label{sec:fouralphas}

For the majority of GLEAM components in the G4Jy Sample, the spectral index fitted over the GLEAM band (`GLEAM\_alpha') is inherited from the EGC. In the time since the publication of \citet{HurleyWalker2017}, we noticed that the spectral-index errors quoted in the parent catalogue were over-estimated by a factor of 5. This has been corrected in a new version of the EGC (v2), available online through VizieR\footnote{http://vizier.u-strasbg.fr/viz-bin/VizieR?-source=VIII/100}. We also include the updated error and reduced-$\chi^2$ columns in the G4Jy catalogue, renaming them `err\_GLEAM\_alpha' and `reduced\_chi2\_GLEAM\_alpha', respectively. 

GLEAM components that were re-fitted for the G4Jy catalogue (refitted\_flag $>0$) have a newly-calculated GLEAM\_alpha value. For this, we fit a power-law spectrum to the integrated flux-densities for multiple sub-bands, in the same way as done for GLEAM components in the EGC. Hence, we also determine consistent errors and reduced-$\chi^2$ values.

Since we are interested in the total GLEAM emission associated with each G4Jy source, we also fit a GLEAM-only spectral index using the total (i.e. summed) integrated flux-densities (Section~\ref{sec:totalGLEAMfluxdensities}). This we refer to as `G4Jy\_alpha', and it will differ from GLEAM\_alpha if the G4Jy source is multi-component in GLEAM. Then, in line with the parent catalogue \citep{HurleyWalker2017}, we mask the GLEAM\_alpha and/or G4Jy\_alpha values in the G4Jy catalogue wherever the corresponding reduced-$\chi^2$ value is $>1.93$. 

In addition, we provide the spectral index calculated using (total) $S_{\mathrm{151\,MHz}}$ and $S_{\mathrm{1400\,MHz}}$ (`G4Jy\_NVSS\_alpha'), for sources at Dec.~$\geq-39.5^{\circ}$, and using $S_{\mathrm{151\,MHz}}$ and $S_{\mathrm{843\,MHz}}$ (`G4Jy\_SUMSS\_alpha'), for sources at Dec.~$<-39.5^{\circ}$. These indices (and their errors) are provided in separate columns, as extrapolating to a common frequency (e.g. 1\,GHz) may obscure the different systematics of the two surveys, and/or conflate potentially-different distributions of spectral curvature. 

We present the mean and median values for each of these four sets of spectral indices in Table~\ref{tab:alphastatistics}. Due to the masking involved for GLEAM\_alpha and G4Jy\_alpha, we also note the number of GLEAM components or G4Jy sources (respectively) for which the spectral index is provided in the catalogue. We direct the reader to Section~\ref{sec:indexanalysis} for an initial discussion of these spectral indices, with further analysis to appear in Papers~III and IV (White et al., in preparation).

\begin{table*}
\centering 
\caption{The mean and median spectral-index, $\alpha$, for each of the four sets of spectral indices provided in the G4Jy catalogue (Section~\ref{sec:fouralphas}). `Number' refers to the number of G4Jy sources for which the statistics apply, except in the case of GLEAM\_alpha, where it is the number of GLEAM components. }
\begin{tabular}{@{}lrrcc@{}}  
 \hline
$\alpha$ name & Number & Frequencies used & Mean   &   Median  \\
 \hline
GLEAM\_alpha &	1670 & 72--231\,MHz &	$-$0.824 $\pm$	0.004 &	$-$0.829 $\pm$	0.006	\\
G4Jy\_alpha &	1603 & 72--231\,MHz &	$-$0.822 $\pm$	0.004	& $-$0.829  $\pm$	0.006	\\
G4Jy\_NVSS\_alpha &	1437 & 151 and 1400\,MHz &	$-$0.786 $\pm$ 	0.005 &	$-$0.786  $\pm$	0.006	\\
G4Jy\_SUMSS\_alpha & 426 & 151 and 843\,MHz &	$-$0.745  $\pm$	0.009	& $-$0.740	$\pm$  0.012	\\
\hline
\label{tab:alphastatistics}
\end{tabular}
\end{table*}

\section{Sample completeness}
\label{sec:completeness}

In order to do high-impact science using the G4Jy Sample, and determine robust statistics on (for example) radio-galaxy properties, it is crucial that the sample is {\it complete}. At this point, we note that there are two situations where we could be missing or misclassifying extended radio-sources. Firstly, our visual inspection, and subsequent investigation, may still miss very extended `double' radio-galaxies, where the individual lobes are separated by more than 30$'$. Secondly, and of more concern, is the possibility of missing a source that has a total flux-density of $S_{\mathrm{151\,MHz}}>4$\,Jy but is resolved into two or more components such that each component has $S_{\mathrm{151\,MHz}}<4$\,Jy. These components would therefore not be included in the initial selection, using the GLEAM catalogue (Section~\ref{sec:sampleselection}). To combat this, we perform checks against numerous bright, radio-source samples in the literature (Section~\ref{sec:literaturecrosschecks}), and also apply criteria designed to identify candidate, extended radio-sources (Section~\ref{sec:internalmatch}). Another important factor, in terms of completeness, is the flux-density scale that we use when defining our sample at a particular frequency. This is considered in Section~7.3, before providing a brief summary of the G4Jy Sample in Section~\ref{sec:summary}.

\subsection{Literature searches for extended sources}
\label{sec:literaturecrosschecks}

We search for missing, multi-component sources by checking the overlap of the GLEAM catalogue with five existing samples: `Southern extragalactic radio-sources' \citep{Jones1992}, `Radio galaxies of the local Universe' \citep{vanVelzen2012}, the original 2-Jy sample \citep{Wall1985}, `Giant radio galaxies' \citep{Malarecki2015}, and the 3CRR sample \citep{Laing1983}. An overview of these samples is presented in Table~\ref{comparisonsamples}. The key part of this work is creating and inspecting several hundreds of {\it extra} overlays, to avoid any assumptions as to what we may expect the radio sources to look like in GLEAM.

\begin{sidewaystable}
\caption{Selection criteria for previous radio-source samples, which we use to check the completeness of the G4Jy Sample (Section~\ref{sec:literaturecrosschecks}). `MRC' is the abbreviation for the Molonglo Reference Catalogue of Radio Sources \citep{Large1981}. The giant radio-galaxies (GRGs) making up the sample assembled by \citet{Malarecki2015} were originally identified in the MRC ($>$\,0.7\,Jy at 408\,MHz) and SUMSS (see Section~\ref{sec:SUMSSdata}).}
\begin{tabular}{@{}lccccc@{}}
 \hline
Sample  & Number & Frequency & Flux-density & Sky coverage & Other  \\
 (reference)& of sources & /\,MHz &  limit  /\,Jy &  &  \\
 \hline
Southern extragalactic radio-sources  & 384 & 408 & 0.7 & $-85^{\circ} < \delta< -30^{\circ}$,  & Flagged as extended or \\
\citep{Jones1992} &&&  & $|b| >3^{\circ}$ & multi-component in the MRC \\
Radio galaxies of the local Universe  & 575  & 1400 & 0.213 & All sky, & $K_{\mathrm{S}} < 11.75$, $E(B-V)<1$, \\
\citep{vanVelzen2012} & & 843 & 0.291 & $|b| >5^{\circ}$ to $|b| >8^{\circ}$ & detected in the $H$ band \\
The original 2-Jy sample  & 233 & 2700 & 2.0 & All sky,  &  \\
\citep{Wall1985} &&&& $|b| >10^{\circ}$ & \\
Giant radio galaxies & 19 & -- & -- & $-85^{\circ} < \delta< -28^{\circ}$,    & Projected linear size $>$ 700\,kpc,  \\
\citep{Malarecki2015} & & && $|b| >3^{\circ}$ & $z<0.15$ \\
The 3CRR sample & 173 & 178 & 10.9$^{\ref{3CRRlimit}}$ & $\delta \geq 10^{\circ}$,  &  \\
\citep{Laing1983} &&&& $|b| \geq 10^{\circ}$ & \\
\hline
\label{comparisonsamples}
\end{tabular}
\end{sidewaystable}

As a result of these cross-checks, we add a total of 15 sources to the G4Jy Sample that otherwise would not have been included. However, extended sources  -- such as B1302$-$492 and B1610$-$608 in \citet{Jones1992} -- are still absent because they lie in one of the GLEAM catalogue's masked regions (notably, the Galactic Plane or the region surrounding Centaurus A). Therefore, we define in Section~\ref{sec:summary} the region over which the sample is complete.

\begin{table*}
\centering
\caption{Radio sources that were missing from the G4Jy Sample, based on the initial selection (Section~\ref{sec:sampleselection}), but are now included as a result of cross-checks against the samples listed in Table~\ref{comparisonsamples} (Section~\ref{sec:literaturecrosschecks}).  Including these radio galaxies gives a total of \nsources G4Jy sources in the sample.
}
\begin{center}
\begin{threeparttable}
\begin{tabular}{@{}llccc@{}}
 \hline
Other name & Source & Corresponding  & Individual & Total \\
 & & GLEAM components   & $S_{\mathrm{151\,MHz}}$\,/\,Jy  &   $S_{\mathrm{151\,MHz}}$\,/\,Jy  \\
 \hline
GIN 049 & G4Jy~131 &  GLEAM~J011257+153048  & $2.73\pm0.04$ & $4.62\pm0.06$  \\
 &  &  GLEAM~J011303+152654  & $1.89\pm0.04$ &   \\ 
B0211$-$479 & G4Jy~234 & GLEAM~J021305$-$474112  & $3.51\pm0.02$ &  $7.27\pm0.03$  \\
  &        & GLEAM~J021311$-$474615   & $3.76\pm0.02$ &    \\ 
NGC~1044 & G4Jy~285  & GLEAM~J024103+084523  & $1.86\pm0.03$ & $4.18\pm0.11$ \\
    &    & GLEAM~J024107+084452  & $2.32\pm0.10$ &  \\ 
GIN~190  & G4Jy~475 & GLEAM~J043409$-$132250  & $2.84\pm0.02$  & $4.07\pm0.05$ \\
  &   & GLEAM~J043415$-$132717  & $1.22\pm0.05$ &   \\  
B0523$-$327 & G4Jy~543 & GLEAM~J052522$-$324121   & $3.44\pm0.02$ & $5.81\pm0.03$ \\
 &         & GLEAM~J052531$-$324357   & $2.37\pm0.02$ &   \\ 
B0546$-$329 & G4Jy~579 & GLEAM~J054825$-$330128  & $2.43\pm0.03$ &  $4.35\pm0.04$ \\
  &        & GLEAM~J054836$-$325458  & $1.91\pm0.02$ &    \\  
PKS~B0616$-$48$^1$ & G4Jy~604 & GLEAM~J061740$-$485426  & $1.01\pm0.04$ & $5.96\pm0.06$ \\
 & & GLEAM~J061803$-$484610  &  $2.21\pm0.02$ &  \\
& & GLEAM~J061812$-$484257 & $2.74\pm0.03$ & \\  
B1137$-$463 & G4Jy~935 & GLEAM~J113943$-$464032  &$1.54\pm0.03$ & $4.12\pm0.04$ \\
   &       & GLEAM~J113956$-$463743   & $2.58\pm0.02$&   \\ 
B1323$-$271$^2$   & G4Jy~1067  &    GLEAM~J132606$-$272641  & $2.81\pm0.03$& $6.27\pm0.05$ \\
   &       &    GLEAM~J132616$-$272632  &$3.46\pm0.04$ &  \\  
PKS~B1834+19   & G4Jy~1496 &  GLEAM J183626+193946 &$2.34\pm0.21$ &  $5.64\pm0.27$ \\
   &    &  GLEAM J183640+194318 & $1.57\pm0.11$&   \\   
   &    &  GLEAM J183649+194105  & $1.73\pm0.13$&  \\   
B1910$-$800 &  G4Jy~1525 & GLEAM~J191905$-$795737  &$3.22\pm0.06$ &  $5.64\pm0.09$ \\
   &       & GLEAM~J191931$-$800128    &$2.42\pm0.07$ &   \\ 
B2026$-$414 & G4Jy~1628 & GLEAM~J202932$-$411755  &$3.58\pm0.03$ &  $5.87\pm0.05$ \\
  &        & GLEAM~J202940$-$412011   & $2.29\pm0.03$ &    \\ 
IC~1347 & G4Jy~1670 & GLEAM~J210135$-$131754  & $3.39\pm0.04$ &  $6.86\pm0.05$ \\
  &  & GLEAM J210154$-$131850  & $3.47\pm0.04$ &   \\  
B2147$-$555 & G4Jy~1732 & GLEAM~J215122$-$552139  & $2.69\pm0.02$ &  $7.23\pm0.05$ \\
 &         & GLEAM~J215123$-$552604  & $1.47\pm0.03$ &    \\
  &        & GLEAM~J215133$-$551636   & $3.07\pm0.04$ &   \\ 
 B2151$-$461 & G4Jy~1741 & GLEAM~J215415$-$455319  & $3.17\pm0.02$ &  $4.23\pm0.03$ \\
  &         & GLEAM~J215435$-$454954   & $1.07\pm0.02$ &   \\ 
\hline
\label{missingsources1}
\end{tabular}
\vspace{-3mm}
\begin{tablenotes}{\footnotesize 
\item{$^1$2MASX~J06181305$-$4844580 and $^2$ESO~509-G003 in \citet{vanVelzen2012}.}}
\end{tablenotes}
\end{threeparttable}
\end{center}
\label{missingsources1}
\end{table*}

\subsubsection{Cross-check using \citet{Jones1992}}
\label{sec:checkagainstJM}

\begin{figure*}
\centering
\subfigure[G4Jy~234 (B0211$-$479)]{
	\includegraphics[scale=1.1]{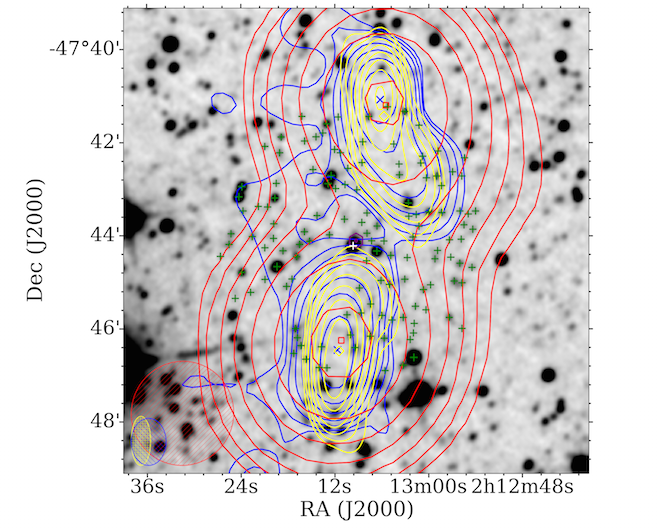}
	}
\subfigure[G4Jy~543 (B0523$-$327)]{
	\includegraphics[scale=1.1]{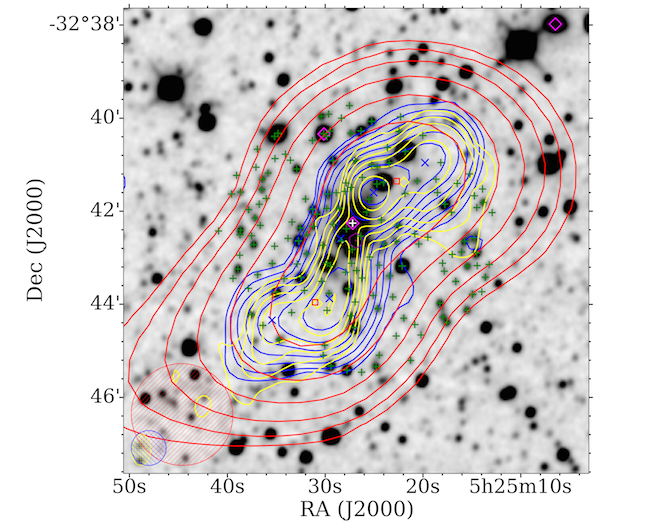}
	}
\subfigure[G4Jy~579 (B0546$-$329)]{
	\includegraphics[scale=1.1]{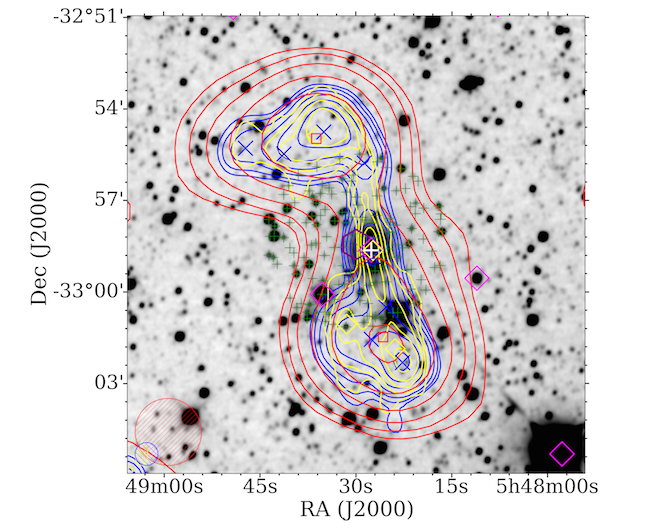}
	} 
\subfigure[G4Jy~935 (B1137$-$463)]{
	\includegraphics[scale=1.1]{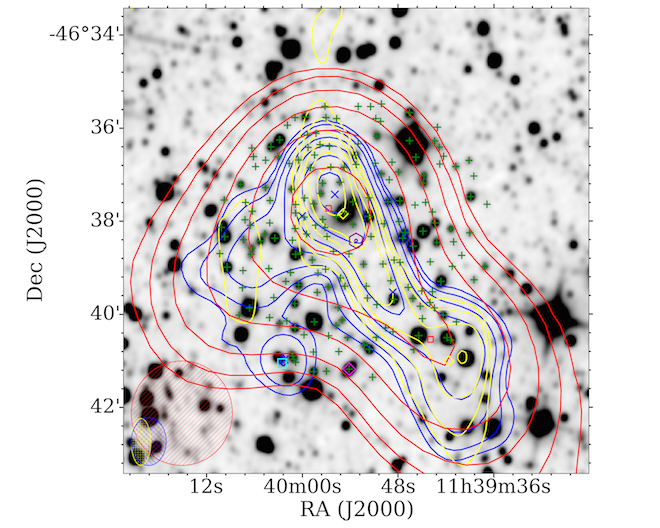}
	}
\subfigure[G4Jy~1525 (B1910$-$800)]{
	\includegraphics[scale=1.1]{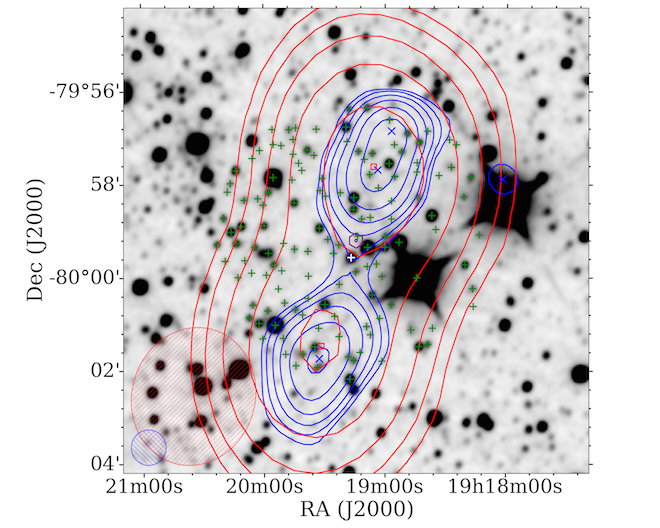}
	}
\subfigure[G4Jy~1628 (B2026$-$414)]{
	\includegraphics[scale=1.1]{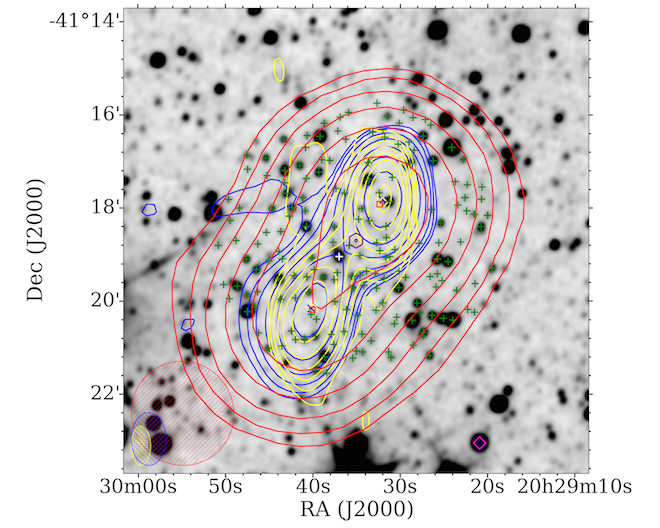}
	} 
\caption{Overlays for six G4Jy sources that were added to the G4Jy Sample following a cross-check against \citet{Jones1992} [Section~\ref{sec:checkagainstJM}]. The datasets, contours, symbols, and beams are the same as those used for Figure~\ref{tiefighterOverlay}, but where blue contours, crosses, and ellipses correspond to NVSS {\it or} SUMSS. In addition, positions from AllWISE are indicated by green plus signs, with host galaxies highlighted in white. }
\label{JMsources1}
\end{figure*}

\begin{figure*}
\centering
\subfigure[G4Jy~1732 (B2147$-$555)]{
	\includegraphics[scale=1.1]{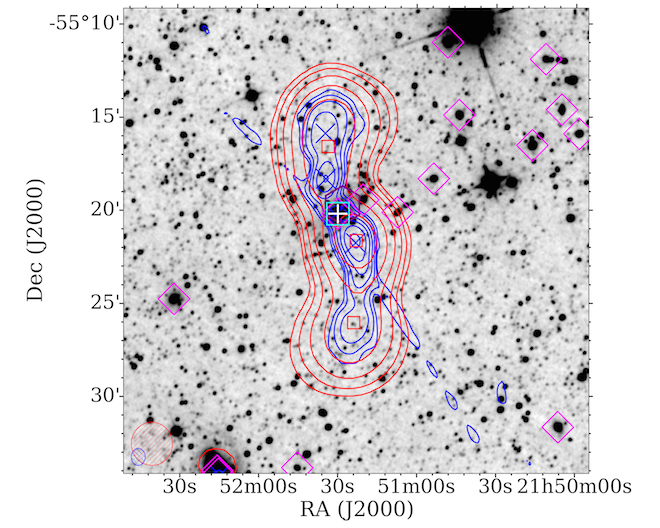}
	} 
\subfigure[G4Jy~1741 (B2151$-$461)]{
	\includegraphics[scale=1.1]{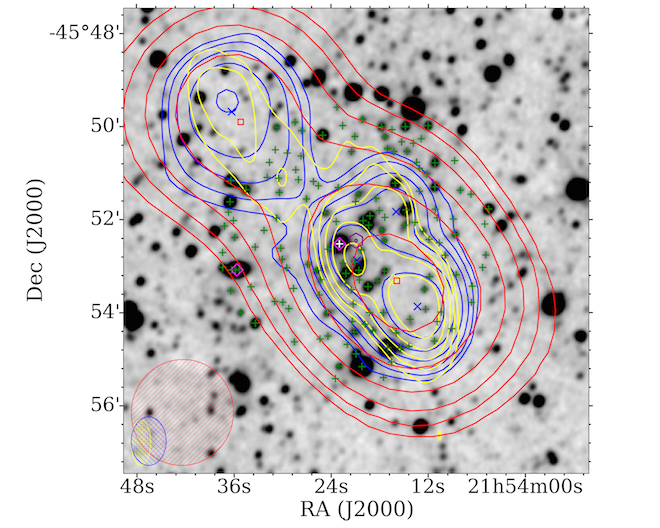}
	} 
\caption{Overlays for two more G4Jy sources that were added to the G4Jy Sample following cross-checks against \citet{Jones1992} [Section~\ref{sec:checkagainstJM}]. The datasets, contours, symbols, and beams are the same as those used for Figure~\ref{JMsources1}. }
\label{JMsources2}
\end{figure*}

Following the comparison with \citet{Jones1992}, we add eight sources to the G4Jy Sample: B0211$-$479, B0523$-$327, B0546$-$329, B1137$-$463, B1910$-$800, B2026$-$414, B2147$-$555, and B2151$-$461 (see Table~\ref{missingsources1}, and Figures~\ref{JMsources1}--\ref{JMsources2}). One of these is an S-shaped source (G4Jy~543; section~4.4.2 of Paper~II), another is a head-tail galaxy (G4Jy~935; section~4.7.2 of Paper~II), and another is a known GRG (G4Jy~1525; section~4.8 of Paper~II). As expected, all of the GLEAM components associated with these sources are individually $< 4$\,Jy at 151\,MHz, but sum to $> 4$\,Jy for their respective sources. In the case of B1910$-$800 (the GRG, G4Jy~1525), our mid-infrared identification differs from the optical position provided by \citet{Jones1992}. Our identification of an obscured host-galaxy is supported by ATCA observations of a radio core at this position \citep{Subrahmanyan1996,Saripalli2005}. 

\subsubsection{Cross-check using \citet{vanVelzen2012}}
\label{sec:checkagainstVV}

\begin{figure*}
\centering
\subfigure[G4Jy~131 (GIN~049)]{
	\includegraphics[scale=1.1]{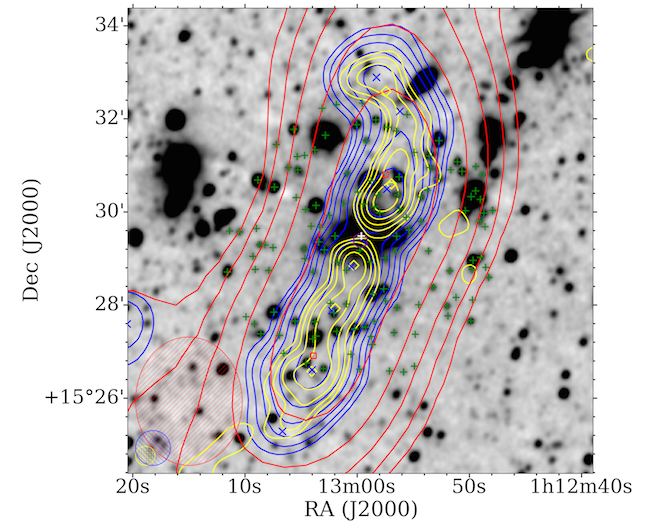}
	}
\subfigure[G4Jy~475 (GIN~190)]{
	\includegraphics[scale=1.1]{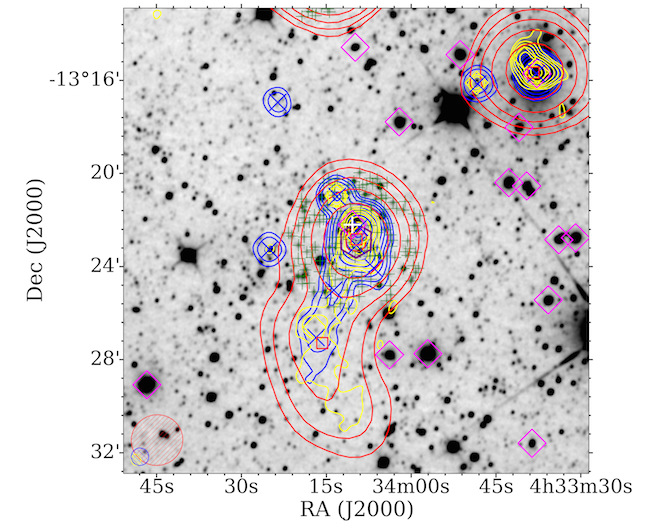}
	}
\subfigure[G4Jy~604 (PKS~B0616$-$48)]{
	\includegraphics[scale=1.1]{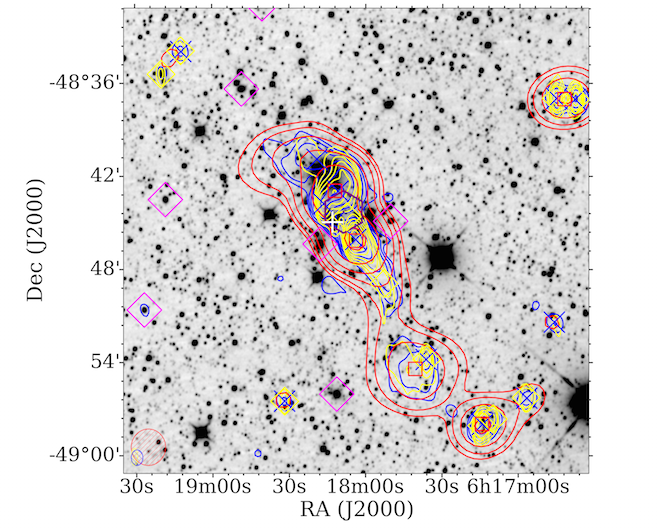}
	} 
\subfigure[G4Jy~1067 (B1323$-$271)]{
	\includegraphics[scale=1.1]{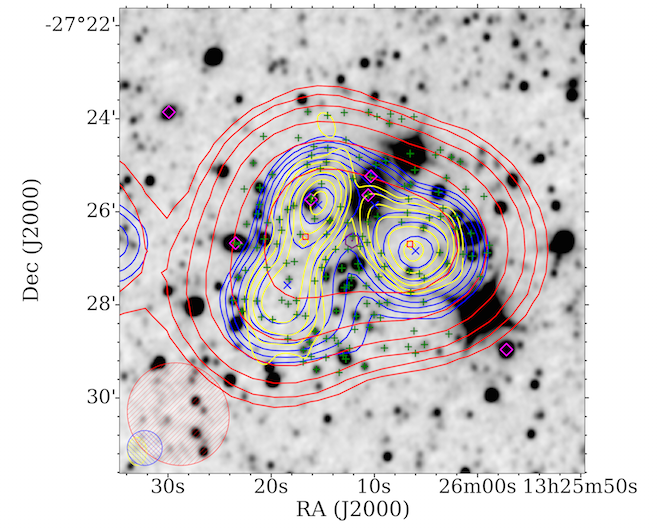}
	}
\subfigure[G4Jy~1496 (PKS~B1834+19)]{
	\includegraphics[scale=1.1]{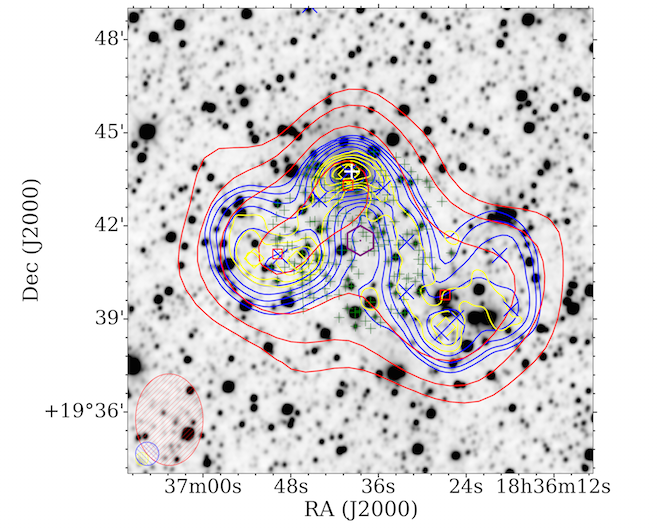}
	}
\subfigure[G4Jy~1670 (IC~1347)]{
	\includegraphics[scale=1.1]{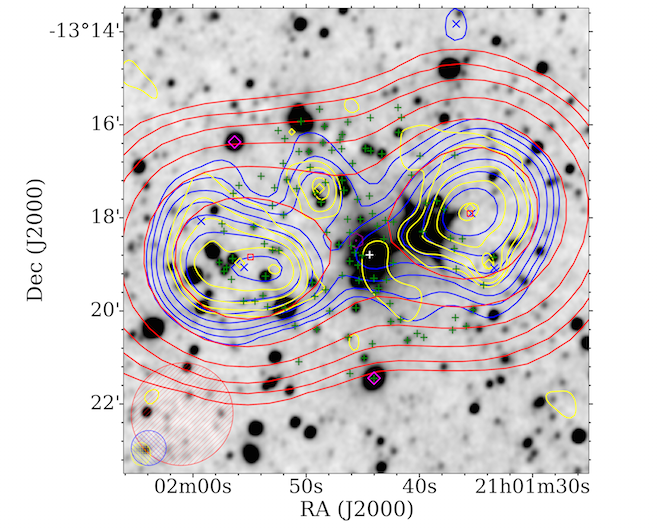}
	}
\caption{Overlays for six G4Jy sources that were added to the G4Jy Sample following a cross-check against \citet{vanVelzen2012} [Section~\ref{sec:checkagainstVV}]. The datasets, contours, symbols, and beams are the same as those used for Figure~\ref{JMsources1}. }
\label{VVsources}
\end{figure*}

Next we cross-check our sample against radio galaxies of the local Universe, as compiled by \citet{vanVelzen2012}. Doing so reveals that we need to add seven sources to our sample: GIN~049, NGC~1044, GIN~190, PKS~B0616$-$48\footnote{We agree with \citet{vanVelzen2012} that all of the extended radio-emission is associated with a single source, as originally identified by \citet{Bajaja1970}. In doing so, we disagree with \citet{Jones1992}, who interpret the morphology of PKS~B0616$-$48 as two unrelated sources. This is understandable given the 843-MHz image that they use.}, B1323$-$271, PKS~B1834+19, and IC~1347 (Table~\ref{missingsources1}). Six of these sources are presented in Figure~\ref{VVsources}, and we refer the reader to figure~17 of Paper~II for the seventh source. The latter is NGC~1044 (G4Jy~285), which has unusual, diffuse, low-frequency emission nearby. GIN~190 (G4Jy~475) is a head-tail galaxy (section~4.7.2 of Paper~II), and B1323$-$271 (G4Jy~1067) and PKS~B1834+19 (G4Jy~1496) are both WAT radio-galaxies (section~4.7.1 of Paper~II).  

As part of our comparison with the catalogue produced by \citet{vanVelzen2012}, we pay closer attention to sources where we significantly disagree as to the flux density measured using NVSS or SUMSS components. In section~4.7.2 of Paper~II we detail discrepancy with respect to G4Jy~325, and we also note that for IC~4296 (G4Jy~1080 in Figure~\ref{tiefighterOverlay}) they measure $S_{\mathrm{1.4\,GHz}}=2.42$\,Jy. This is the total flux-density when summing over the NVSS components for the inner jets, but not including the NVSS components associated with the well-separated lobes. We do include the latter, and calculate $S_{\mathrm{1.4\,GHz}}=4.91$\,Jy over 23 NVSS components. 

\begin{figure}
\centering
\includegraphics[scale=1.1]{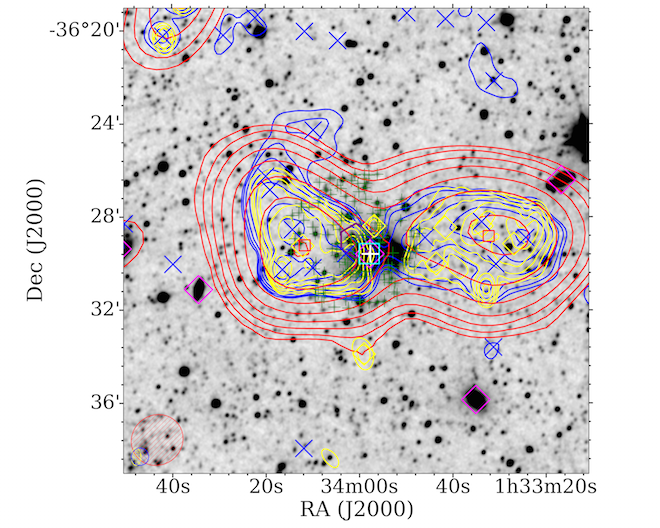}
\caption{An overlay for G4Jy~171 (Section~\ref{sec:checkagainstVV}), where the datasets, contours, symbols, and beams are the same as those used for Figure~\ref{tiefighterOverlay}. In addition, positions from AllWISE are indicated by green plus signs, with the host galaxy highlighted in white. }
\label{updatedcomponents}
\end{figure}

Comparing our NVSS/SUMSS flux-densities also highlighted that we had used the wrong number of components for NGC~253 (G4Jy~86 in figure~3a of Paper~II) and NGC~612 (G4Jy~171 in Figure~\ref{updatedcomponents}). We re-calculate their flux density and brightness-weighted centroid position, and duly update the centroid flags to `1' (Section~\ref{sec:recentroiding}). 

\subsubsection{Cross-check using \citet{Wall1985}}
\label{sec:checkagainst2Jy}

Our cross-check against the 2-Jy sample \citep{Wall1985} proved to be unfruitful, in terms of identifying extended radio-sources that may have been missed. All radio sources found to have multiple components in GLEAM were already in the G4Jy Sample via the initial selection (Section~\ref{sec:sampleselection}). This is unsurprising given that $S_{\mathrm{2.7\,GHz}}=2$\,Jy corresponds to $S_{\mathrm{151\,MHz}}=15$\,Jy, assuming a standard power-law function ($S_{\nu} \propto \nu^{\alpha}$) with spectral index, $\alpha = -0.7$. Nonetheless, the extra inspection was performed in case any of these bright sources showed evidence of fainter, remnant radio-emission within 30$'$ (since our larger overlays are 1\arcdeg\ across).

\subsubsection{Cross-check using \citet{Malarecki2015}}
\label{sec:checkagainstGRGs}

Our fourth cross-check is against a sample of 19 GRGs, used by \citet{Malarecki2015} to trace large-scale structure. They select this sample on the basis of proximity and size, and so these radio galaxies may be both brighter than 4\,Jy and resolved into multiple GLEAM components. Upon inspection we find that either the GRG is already in the G4Jy Sample, or it has a total, integrated flux-density at 151 MHz that is below the 4-Jy threshold. However, we question whether B0703$-$451 (G4Jy~641) is truly a GRG, due to insufficient evidence regarding its host-galaxy identification (see section~5.2 of Paper~II). We also do not consider B0707$-$359 (G4Jy~644) to be a GRG, as its projected, linear size is $<1$\,Mpc [instead satisfying the $>0.7$\,Mpc criterion that \citet{Malarecki2015} use to define a GRG].

\subsubsection{Cross-check using \citet{Laing1983}}
\label{sec:checkagainst3CRR}

Finally, we look at where the 3CRR sample \citep{Laing1983} overlaps with GLEAM, which corresponds to sources at $10^{\circ}<$\,Dec.\,$< 30^{\circ}$. Within this declination range are 79 3CRR sources, 67 of which are already in the G4Jy Sample. These are analysed in further detail in Section~7.3. The other 12 3CRR sources, listed in Table~\ref{absent3CRRsources}, are absent from our sample due to the GLEAM data not being of sufficient quality for obtaining 20 sub-band measurements. With the exception of 3C~433 (GLEAM~J212344+250412), these sources lie in masked regions of the EGC. Meanwhile, 3C~433 is not included in the G4Jy Sample because the source resides in an area of the sky ($-18.3^{\circ} < b < -10.0^{\circ}$, $65.4^{\circ} < l < 81.1^{\circ}$, Dec.\,$<30.0^{\circ}$) that is difficult to calibrate, due to the influence of Cygnus A [at R.A. = 19:59:28.36, Dec. = +40:44:02.1 (J2000)]. Although the rms noise is too high for the point-spread function to be characterised at each of GLEAM's 20 sub-bands, the noise in the wide-band image (170--231\,MHz) is sufficiently low for wide-band measurements (and the source-fitting that follows). Hence, the GLEAM catalogue contains 363 components in this region that have wide-band flux-densities {\it but no sub-band flux-densities} (plus another 7 components with negative-value $S_{\mathrm{151\,MHz}}$). This means that there may be additional sources brighter than 4\,Jy at 151\,MHz (our sub-band of interest). We take this into account (in Section~\ref{sec:summary}) by considering our completeness over the EGC footprint {\it minus} the region defined above.



\begin{table}
\centering 
\caption{A list of 3CRR sources \citep{Laing1983} that are not in the G4Jy Sample, despite being at Dec.\,$< 30^{\circ}$. Their absence is due to each of them having poor-quality data in the GLEAM Survey, and so -- with the exception of 3C~433 -- the region in which they lie is masked \citep{HurleyWalker2017}. An explanation of why 3C~433 is present in the GLEAM catalogue, yet absent from the G4Jy Sample, can be found in Section~\ref{sec:checkagainst3CRR}. Below, we use `Cen A' as shorthand for `Centaurus A'.}
\begin{tabular}{@{}lcc@{}}
\hline
Source  & B1950 name & Reason for absence  \\
\hline
3C 274 & B1228+126 & Virgo A (see Section~\ref{sec:AteamOrion}) \\
3C 284 & B1308+277 & Sidelobe reflection of Cen A \\
3C 287 & B1328+254 & Sidelobe reflection of Cen A \\
3C 433 & B2121+248 & Affected by Cygnus A \\
3C 441 & B2203+292 & Ionospherically distorted \\
3C 442A & B2212+135 & Ionospherically distorted \\
NGC 7385 & B2247+113 & Ionospherically distorted \\
3C 454 & B2249+185 & Ionospherically distorted \\
3C 454.3 & B2251+158 & Ionospherically distorted \\
3C 455 & B2252+129 & Ionospherically distorted \\
3C 457 & B2309+184 & Ionospherically distorted \\
3C 465 & B2335+267 & Ionospherically distorted \\
\hline
\label{absent3CRRsources}
\end{tabular}
\end{table}

\subsection{Internal matching of the EGC}
\label{sec:internalmatch}

We also search for missing sources by cross-matching the GLEAM components into pairs/groups and then selecting potential, extended G4Jy sources for visual inspection. The two methods we use for selecting these candidate sources are described below. Following inspection of internal matches, we add a total of nine sources to the G4Jy Sample that would otherwise be absent.

\subsubsection{Applying a 4-arcmin matching radius}
\label{sec:internalmatchTOPCAT}

For the first method, we use {\sc topcat} to apply a friend-of-friends internal match, which is based purely on GLEAM positions. This allows us to include `chains' of low-frequency radio emission in our selection, and we set 4$'$ as the maximum separation between two adjacent GLEAM components. This matching radius is chosen bearing in mind the resolution of GLEAM ($\sim$2$'$), and that if the GLEAM components are well-separated, it becomes more difficult to tell whether the radio emission is associated or not. Furthermore, a 1-Mpc source at $z\sim0.5$ is $<3'$ in angular size, and we do not expect an overabundance of GRGs at low redshift.

\begin{figure*}
\centering
\subfigure[G4Jy~189]{
	\includegraphics[scale=1.1]{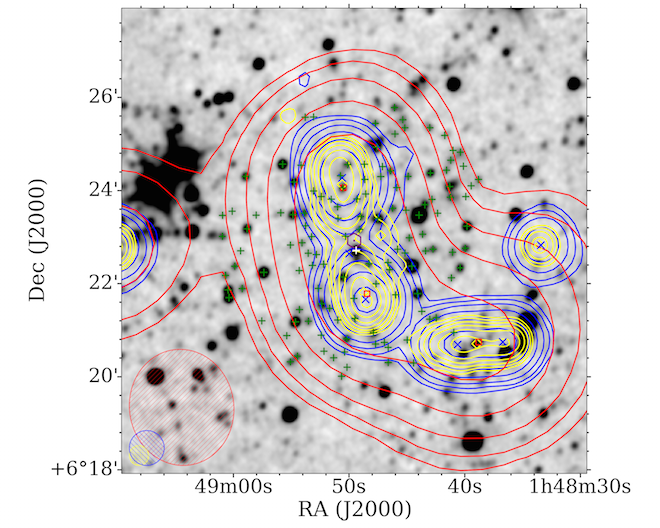}
	}
\subfigure[G4Jy~270]{
	\includegraphics[scale=1.1]{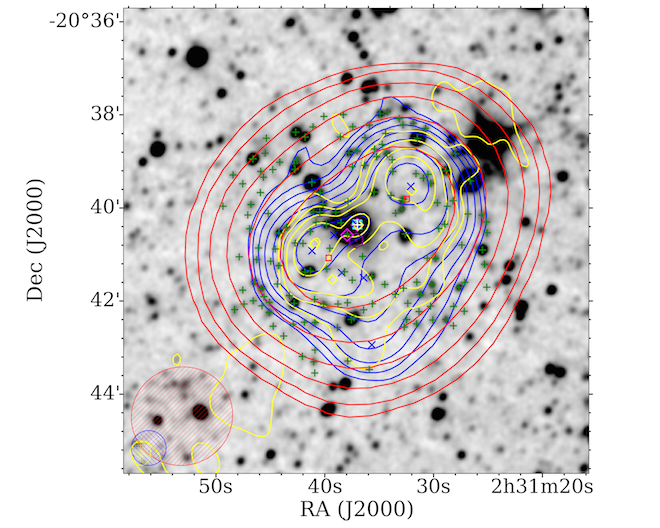}
	}
\subfigure[G4Jy~318]{
	\includegraphics[scale=1.1]{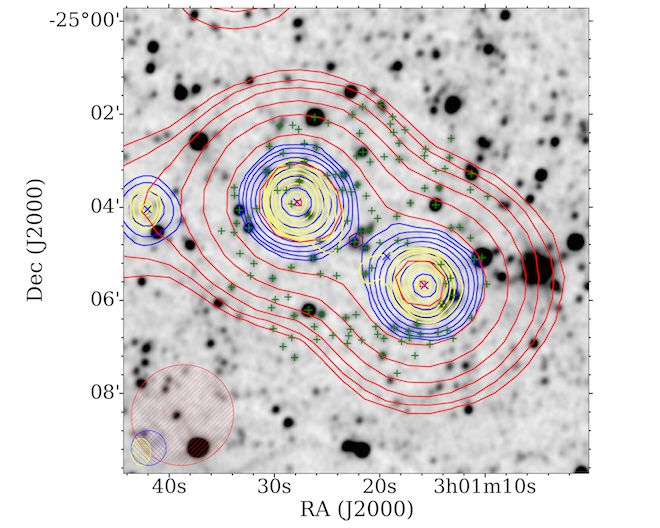}
	}
\subfigure[G4Jy~447]{
	\includegraphics[scale=1.1]{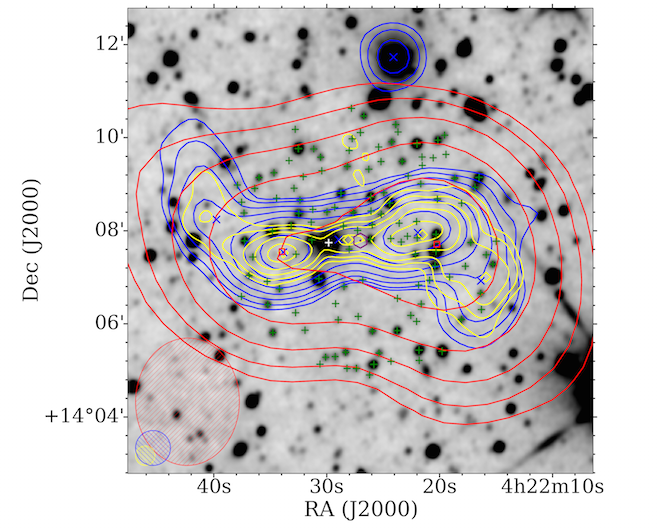}
	}
\subfigure[G4Jy~729]{
	\includegraphics[scale=1.1]{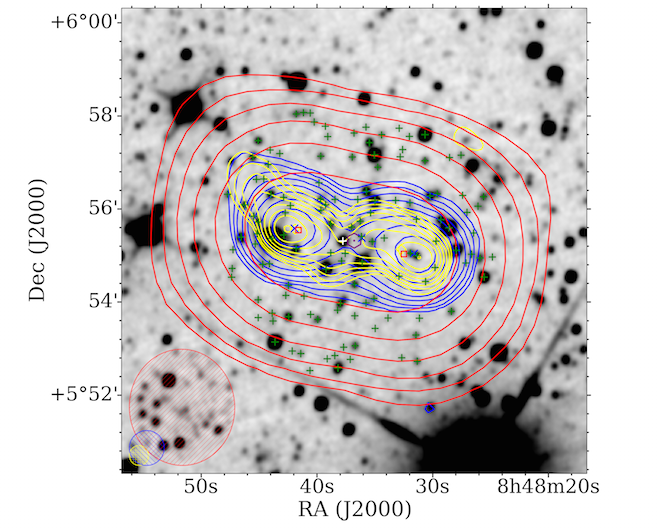}
	}
\subfigure[G4Jy~1021]{
	\includegraphics[scale=1.1]{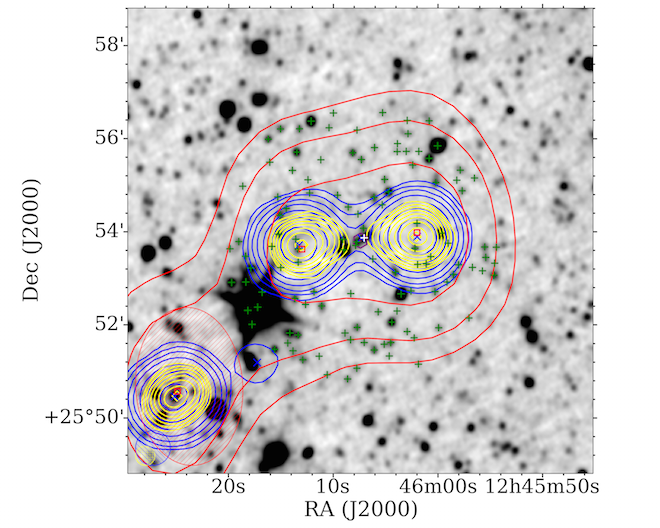}
	}
\caption{Overlays for six G4Jy sources that were added to the G4Jy Sample following internal matching (Section~\ref{sec:internalmatchTOPCAT}). The datasets, contours, symbols, and beams are the same as those used for Figure~\ref{JMsources1}. }
\label{IMsources1}
\end{figure*}

\begin{figure}
\centering
\subfigure[G4Jy~1428]{
	\includegraphics[scale=1.1]{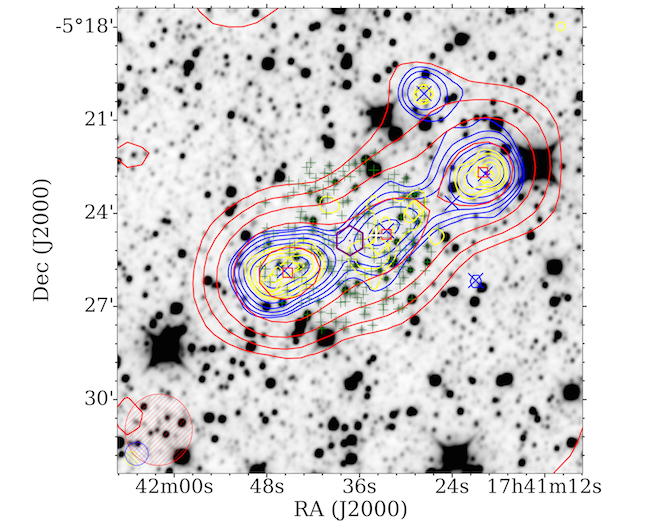}
	}
\subfigure[G4Jy~1480]{
	\includegraphics[scale=1.1]{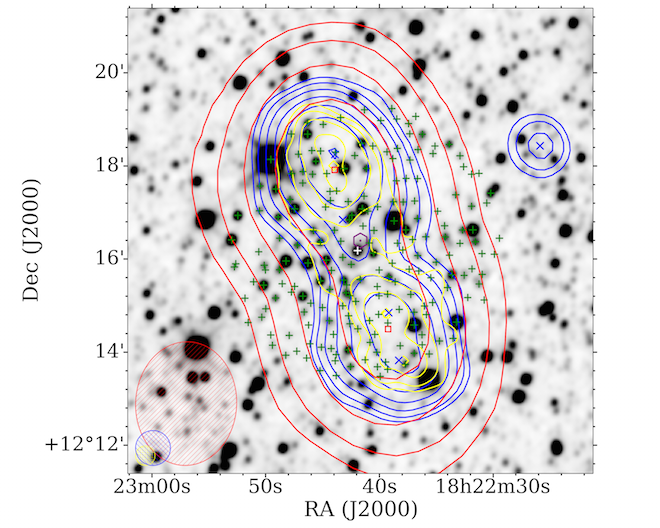}
	}
\subfigure[G4Jy~1718]{
	\includegraphics[scale=1.1]{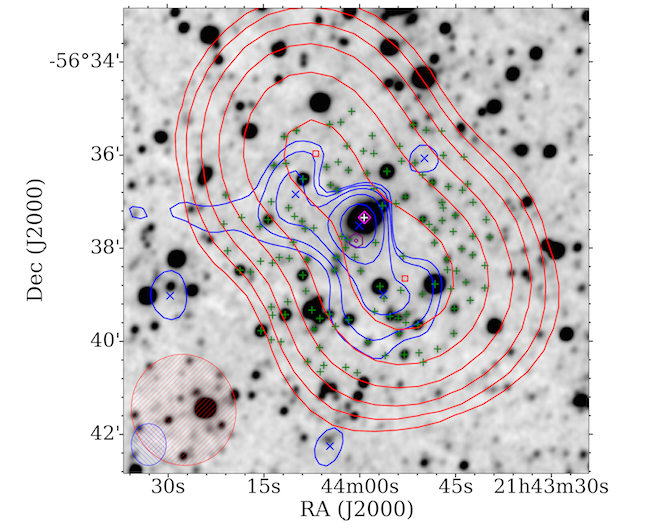}
	}
\caption{Overlays for three more G4Jy sources that were added to the G4Jy Sample following internal matching (Section~\ref{sec:internalmatchTOPCAT}). Datasets, contours, symbols, and beams are the same as for Figure~\ref{JMsources1}. }
\label{IMsources2}
\end{figure}

After removing groups with at least one GLEAM component brighter than 4\,Jy (since these have already been inspected), there remains 14,441 groupings. We then sum $S_{\mathrm{151\,MHz}}$ for each group and retain those for which the total is $>4$\,Jy. This drastically cuts the number of GLEAM components down to 88. A further 10 are removed, as they have already been identified as belonging to the G4Jy Sample via our cross-checks against the literature (Section~\ref{sec:literaturecrosschecks}). For the remaining 78 GLEAM components (forming 37 groups), we download the relevant images and create new overlays for visual inspection. We find that the majority of these groups are the result of unrelated GLEAM components being close to one another, whilst the remainder are groups that genuinely represent associated low-frequency emission. This prompts us to add another nine radio galaxies to the G4Jy Sample, which are listed in Table~\ref{missingsources2} and presented in Figures~\ref{IMsources1}--\ref{IMsources2}. Of particular note is the S-shaped source, G4Jy~447 (section~4.4.2 of Paper~II). 

Consideration of a tenth source for the sample was more convoluted but, ultimately, short-lived. This NVSS `triple' is blended with an unrelated point-source, together having their low-frequency emission characterised by GLEAM~J215506$-$321945 and GLEAM~J215519$-$321841. The {\it relative} TGSS flux-densities of the two sources suggest that the `triple' exceeds the 4-Jy threshold (cf. Appendix~\ref{sec:removedcomponents}), so we attempted to de-blend their GLEAM emission via priorised re-fitting (cf. Appendix~D.3). However, the re-fitted $S_{\mathrm{151\,MHz}}$ for the `triple' is below 4\,Jy, and so the source is not considered any further. 

\begin{table*}
\centering 
\caption{Radio sources that are now included in the G4Jy Sample, having been identified through a friends-of-friends match using the GLEAM extragalactic catalogue (Section~\ref{sec:internalmatchTOPCAT}). Including these radio galaxies gives a total of \nsources G4Jy sources in the sample.
}
\begin{tabular}{@{}lccc@{}}
 \hline
Source & Corresponding  & Individual & Total \\
& GLEAM components   &  $S_{\mathrm{151\,MHz}}$\,/\,Jy  &   $S_{\mathrm{151\,MHz}}$\,/\,Jy  \\
 \hline
   G4Jy~189 & GLEAM~J014848+062147 & $2.16\pm0.03$ & $4.43\pm0.04$ \\
   &  GLEAM~J014850+062403 & $2.27\pm0.03$ &\\ 
     G4Jy~270 &  GLEAM~J023132$-$203948 & $2.20\pm0.03$ & $4.38\pm0.04$\\
   &  GLEAM~J023139$-$204104 & $2.18\pm0.03$ &\\ 
   G4Jy~318 & GLEAM~J030115$-$250538 & $1.85\pm0.02$ & $4.11\pm0.03$ \\
   &  GLEAM~J030127$-$250354 & $2.27\pm0.02$ &\\ 
  G4Jy~447 &  GLEAM~J042220+140742 & $3.00\pm0.04$ & $4.47\pm0.06$\\
   &  GLEAM~J042233+140733 & $1.47\pm0.04$ &\\ 
 G4Jy~729 & GLEAM~J084832+055502 & $3.40\pm0.03$ & $6.88\pm0.04$\\
 & GLEAM~J084841+055532 & $3.48\pm0.03$ &\\
    G4Jy~1021 & GLEAM~J124602+255359 & $2.45\pm0.07$ & $4.30\pm0.09$\\
   &  GLEAM~J124612+255337 & $1.85\pm0.07$ &\\ 
   G4Jy~1428 &  GLEAM~J174120$-$052240 & $1.50\pm0.05$ & $6.20\pm0.10$\\ 
    &  GLEAM~J174132$-$052440 & $1.89\pm0.07$ &\\ 
       & GLEAM~J174145$-$052554 & $2.81\pm0.06$ & \\
    G4Jy~1480 & GLEAM~J182239+121429  & $1.83\pm0.09$ & $4.49\pm0.12$\\
   &  GLEAM~J182243+121754 & $2.66\pm0.08$ &\\ 
    G4Jy~1718 & GLEAM~J214352$-$563839 & $3.56\pm0.03$ & $6.28\pm0.04$ \\
   & GLEAM~J214406$-$563558 & $2.72\pm0.03$ & \\ 
\hline
\label{missingsources2}
\end{tabular}
\end{table*}

\subsubsection{Applying empirically-derived criteria}
\label{sec:NickandStuart}

 For our second method, we use two criteria: a flux-density ratio of $0.5<S_{1}/S_{2}< 2$, and a normalised separation (in degrees/$\sqrt{\mathrm{Jy}}$) of $\theta/\sqrt(S_{1}+S_{2})<0.13$, where $S_{1}$ and $S_{2}$ are the 151\,MHz flux-densities in Jy. This parameter space was chosen from examination of flux-density ratio plotted against normalised separation, for sources from the FIRST survey \citep{White1997}. That analysis has been done as part of the follow-up to an automated cross-identification paper based on the likelihood ratio \citep[{\sc lrpy};][]{Weston2018}. The goal was to select pairs of radio components that could potentially be true `doubles' and then see whether the {\sc lrpy} code would be more likely to select a counterpart to the radio source if it was a `double', compared to if it were two `single' sources. The normalised separation comes from the assumption that the luminosity is constant and therefore distance is proportional to $1/\sqrt S$, and that the linear size is also constant. Neither of these assumptions is true, but we note that when plotting these parameters against each other using FIRST, there is a cloud of sources within the region bound by the two criteria, as well as a larger (partially-overlapping) distribution. The cloud of sources most likely corresponds to true pairs (i.e. the two lobes of one radio source) and the more-widespread distribution (going to higher flux-density ratios and normalised separations) is due to random matches. This separation of `true pairs' and `random pairs' is confirmed by visual inspection of a randomly-selected sub-sample. We note that this parameter space happens to be equivalent to that from \citet{Magliocchetti1998}, albeit derived differently.

Through this analysis we find 47 potential `doubles' where both GLEAM components are brighter than 4\,Jy. Of these, 18 are confirmed through visual inspection to be true `doubles', and one is a `triple'. (All 19 had already been identified as multi-GLEAM-component sources.) The remaining 28 candidate `doubles' are found to be pairs of GLEAM components that are {\it not} associated with one another. Furthermore, there are 14 other previously-confirmed `doubles'/`triples' (Section~\ref{sec:morphology}) that are {\it not} selected via the criteria above. Therefore, for this subset ($S_{1}$, $S_{2}>4$\,Jy), we estimate the reliability of the algorithm as $19/47=0.40$, and its completeness as $19/33=0.58$.

Meanwhile, out of 8 potential `doubles' where both GLEAM components are {\it fainter} than 4\,Jy, we find that one of them is a true `double' with total flux-density $>4\,$Jy. This is G4Jy~729 (Figure~\ref{IMsources1}e), which was found via our previous method (Section~\ref{sec:internalmatchTOPCAT}). That method led to us identifying eight other radio galaxies, but none of these are selected via the criteria based on flux-density ratio and normalised separation. Hence, for this subset ($S_{1}$, $S_{2}<4$\,Jy; $S_{1}+S_{2}>4$\,Jy), we estimate the reliability of the algorithm as $1/8=0.13$, and its completeness as $1/9=0.11$. 

The main reason why applying this method to GLEAM data did not work very well (considering the reliability and completeness) is likely that the selection criteria were derived for radio sources in FIRST. Radio-galaxy populations from this survey are incomplete, and biased towards sources with their radio axis close to our line-of-sight. Furthermore, FIRST is sensitive to bright, compact emission, and so would pick up a greater proportion of radio galaxies that have distinct hotspots (typical of `double', FR-II morphology). In GLEAM we see radio sources with more-diffuse emission and a wider range of morphology, for which the selection criteria are less appropriate. However, this method {\it was} able to select very extended radio-sources with GLEAM-component separations, $\theta'$, of up to 8$'$. Those with separations of $\theta' < 4'$ (11 sources) would have been selected for visual inspection via the approach taken in Section~\ref{sec:internalmatchTOPCAT}, while the remaining 9 sources (with $4' < \theta' < 8'$) would not have been. Nonetheless, the latter subset each have at least one GLEAM component with $S_{\mathrm{151\,MHz}}>4$\,Jy, and so were already visually inspected following the initial selection (Section~\ref{sec:sampleselection}).



\subsection{Flux-density comparison with 3CRR}   
\label{sec:3CRRcomparison}

The G4Jy Sample will enable further investigation of relativistic jets and their interaction with the environment, as already explored using the well-studied 3CRR sample. Given the prominence of the latter, we conduct a flux-density analysis for 67 G4Jy sources that overlap with 3CRR (following on from Section~\ref{sec:checkagainst3CRR}). These are listed in Table~\ref{tab:3CRRoverlap}.

\begin{landscape}
\begin{table}
\centering 
\caption{The 67 G4Jy sources that are also in the 3CRR sample. `No. comp.' refers to the number of GLEAM components associated with the G4Jy source, and `3CRR ref.' indicates the origin of the 3CRR 178-MHz flux-density: 1 -- 4CT \citep{Williams1968,Caswell1969,Kellermann1969}; 2 -- 4C \citep{Clarke1965,Wills1966}; 3 -- 4C \citep{Pilkington1965,Gower1967}; 4 -- 3CR \citep{Bennett1962}; 5 -- corrected 3CR \citep{Veron1977}; 6 -- interpolation or extrapolation. The references provide expressions for the corresponding beamsize, which we evaluate at the relevant declination, and present in the next column. These `3CRR beams' are applied to GLEAM images (Section~7.3), from which we derive the $S_{\mathrm{178\,MHz}}$ shown in column 7. $S_{\mathrm{178\,MHz}}$ values in column 8 are calculated by extrapolating from 181\,MHz to 178\,MHz using the G4Jy\_alpha value (Section~\ref{sec:fouralphas}), or the spectral index from the 3CRR catalogue (as indicated by `$\alpha$ flag' = 1). Due to space considerations, we note here that columns 4, 7, and 8 are in units of Jy. The `original ratio' is the extrapolated, GLEAM $S_{\mathrm{178\,MHz}}$ (column 8) divided by the 3CRR $S_{\mathrm{178\,MHz}}$. For the `re-scaled ratio' we instead divide by a re-scaled version of the 3CRR $S_{\mathrm{178\,MHz}}$, as described in Section~7.3.} 
\begin{tabular}{@{}rcccccrrccc@{}} 
\hline 
(1) & (2)  & (3)  & (4)  & (5)  & (6) & (7) & (8)  & (9)  & (10) & (11) \\
 Source & No.  & 3CRR  & 3CRR                    & 3CRR  & NS $\times$ EW beam & GLEAM $S_{\mathrm{178\,MHz}}$ & GLEAM $S_{\mathrm{178\,MHz}}$        & $\alpha$         & `Original' &  `Re-scaled'   \\ 
& comp. & name  & $S_{\mathrm{178\,MHz}}$ & ref.  & / arcmin$^2$   & within 3CRR beam              & from (EGC) $S_{\mathrm{181\,MHz}}$  & flag  & ratio      &  ratio \\ 
\hline 
G4Jy 18   & 1 & 4C +12.03 & 10.9 & 1 &  24.6 $\times$ 24.0 &  8.854 $\pm$ 0.316  &  7.825 $\pm$ 0.055  &  0  & 0.718 $\pm$ 0.005 & 0.812 $\pm$ 0.030 \\ 
G4Jy 38   & 1 & 3C  9    & 19.4 & 1 &  23.6 $\times$ 24.0 &  14.446 $\pm$ 0.515  &  14.021 $\pm$ 0.047  &  0  & 0.723 $\pm$ 0.002 & 0.745 $\pm$ 0.027 \\ 
G4Jy 65   & 1 & 3C 14    & 11.3 & 3 &  37.2 $\times$ 23.0 &  9.506 $\pm$ 0.339  &  9.742 $\pm$ 0.038  &  0  & 0.862 $\pm$ 0.003 & 0.841 $\pm$ 0.030 \\ 
G4Jy 68   & 1 & 3C 16    & 12.2 & 1 &  24.4 $\times$ 24.0 &  9.915 $\pm$ 0.354  &  9.560 $\pm$ 0.031  &  0  & 0.784 $\pm$ 0.003 & 0.813 $\pm$ 0.029 \\ 
G4Jy 99   & 1 & 3C 28    & 17.8 & 3 &  34.4 $\times$ 23.0 &  19.671 $\pm$ 0.702  &  15.768 $\pm$ 0.053  &  0  & 0.886 $\pm$ 0.003 & 1.105 $\pm$ 0.040 \\ 
G4Jy 126  & 2 & 3C 33    & 59.3 & 4 &  276.0 $\times$ 13.6 & - & - & - & - & - \\ 
G4Jy 159  & 1 & 3C 42    & 13.1 & 2 &  20.7 $\times$ 15.0 &  13.192 $\pm$ 0.471  &  11.121 $\pm$ 0.052  &  0  & 0.849 $\pm$ 0.004 & 1.007 $\pm$ 0.037 \\ 
G4Jy 161  & 1 & 3C 43    & 12.6 & 1 &  21.6 $\times$ 24.0 &  11.128 $\pm$ 0.397  &  10.691 $\pm$ 0.038  &  0  & 0.848 $\pm$ 0.003 & 0.883 $\pm$ 0.032 \\ 
G4Jy 174  & 1 & 3C 47    & 28.8 & 3 &  36.2 $\times$ 23.0 &  28.122 $\pm$ 1.003  &  25.458 $\pm$ 0.040  &  0  & 0.884 $\pm$ 0.001 & 0.976 $\pm$ 0.035 \\ 
G4Jy 179  & 1 & 3C 49    & 11.2 & 1 &  24.2 $\times$ 24.0 &  9.281 $\pm$ 0.331  &  8.682 $\pm$ 0.026  &  0  & 0.775 $\pm$ 0.002 & 0.829 $\pm$ 0.030 \\ 
G4Jy 206  & 1 & 3C 55    & 23.4 & 5 &  20.7 $\times$ 15.0 &  19.203 $\pm$ 0.685  &  16.590 $\pm$ 0.058  &  0  & 0.709 $\pm$ 0.002 & 0.821 $\pm$ 0.030 \\ 
G4Jy 258  & 1 & 3C 67    & 10.9 & 2 &  20.9 $\times$ 15.0 &  10.321 $\pm$ 0.368  &  9.782 $\pm$ 0.049  &  0  & 0.897 $\pm$ 0.004 & 0.947 $\pm$ 0.034 \\ 
G4Jy 321  & 1 & 3C 76.1  & 13.3 & 1 &  23.4 $\times$ 24.0 &  9.616 $\pm$ 0.343  &  9.512 $\pm$ 0.039  &  0  & 0.715 $\pm$ 0.003 & 0.723 $\pm$ 0.026 \\ 
G4Jy 329  & 1 & 3C 79    & 33.2 & 1 &  23.2 $\times$ 24.0 &  27.601 $\pm$ 0.985  &  24.478 $\pm$ 0.044  &  0  & 0.737 $\pm$ 0.001 & 0.831 $\pm$ 0.030 \\ 
G4Jy 400  & 2 & 3C 98    & 51.4 & 1 &  25.5 $\times$ 24.0 & - & - & - & - & - \\ 
G4Jy 431  & 1 & 3C 109   & 23.5 & 1 &  25.2 $\times$ 24.0 &  20.048 $\pm$ 0.715  &  19.609 $\pm$ 0.034  &  0  & 0.834 $\pm$ 0.001 & 0.853 $\pm$ 0.031 \\ 
G4Jy 432  & 1 & 4C +14.11 & 12.1 & 1 &  24.1 $\times$ 24.0 &  8.985 $\pm$ 0.321  &  8.768 $\pm$ 0.035  &  0  & 0.725 $\pm$ 0.003 & 0.743 $\pm$ 0.027 \\ 
G4Jy 480  & 1 & 3C 123   & 206.0 & 4 &  276.0 $\times$ 13.6 &  222.973 $\pm$ 7.954  &  197.503 $\pm$ 0.284  &  0  & 0.959 $\pm$ 0.001 & 1.082 $\pm$ 0.039 \\ 
G4Jy 511  & 1 & 3C 132   & 14.9 & 1 &  21.8 $\times$ 24.0 &  12.311 $\pm$ 0.439  &  12.256 $\pm$ 0.055  &  0  & 0.823 $\pm$ 0.004 & 0.826 $\pm$ 0.030 \\ 
G4Jy 537  & 1 & 3C 138   & 24.2 & 1 &  23.3 $\times$ 24.0 &  16.650 $\pm$ 0.594  &  16.505 $\pm$ 0.045  &  1  & 0.682 $\pm$ 0.002 & 0.688 $\pm$ 0.025 \\ 
G4Jy 638  & 1 & 3C 172   & 16.5 & 1 &  21.3 $\times$ 24.0 &  13.287 $\pm$ 0.474  &  12.596 $\pm$ 0.049  &  0  & 0.763 $\pm$ 0.003 & 0.805 $\pm$ 0.029 \\ 
G4Jy 648  & 1 & 3C 175   & 19.2 & 1 &  24.9 $\times$ 24.0 &  16.610 $\pm$ 0.593  &  16.512 $\pm$ 0.029  &  0  & 0.860 $\pm$ 0.002 & 0.865 $\pm$ 0.031 \\ 
G4Jy 649  & 1 & 3C 175.1 & 12.4 & 1 &  24.0 $\times$ 24.0 &  8.501 $\pm$ 0.303  &  7.127 $\pm$ 0.030  &  0  & 0.575 $\pm$ 0.002 & 0.686 $\pm$ 0.025 \\ 
G4Jy 661  & 1 & 3C 181   & 15.8 & 1 &  24.0 $\times$ 24.0 &  10.182 $\pm$ 0.363  &  10.112 $\pm$ 0.028  &  0  & 0.640 $\pm$ 0.002 & 0.644 $\pm$ 0.023 \\ 
G4Jy 679  & 1 & 3C 190   & 16.4 & 1 &  24.1 $\times$ 24.0 &  14.676 $\pm$ 0.524  &  14.531 $\pm$ 0.032  &  0  & 0.886 $\pm$ 0.002 & 0.895 $\pm$ 0.032 \\ 
G4Jy 682  & 1 & 3C 191   & 14.2 & 2 &  25.5 $\times$ 15.0 &  13.314 $\pm$ 0.475  &  12.405 $\pm$ 0.027  &  0  & 0.874 $\pm$ 0.002 & 0.938 $\pm$ 0.034 \\ 
G4Jy 683  & 1 & 3C 192   & 23.0 & 1 &  21.5 $\times$ 24.0 &  21.322 $\pm$ 0.761  &  20.552 $\pm$ 0.047  &  0  & 0.894 $\pm$ 0.002 & 0.927 $\pm$ 0.033 \\ 
G4Jy 707  & 1 & 3C 200   & 12.3 & 2 &  20.6 $\times$ 15.0 &  12.045 $\pm$ 0.430  &  10.592 $\pm$ 0.048  &  1  & 0.861 $\pm$ 0.004 & 0.979 $\pm$ 0.035 \\ 
G4Jy 714  & 1 & 4C +14.27 & 11.2 & 1 &  24.1 $\times$ 24.0 &  7.303 $\pm$ 0.261  &  6.732 $\pm$ 0.026  &  0  & 0.601 $\pm$ 0.002 & 0.652 $\pm$ 0.024 \\ 
G4Jy 722  & 1 & 3C 207   & 14.8 & 1 &  24.4 $\times$ 24.0 &  11.576 $\pm$ 0.413  &  10.432 $\pm$ 0.025  &  0  & 0.705 $\pm$ 0.002 & 0.782 $\pm$ 0.028 \\ 
G4Jy 731  & 1 & 3C 208   & 18.3 & 2 &  24.2 $\times$ 15.0 &  16.132 $\pm$ 0.575  &  15.490 $\pm$ 0.030  &  0  & 0.846 $\pm$ 0.002 & 0.882 $\pm$ 0.032 \\ 
\hline 
\label{tab:3CRRoverlap} 
\end{tabular}
\end{table} 
\end{landscape} 

\setcounter{table}{7} 

\begin{landscape} 
\begin{table} 
\centering 
\caption{{\it Continued} -- The 67 G4Jy sources that are also in the 3CRR sample. Note that beam dimensions cannot be provided for 3CRR sources that have an interpolated/extrapolated $S_{\mathrm{178\,MHz}}$ (3CRR ref. = 6). Hence, the remaining columns for these sources are also left unfilled. Due to space considerations, we note here that columns 4, 7, and 8 are in units of Jy.} 
\begin{tabular}{@{}rcccccrrccc@{}} 
\hline
(1) & (2)  & (3)  & (4)  & (5)  & (6) & (7) & (8)  & (9)  & (10) & (11) \\
 Source & No.  & 3CRR  & 3CRR                    & 3CRR  & NS $\times$ EW beam & GLEAM $S_{\mathrm{178\,MHz}}$ & GLEAM $S_{\mathrm{178\,MHz}}$        & $\alpha$         & `Original' &  `Re-scaled'   \\ 
        & comp. & name  & $S_{\mathrm{178\,MHz}}$ & ref.  & / arcmin$^2$   & within 3CRR beam              & from (EGC) $S_{\mathrm{181\,MHz}}$  & flag  & ratio      &  ratio   \\ 
\hline 
G4Jy 742  & 1 & 3C 212   & 16.5 & 1 &  24.1 $\times$ 24.0 &  13.234 $\pm$ 0.472  &  11.667 $\pm$ 0.026  &  1  & 0.707 $\pm$ 0.002 & 0.802 $\pm$ 0.029 \\ 
G4Jy 751  & 1 & 3C 215   & 12.4 & 1 &  23.3 $\times$ 24.0 &  10.500 $\pm$ 0.375  &  9.841 $\pm$ 0.027  &  1  & 0.794 $\pm$ 0.002 & 0.847 $\pm$ 0.030 \\ 
G4Jy 783  & 1 & 3C 225B  & 23.2 & 6 &  276.0 $\times$ 13.6 &  24.922 $\pm$ 0.889  &  15.718 $\pm$ 0.031  &  1  & 0.677 $\pm$ 0.001 & 1.074 $\pm$ 0.038 \\ 
G4Jy 788  & 1 & 3C 226   & 16.4 & 1 &  25.7 $\times$ 24.0 &  17.297 $\pm$ 0.617  &  14.488 $\pm$ 0.027  &  1  & 0.883 $\pm$ 0.002 & 1.055 $\pm$ 0.038 \\ 
G4Jy 797  & 1 & 3C 228   & 23.8 & 1 &  24.1 $\times$ 24.0 &  16.661 $\pm$ 0.594  &  14.839 $\pm$ 0.031  &  1  & 0.624 $\pm$ 0.001 & 0.700 $\pm$ 0.025 \\ 
G4Jy 813  & 1 & 3C 234   & 34.2 & 1 &  20.7 $\times$ 24.0 &  35.667 $\pm$ 1.272  &  33.106 $\pm$ 0.070  &  1  & 0.968 $\pm$ 0.002 & 1.043 $\pm$ 0.037 \\ 
G4Jy 840  & 1 & 3C 241   & 12.6 & 1 &  22.0 $\times$ 24.0 &  12.435 $\pm$ 0.444  &  10.290 $\pm$ 0.035  &  1  & 0.817 $\pm$ 0.003 & 0.987 $\pm$ 0.036 \\ 
G4Jy 865  & 1 & 3C 245   & 15.7 & 1 &  24.8 $\times$ 24.0 &  12.725 $\pm$ 0.454  &  11.405 $\pm$ 0.027  &  0  & 0.726 $\pm$ 0.002 & 0.811 $\pm$ 0.029 \\ 
G4Jy 945  & 1 & 3C 263.1 & 19.8 & 1 &  22.0 $\times$ 24.0 &  14.977 $\pm$ 0.534  &  15.641 $\pm$ 0.060  &  0  & 0.790 $\pm$ 0.003 & 0.756 $\pm$ 0.027 \\ 
G4Jy 949  & 1 & 3C 264   & 28.3 & 1 &  22.5 $\times$ 24.0 &  24.955 $\pm$ 0.890  &  19.300 $\pm$ 0.069  &  1  & 0.682 $\pm$ 0.002 & 0.882 $\pm$ 0.032 \\ 
G4Jy 959  & 1 & 3C 267   & 15.9 & 4 &  276.0 $\times$ 13.6 &  9.576 $\pm$ 0.342  &  11.353 $\pm$ 0.040  &  0  & 0.714 $\pm$ 0.002 & 0.602 $\pm$ 0.022 \\ 
G4Jy 998  & 1 & 3C 272.1 & 21.1 & 4 &  276.0 $\times$ 13.6 &  16.747 $\pm$ 0.597  &  15.948 $\pm$ 0.092  &  1  & 0.756 $\pm$ 0.004 & 0.794 $\pm$ 0.029 \\ 
G4Jy 1004 & 1 &  1227+119  & 12.5 & 6 &  - &  -  &  -  &  - & - & - \\ 
G4Jy 1008 & 1 & 3C 274.1 & 18.0 & 1 &  22.1 $\times$ 24.0 &  12.931 $\pm$ 0.461  &  15.641 $\pm$ 0.106  &  1  & 0.869 $\pm$ 0.006 & 0.718 $\pm$ 0.027 \\ 
G4Jy 1019 & 1 & 3C 275.1 & 19.9 & 1 &  23.4 $\times$ 24.0 &  13.610 $\pm$ 0.485  &  14.015 $\pm$ 0.084  &  1  & 0.704 $\pm$ 0.004 & 0.684 $\pm$ 0.025 \\ 
G4Jy 1031 & 1 & 3C 277.2 & 13.1 & 1 &  23.6 $\times$ 24.0 &  11.862 $\pm$ 0.423  &  9.734 $\pm$ 0.093  &  1  & 0.743 $\pm$ 0.007 & 0.905 $\pm$ 0.035 \\ 
G4Jy 1137 & 1 & 3C 296   & 14.2 & 1 &  25.3 $\times$ 24.0 &  10.481 $\pm$ 0.374  &  10.783 $\pm$ 0.069  &  0  & 0.759 $\pm$ 0.005 & 0.738 $\pm$ 0.027 \\ 
G4Jy 1153 & 1 & 3C 300   & 19.5 & 1 &  22.5 $\times$ 24.0 & - & - & - & - & - \\ 
G4Jy 1216 & 1 & 3C 310   & 60.1 & 4 &  276.0 $\times$ 13.6 &  53.226 $\pm$ 1.899  &  55.838 $\pm$ 0.257  &  1  & 0.929 $\pm$ 0.004 & 0.886 $\pm$ 0.032 \\ 
G4Jy 1233 & 1 & 3C 315   & 19.4 & 5 &  21.1 $\times$ 15.0 &  22.833 $\pm$ 0.815  &  19.983 $\pm$ 0.134  &  1  & 1.030 $\pm$ 0.007 & 1.177 $\pm$ 0.043 \\ 
G4Jy 1244 & 1 & 3C 318   & 13.4 & 1 &  22.4 $\times$ 24.0 &  16.480 $\pm$ 0.588  &  12.337 $\pm$ 0.071  &  1  & 0.921 $\pm$ 0.005 & 1.230 $\pm$ 0.045 \\ 
G4Jy 1265 & 2 & 3C 321   & 14.7 & 4 &  276.0 $\times$ 13.6 & - & - & - & - & - \\ 
G4Jy 1280 & 1 & 3C 324   & 17.2 & 1 &  22.1 $\times$ 24.0 &  17.166 $\pm$ 0.612  &  12.516 $\pm$ 0.072  &  0  & 0.728 $\pm$ 0.004 & 0.998 $\pm$ 0.037 \\ 
G4Jy 1282 & 3 & 3C 326   & 22.2 & 6 &  276.0 $\times$ 13.6 & - & - & - & - & - \\ 
G4Jy 1323 & 1 & 3C 334   & 11.9 & 1 &  23.1 $\times$ 24.0 &  10.699 $\pm$ 0.382  &  11.544 $\pm$ 0.069  &  0  & 0.970 $\pm$ 0.006 & 0.899 $\pm$ 0.033 \\ 
G4Jy 1332 & 1 & 3C 336   & 12.5 & 2 &  21.6 $\times$ 15.0 &  12.398 $\pm$ 0.442  &  12.029 $\pm$ 0.088  &  0  & 0.962 $\pm$ 0.007 & 0.992 $\pm$ 0.037 \\ 
G4Jy 1339 & 1 & 3C 341   & 11.8 & 1 &  20.9 $\times$ 24.0 &  14.749 $\pm$ 0.526  &  10.855 $\pm$ 0.114  &  0  & 0.920 $\pm$ 0.010 & 1.250 $\pm$ 0.048 \\ 
G4Jy 1341 & 1 & 3C 340   & 11.0 & 1 &  21.7 $\times$ 24.0 &  8.102 $\pm$ 0.289  &  10.730 $\pm$ 0.089  &  0  & 0.975 $\pm$ 0.008 & 0.737 $\pm$ 0.028 \\ 
G4Jy 1358 & 1 & 3C 346   & 11.9 & 1 &  23.2 $\times$ 24.0 &  11.854 $\pm$ 0.423  &  11.108 $\pm$ 0.072  &  0  & 0.933 $\pm$ 0.006 & 0.996 $\pm$ 0.037 \\ 
G4Jy 1419 & 1 & 4C +16.49 & 11.4 & 3 &  38.4 $\times$ 23.0 &  10.086 $\pm$ 0.360  &  9.499 $\pm$ 0.073  &  0  & 0.833 $\pm$ 0.006 & 0.885 $\pm$ 0.033 \\ 
G4Jy 1456 & 1 & 4C +13.66 & 12.3 & 1 &  24.2 $\times$ 24.0 &  6.598 $\pm$ 0.235  &  9.498 $\pm$ 0.060  &  1  & 0.772 $\pm$ 0.005 & 0.536 $\pm$ 0.020 \\ 
G4Jy 1460 & 1 & 3C 368   & 15.0 & 1 &  25.2 $\times$ 24.0 &  16.236 $\pm$ 0.579  &  13.492 $\pm$ 0.059  &  0  & 0.899 $\pm$ 0.004 & 1.082 $\pm$ 0.039 \\ 
G4Jy 1499 & 1 & 3C 386   & 26.1 & 1 &  23.2 $\times$ 24.0 &  25.886 $\pm$ 0.923  &  24.399 $\pm$ 0.124  &  0  & 0.935 $\pm$ 0.005 & 0.992 $\pm$ 0.036 \\ 
G4Jy 1690 & 1 & 3C 432   & 12.0 & 2 &  23.2 $\times$ 15.0 &  11.273 $\pm$ 0.402  &  10.474 $\pm$ 0.057  &  1  & 0.873 $\pm$ 0.005 & 0.939 $\pm$ 0.034 \\ 
G4Jy 1719 & 1 & 3C 436   & 19.4 & 2 &  20.8 $\times$ 15.0 &  22.127 $\pm$ 0.789  &  17.100 $\pm$ 0.097  &  1  & 0.881 $\pm$ 0.005 & 1.141 $\pm$ 0.042 \\ 
G4Jy 1724 & 1 & 3C 437   & 15.9 & 1 &  23.7 $\times$ 24.0 &  13.270 $\pm$ 0.473  &  11.701 $\pm$ 0.067  &  1  & 0.736 $\pm$ 0.004 & 0.835 $\pm$ 0.031 \\
\hline 
\label{tab:3CRRoverlap2} 
\end{tabular}
\end{table} 
\end{landscape}

\begin{figure}
\centering
\includegraphics[scale=0.4]{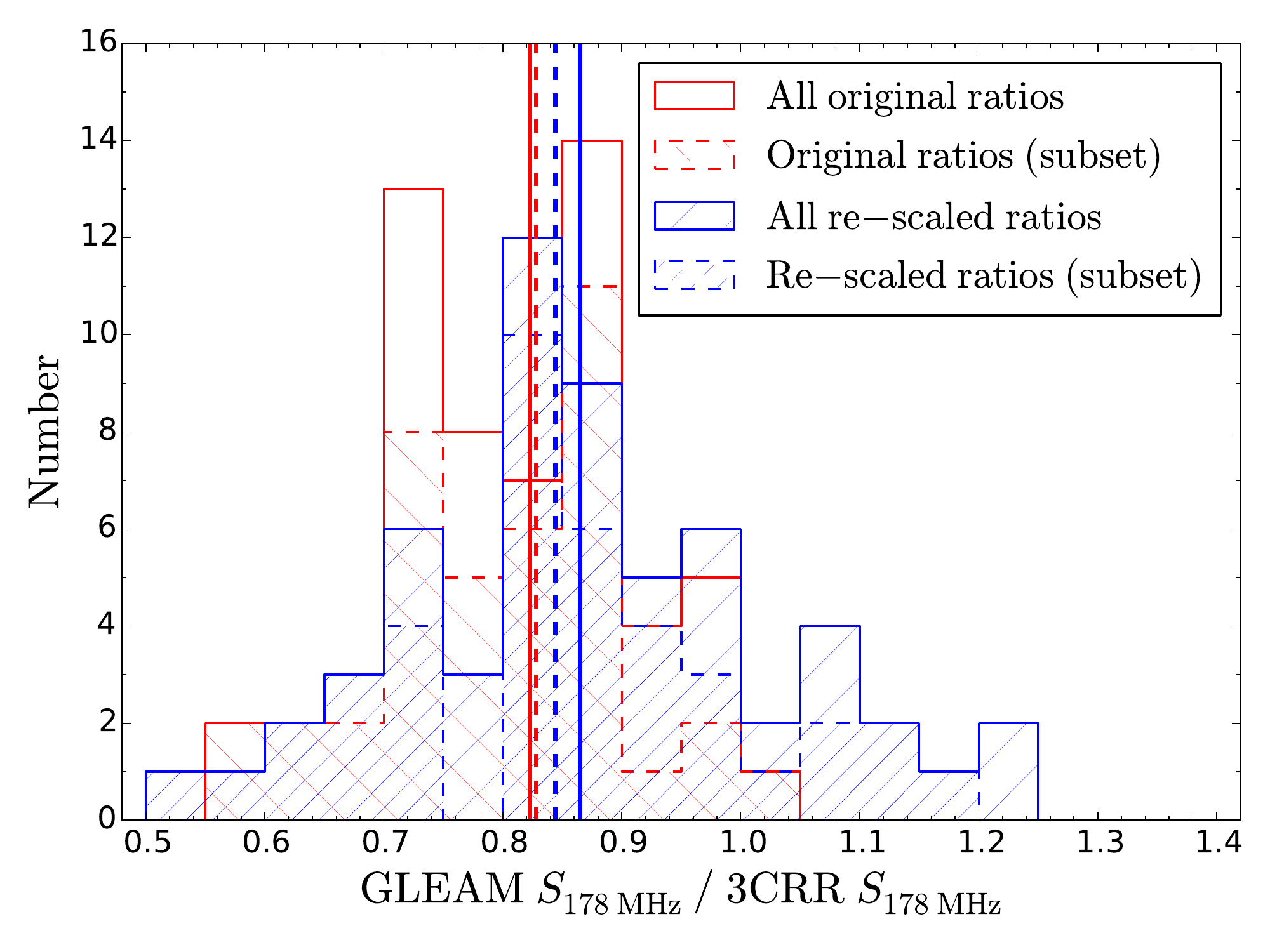}
\caption{The ratio of $S_{\mathrm{178\,MHz}}$ measured using GLEAM data, to $S_{\mathrm{178\,MHz}}$ using the 3CRR catalogue \citep{Laing1983}. These are for 60 of the 67 3CRR sources that overlap with the G4Jy Sample, where `original ratios' refers to the 3CRR $S_{\mathrm{178\,MHz}}$ being the value provided in the 3CRR catalogue. The median original-ratio is 0.82, and is indicated by a thick, vertical, red, solid line. `Re-scaled ratios' are those where the 3CRR $S_{\mathrm{178\,MHz}}$ value has had its corresponding beam-size (Table~\ref{tab:3CRRoverlap}) taken into account, leading to re-scaling of this flux density (see Section~7.3 for details). The median rescaled-ratio is 0.87, and is indicated by a thick, vertical, blue, solid line. For both sets of ratios, the GLEAM $S_{\mathrm{178\,MHz}}$ value is extrapolated from the $S_{\mathrm{181\,MHz}}$ measurement in the EGC \citep{HurleyWalker2017}. Meanwhile, `subset' (see legend) refers to the G4Jy sources for which we are able to use the G4Jy spectral-index for extrapolating flux densities from one frequency to another (as indicated by $\alpha$~flag = `0' in Table~\ref{tab:3CRRoverlap}). The thick, vertical, {\it dashed} lines indicate the median values for this subset, with respect to the original ratios (median = 0.83; red) and re-scaled ratios (median = 0.84; blue). }
\label{3CRRhistogram}
\end{figure}

Firstly, the closest GLEAM flux-density measurement we have to $S_{\mathrm{178\,MHz}}$ is that for the 181-MHz sub-band. We extrapolate this to 178\,MHz by assuming a power-law description of the radio emission, and using the spectral index (G4Jy\_alpha) fitted to the G4Jy total flux-densities (Sections~\ref{sec:totalGLEAMfluxdensities} and \ref{sec:fouralphas}). Where the associated reduced-$\chi^2$ value is $>1.93$, we instead use the spectral index provided in the 3CRR catalogue (obtained through VizieR) for this extrapolation. The ratio of GLEAM $S_{\mathrm{178\,MHz}}$ to 3CRR $S_{\mathrm{178\,MHz}}$ is presented in Table~\ref{tab:3CRRoverlap} as the `original' ratio. Looking at the distribution of this ratio (Figure~\ref{3CRRhistogram}), we see that the GLEAM flux-density appears to be systematically lower than that measured for 3CRR. Note that the 3CRR catalogue does not provide errors for $S_{\mathrm{178\,MHz}}$ (nor for the spectral index), but it uses the RBC scale of \citet{Roger1973}$^{\ref{3CRRlimit}}$, which is known to differ by $\sim9$\% from the KPW scale \citep{Kellermann1969} at 178\,MHz \citep{Laing1980,Laing1983}. Meanwhile, the EGC has an uncertainty in the flux-density scale of $8.0\pm0.5$\% for sources at $-72.0^{\circ}<$~Dec.\,$<18.5^{\circ}$, and $11\pm2$\% for sources at Dec.\,$>18.5^{\circ}$ \citep{HurleyWalker2017}. However, a combination of the two flux-density-scale errors (where we consider the extreme fractional-errors of 0.09 and 0.13) is not enough to explain the discrepancy for the 23 G4Jy--3CRR sources with ratio\,$<0.78$. 

Next, we note that the 178-MHz flux-densities in the 3CRR catalogue are a compilation from numerous surveys: 4CT, 4C and 3CR (see references in caption of Table~\ref{tab:3CRRoverlap}). Each of these surveys were conducted using a beam ranging from 19 to 235 times larger (by area) than that of the MWA, prompting us to investigate whether confusion/unrelated emission could account for the 3CRR flux-densities being systematically higher than those from GLEAM. 

As {\sc Aegean} may under-estimate the flux density for extended sources, we only consider for the analysis G4Jy--3CRR sources that are characterised by a single GLEAM component. For each of these sources we apply an ellipse, of the relevant beam dimensions (column 6 of Table~\ref{tab:3CRRoverlap}), to the 181-MHz sub-band image from GLEAM\footnote{Available through the GLEAM Postage Stamp Service: http://mwa-web.icrar.org/gleam\_postage/q/form.}. We then use the solid angle of the MWA beam to calculate the integrated flux-density over this ellipse (which involves summing over all pixel values within the ellipse and normalising with respect to the beam). In addition, we wish to characterise how well this method is able to reproduce the 181-MHz, integrated flux-density measurement that is provided in the EGC. For this, we apply an ellipse of dimensions fitted by {\sc Aegean} (i.e. the semi-major and semi-major axes, a\_181 and b\_181, respectively), along with the fitted position angle (pa\_181), and again calculate the integrated flux-density over the ellipse. The mean ratio of this ellipse-derived $S_{\mathrm{181\,MHz}}$ to the original EGC $S_{\mathrm{181\,MHz}}$ then allows us to `correct' other flux densities determined in a similar way. Similarly, we use the standard deviation in the ratio to estimate the error in the integrated flux-density calculated over an ellipse. The result is a corrected GLEAM $S_{\mathrm{181\,MHz}}$ measured within the `3CRR' beam, which we then extrapolate to 178\,MHz (again using either G4Jy\_alpha or the 3CRR spectral-index, as previously). The extrapolated value is what appears in column 7 of Table~\ref{tab:3CRRoverlap}.


How much extra emission is detected through the large beam associated with 3CRR, compared to the fitted GLEAM measurement, is apparent from comparing columns 7 and 8 of Table~\ref{tab:3CRRoverlap}. Dividing the latter by the former then gives a `corrective factor', which we use to re-scale the 3CRR $S_{\mathrm{178\,MHz}}$. The `re-scaled' ratio (column 11) is the ratio of the GLEAM $S_{\mathrm{178\,MHz}}$ (column 8) to this {\it re-scaled} 3CRR $S_{\mathrm{178\,MHz}}$. As shown in Figure~\ref{3CRRhistogram}, these re-scaled ratios (mean $=  0.87 \pm 0.16$) are closer to 1.0 than the original ratios (mean $=  0.81 \pm 0.11$), but more-widely distributed, and now as large as 1.25. Therefore, whilst unrelated emission may still play a part, we cannot conclude that it is the main reason for the offset between GLEAM and 3CRR flux-densities. Furthermore, a two-sample Kolmogorov--Smirnov test gives $D$ = 0.27 (where 0.30 is the $D$ statistic to exceed) and $p$-value = 0.02, indicating that the two distributions are not significantly distinct. 

\begin{figure*}
\vspace{-0.7cm}
\centering
\subfigure[No correlation with declination]{
	\includegraphics[scale=0.4]{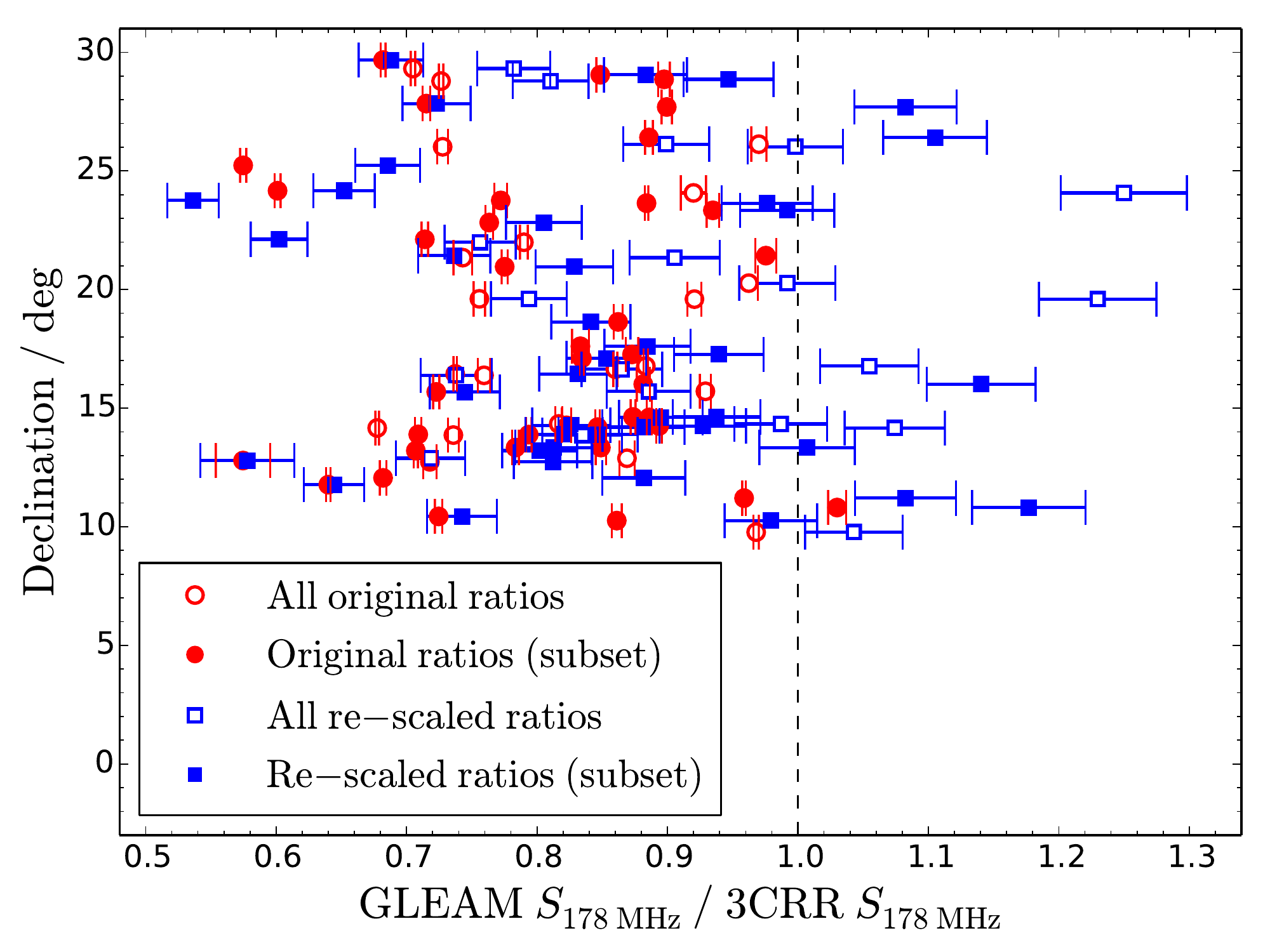}
	} 
\subfigure[No correlation with flux density]{
	\includegraphics[scale=0.4]{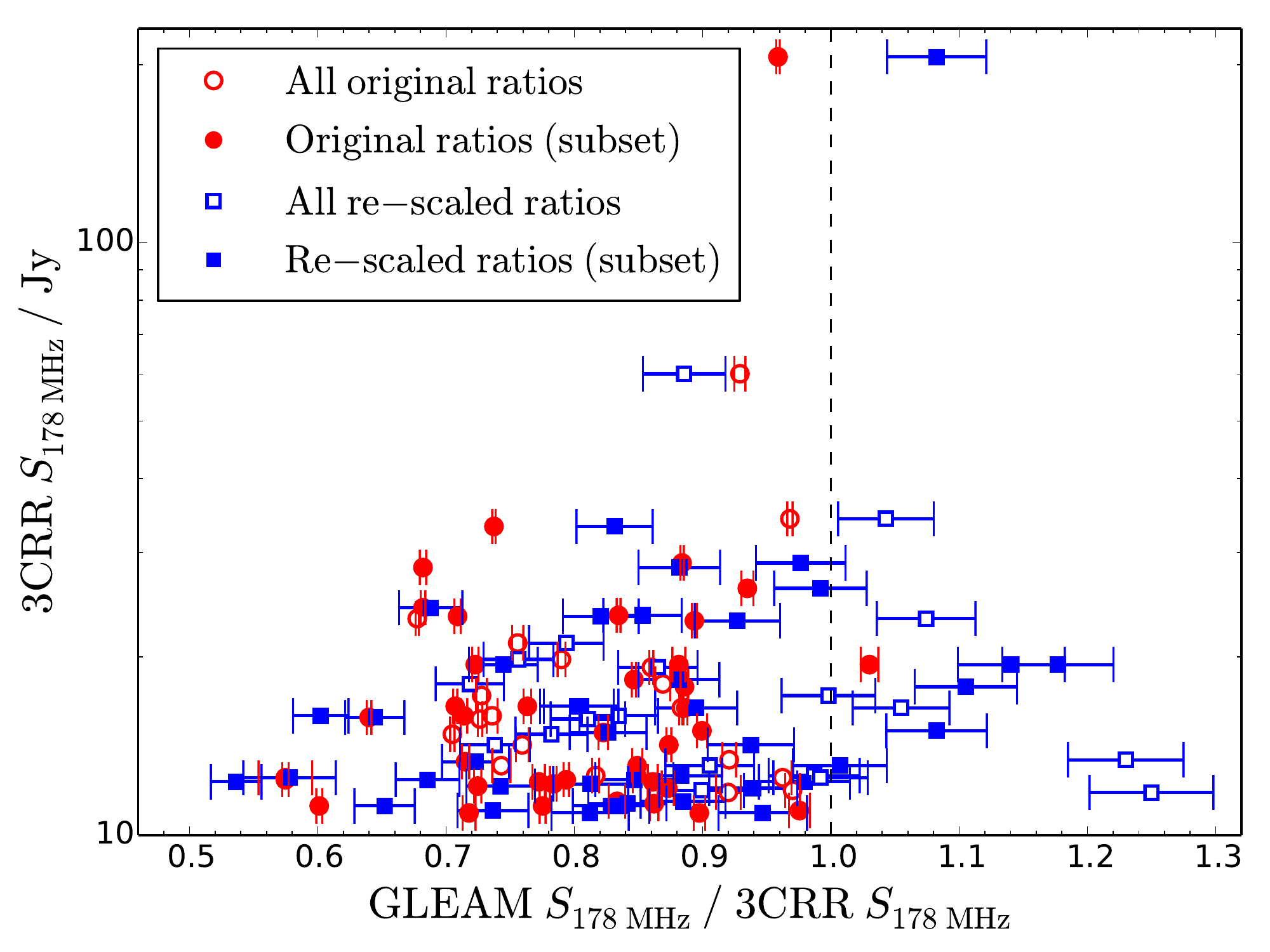} 
	} 
\caption{The GLEAM $S_{\mathrm{178\,MHz}}$/3CRR $S_{\mathrm{178\,MHz}}$ ratio plotted against (a) declination, and (b) 3CRR $S_{\mathrm{178\,MHz}}$. These are for 60 of the 67 3CRR sources that overlap with the G4Jy Sample, where `original ratios' refers to the 3CRR $S_{\mathrm{178\,MHz}}$ value being that provided in the 3CRR catalogue. `Re-scaled ratios' are those where the 3CRR $S_{\mathrm{178\,MHz}}$ value has had its corresponding beam-size (Table~\ref{tab:3CRRoverlap}) taken into account, leading to re-scaling of this flux density (see Section~7.3 for details). As in Figure~\ref{3CRRhistogram}, `subset' (see legends) refers to the G4Jy sources for which we are able to use the G4Jy spectral-index for extrapolating flux densities from one frequency to another (as indicated by $\alpha$~flag = `0' in Table~\ref{tab:3CRRoverlap}). For both panels, the vertical, black, dashed line is where the ratio is equal to 1.0, to guide the eye. \label{3CRRscatter}}
\end{figure*}

We now return to an explanation that has already been touched upon in this sub-section: that the flux-density calibration of GLEAM components is worse at the highest declinations. Whilst this is true for the GLEAM catalogue as a whole, we see no trend in the GLEAM $S_{\mathrm{178\,MHz}}$/3CRR $S_{\mathrm{178\,MHz}}$ ratio with declination, as shown in Figure~\ref{3CRRscatter}a. We also see no trend in the ratio with the integrated flux-density of the source (Figure~\ref{3CRRscatter}b), but cannot use this to rule out non-linearity in the flux-scale calibration. This is because the overlap of the G4Jy Sample with 3CRR restricts our investigation to $S_{\mathrm{178\,MHz}}\gtrsim10.9$\,Jy sources, whereas one of the criteria used to select sources for setting the GLEAM flux-density scale   \citep[$S_{\mathrm{74\,MHz}}>2$\,Jy;][]{HurleyWalker2017} corresponds to  $S_{\mathrm{178\,MHz}}\gtrsim1$\,Jy (assuming $S_{\nu} \propto \nu^{-0.7}$). It is possible that the flux-density scale starts to show non-linearity \citep{Scott1971,Laing1980} as the 10.9-Jy threshold is approached, but further investigation is beyond the scope of this work. In addition, Eddington bias may be affecting 3CRR sources detected at lower signal-to-noise (as weakly suggested by Figure~\ref{3CRRscatter}b), leading to the 3CRR $S_{\mathrm{178\,MHz}}$ being over-estimated. However, this is difficult to explore further due to the absence of $S_{\mathrm{178\,MHz}}$ uncertainties in the 3CRR catalogue.

Next, we plot the total, integrated flux-densities across the GLEAM band (72--231\,MHz) alongside previous multi-frequency measurements for the 3CRR sample. For this, we include the G4Jy--3CRR sources that have multiple GLEAM components associated with them. The resulting SEDs (spanning 10\,MHz to 15\,GHz) are available as online supplementary material (Appendix~\ref{app:sedsbyJoe}), and confirm that the G4Jy flux-densities are consistently lower for this subset of very bright sources at high declination. Although we are unable to conclusively state the reason(s) for this offset, we point out that the G4Jy catalogue inherits from the EGC the internal-calibration consistency of $\leq$\,3\%, as determined over 245,457 GLEAM components \citep{HurleyWalker2017}. Furthermore, recent P-band (230--470\,MHz) VLA observations of $\sim$40 unresolved sources suggest that the GLEAM flux-density scale is $\sim$3\% too low (Callingham et al., in preparation). This is based on a comparison of the $S_{\mathrm{230\,MHz}}$ measurement from GLEAM and the expected flux-density at 230\,MHz, following spectral fitting of the GLEAM and P-band data. The latter are tied to the flux-density scale of \citet{Perley2017}, over 50\,MHz to 50\,GHz. 

\subsection{Summary of the results from visual inspection and checks for completeness}
\label{sec:summary}

\begin{figure}
\centering
\includegraphics[scale=0.4]{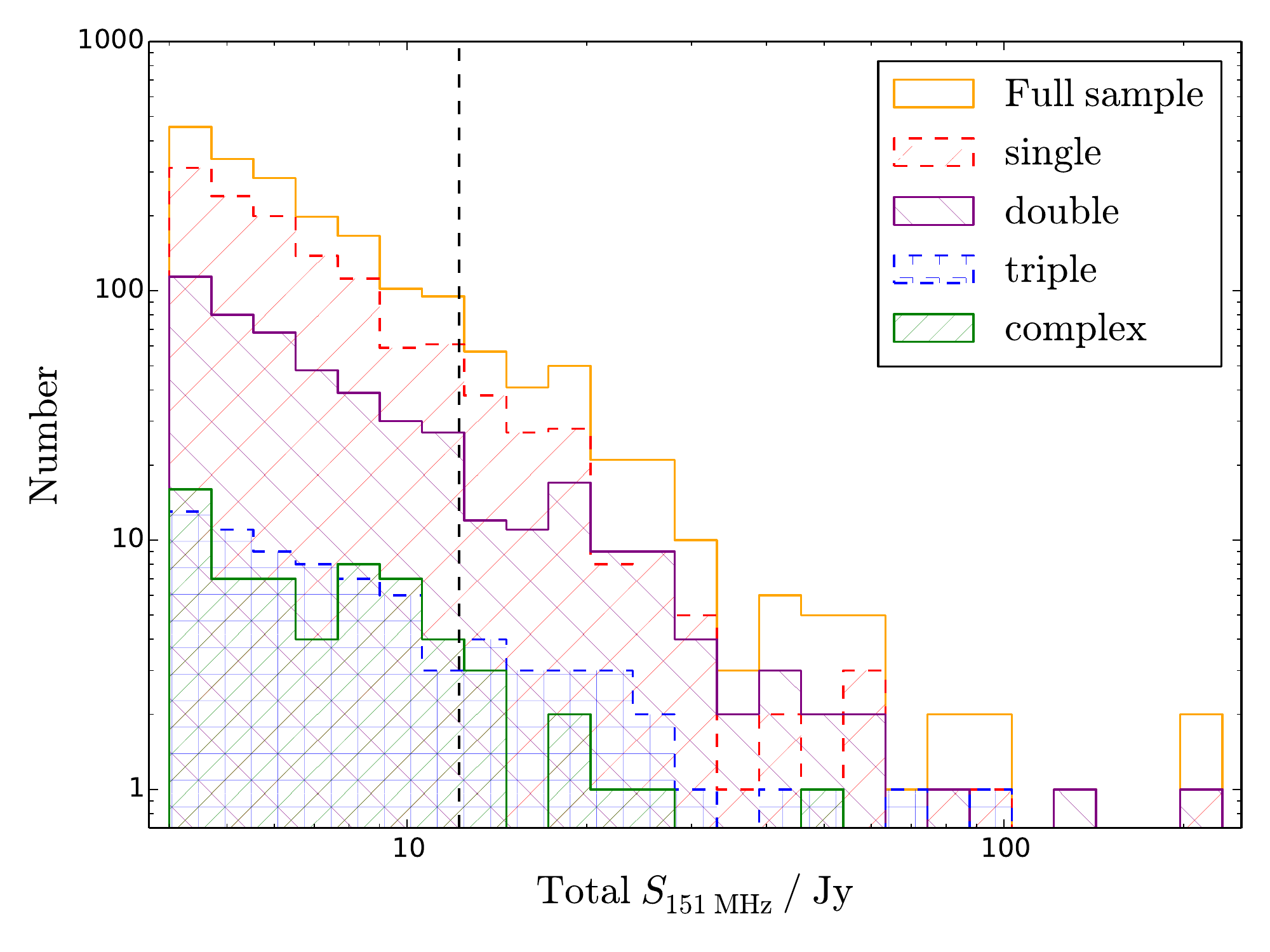} 
\caption{The distribution in $S_{\mathrm{151\,MHz}}$ for the full sample, and when split by morphology (`single', `double', `triple', and `complex') in NVSS/SUMSS/TGSS (Sections~\ref{sec:morphology} and \ref{sec:summary}). The vertical line is where  $S_{\mathrm{151\,MHz}} = 12.2$\,Jy, which corresponds to $S_{\mathrm{178\,MHz}} = 10.9$\,Jy (assuming a power-law radio spectrum with spectral index, $\alpha = -0.7$). Therefore, the G4Jy sources to the right of the vertical line are akin to those in the 3CRR sample \citep{Laing1983}.  }
\label{S151distribution}
\end{figure}

As a result of our visual inspection and further checks, the original list of 1,879 GLEAM {\it components} becomes a list of 1,960 components. In conjunction, this is reduced to a list of 1,863 GLEAM {\it sources}, and it is this source list that we refer to as the G4Jy Sample. 67\% of the G4Jy sources are `singles', 26\% are `doubles', 4\% are `triples', and 3\% have `complex' morphology (Table~\ref{numberG4Jy}). In Figure~\ref{S151distribution} we show the distributions in $S_{\mathrm{151\,MHz}}$ for these subsets, in addition to that for the full sample. We note that there are 233 G4Jy sources brighter than 12.2\,Jy at 151\,MHz, which corresponds to the flux-density limit for the 3CRR sample (173 sources). Of these sources, the fraction that are a `double' or a `triple' is 41\%, whilst for the 1,630 G4Jy sources below this threshold, the fraction falls to 28\%. Without redshifts to consider the radio luminosities, we hypothesise that the brighter sources are likely to be closer and more extended, and therefore resolved in NVSS/SUMSS/TGSS (which are the surveys we use to determine the morphology). Meanwhile, it is important to note that 21\% (383) of the G4Jy sources are affected by confusion (Section~\ref{sec:morphology}), and so the GLEAM flux-densities will need to be updated in the future. 

\begin{table*}
\centering 
\caption{Characteristics of the G4Jy Sample (Section~\ref{sec:summary}), in terms of the number of GLEAM components associated with an individual source, and the morphology of the NVSS/SUMSS emission (Section~\ref{sec:morphology}).
}
\begin{tabular}{@{}lccccr@{}} 
 \hline
Number of associated & \multicolumn{4}{c}{  Description of the NVSS/SUMSS emission} &  {\bf Total number} \\
GLEAM components & `single' & `double' & `triple' &  `complex'  & {\bf of sources} \\
 \hline
One & 1,245 & 432 & 50 & 58 & 1,785 \\
Two  & 0 & 42 & 20 & 2 & 64 \\
Three  & 0 & 4 & 6 & 0 & 10 \\
Four  & 0 & 1 & 1 & 1 & 3 \\
Five  & 0 & 0 & 0 & 1 & 1 \\
\hline
{\bf Total number of sources} & 1,245 & 479 & 77 & 62 & 1,863  \\
\hline
\label{numberG4Jy}
\end{tabular}
\end{table*}

Following our best efforts, the G4Jy Sample of $S_{\mathrm{151\,MHz}}>4$\,Jy radio-sources is effectively {\it complete} over the footprint of the EGC \citep{HurleyWalker2017} minus the region defined as $-18.3^{\circ} < b < -10.0^{\circ}$, $65.4^{\circ} < l < 81.1^{\circ}$, Dec.\,$<30.0^{\circ}$ (Section~\ref{sec:checkagainst3CRR}). This covers 24,731 deg$^2$ (i.e. 60\% of the entire sky), all of which is accessible to the SKA and its precursor telescopes. The sky density of the G4Jy Sample is therefore one source per $\sim13$\,deg$^2$. Within this total sky area we know that at least one $>4$\,Jy source is absent (Orion A; see Section~\ref{sec:AteamOrion}) and acknowledge that we may still be missing a few extended radio-sources. Hence, we estimate that the G4Jy Sample is 99.50--99.95\% complete. In addition, the brightest G4Jy source (G4Jy~1402; 3C~353) is at $S_{\mathrm{151\,MHz}}=232$\,Jy, with our sample being biased against sources that are brighter than this (namely, `A-team' sources, which are masked for the EGC; see Table~\ref{GLEAMexclusions}).

Finally, in light of a $\sim$3\% offset (Callingham et al., in preparation; Section~7.3) between the flux-density scale of GLEAM and that of \citet{Perley2017}, we consider the effect that this would have on the completeness of the G4Jy Sample. We find that there are 88 GLEAM components in the EGC at $3.89 < S_{\mathrm{151\,MHz}}\mathrm{/Jy}< 4.00$ that would then need to be considered for inclusion. One of these is GLEAM~J151635+001603, which is already in the sample as being associated with GLEAM~J151643+001410 (together composing G4Jy~1238). Similarly, some of the remaining 87 components may be associated with each other, but equally, some components $<$\,3.89\,Jy may sum together to $>$\,3.89\,Jy (cf. Sections~\ref{sec:literaturecrosschecks} and \ref{sec:internalmatch}, regarding the 4-Jy threshold). Therefore, for simplicity (and remembering that Orion~A is absent from the considered footprint), we estimate the completeness to be: $1863 / (1863 + 87 +1) = 95.5$\% on the flux-density scale of  \citet{Perley2017}.

\section{Initial and future analyses} \label{sec:discussion}

Having produced a thorough compilation of the brightest radio-sources in the southern sky, we use the G4Jy catalogue to perform some initial analysis, which we describe in this section. This will be followed by full-sample multi-wavelength analysis in Paper~III of the G4Jy series, and investigation into broadband radio spectra (covering 72\,MHz to 20\,GHz) to be presented in Paper~IV. In addition, we will compare our association of components observed at $\sim$1\,GHz (White \& Line, in preparation) with the results obtained using the Positional Update and Matching Algorithm \citep[PUMA;][]{Line2017}. We envisage that this can be extended to using the G4Jy Sample as a training set for machine-learning algorithms, and so leveraging the effort that we have put into host-galaxy identification (which is summarised in Section~\ref{sec:identifyhost}, and detailed in Paper~II).

\subsection{Angular-size information}
\label{sec:sizeanalysis}

An overview of linear, physical sizes for G4Jy sources will be provided in Paper~III, as calculating these sizes is reliant on redshifts being compiled for the sample. These may reveal that additional sources are GRGs, further to those listed in table~3 of Paper~II.

\begin{figure*}
\vspace{-0.7cm}
\centering
\subfigure[Angular-size distribution on a $\log$ scale]{
	\includegraphics[width=0.38\linewidth]{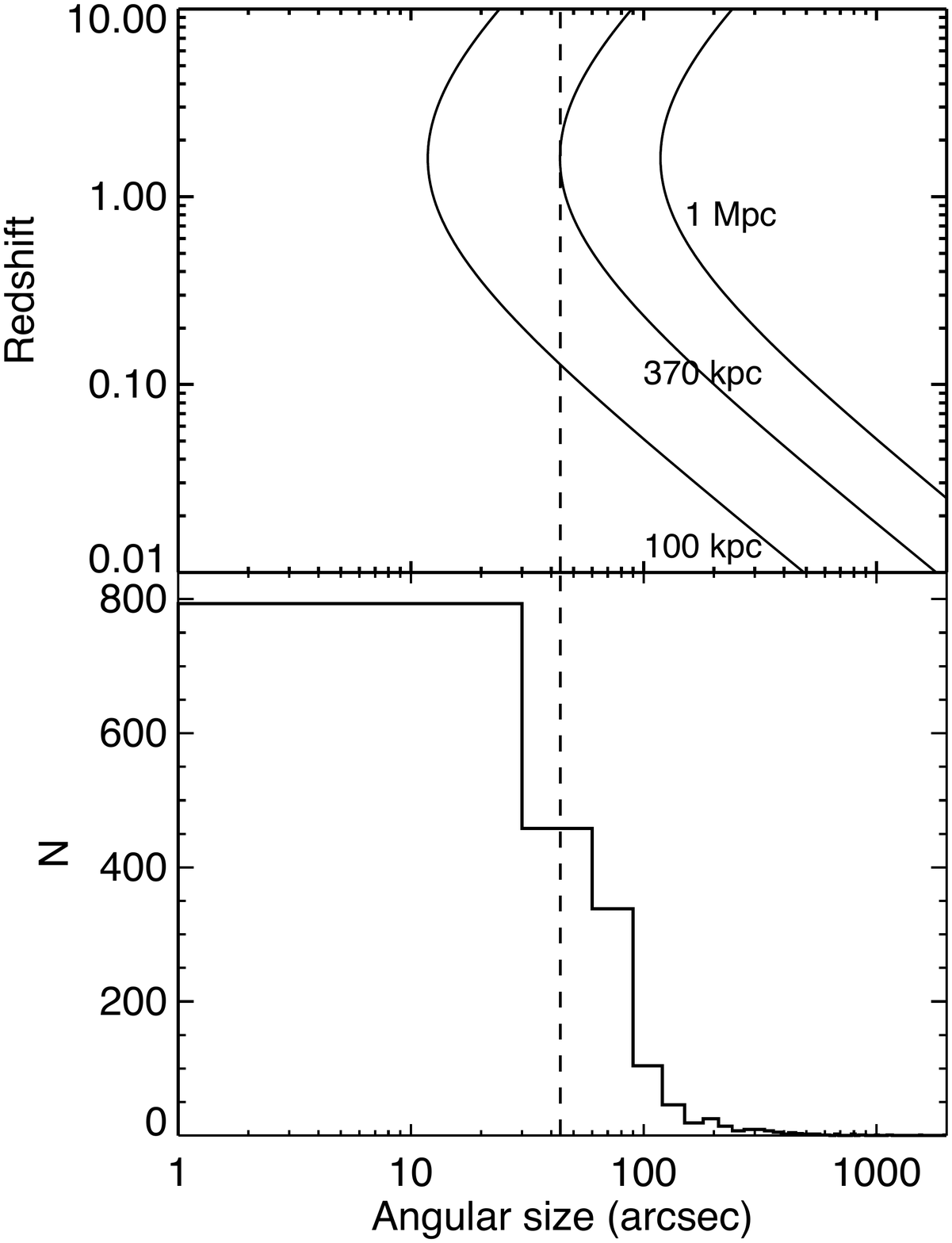}
	} 
\subfigure[Angular-size distribution on a linear scale]{
	\includegraphics[width=0.38\linewidth]{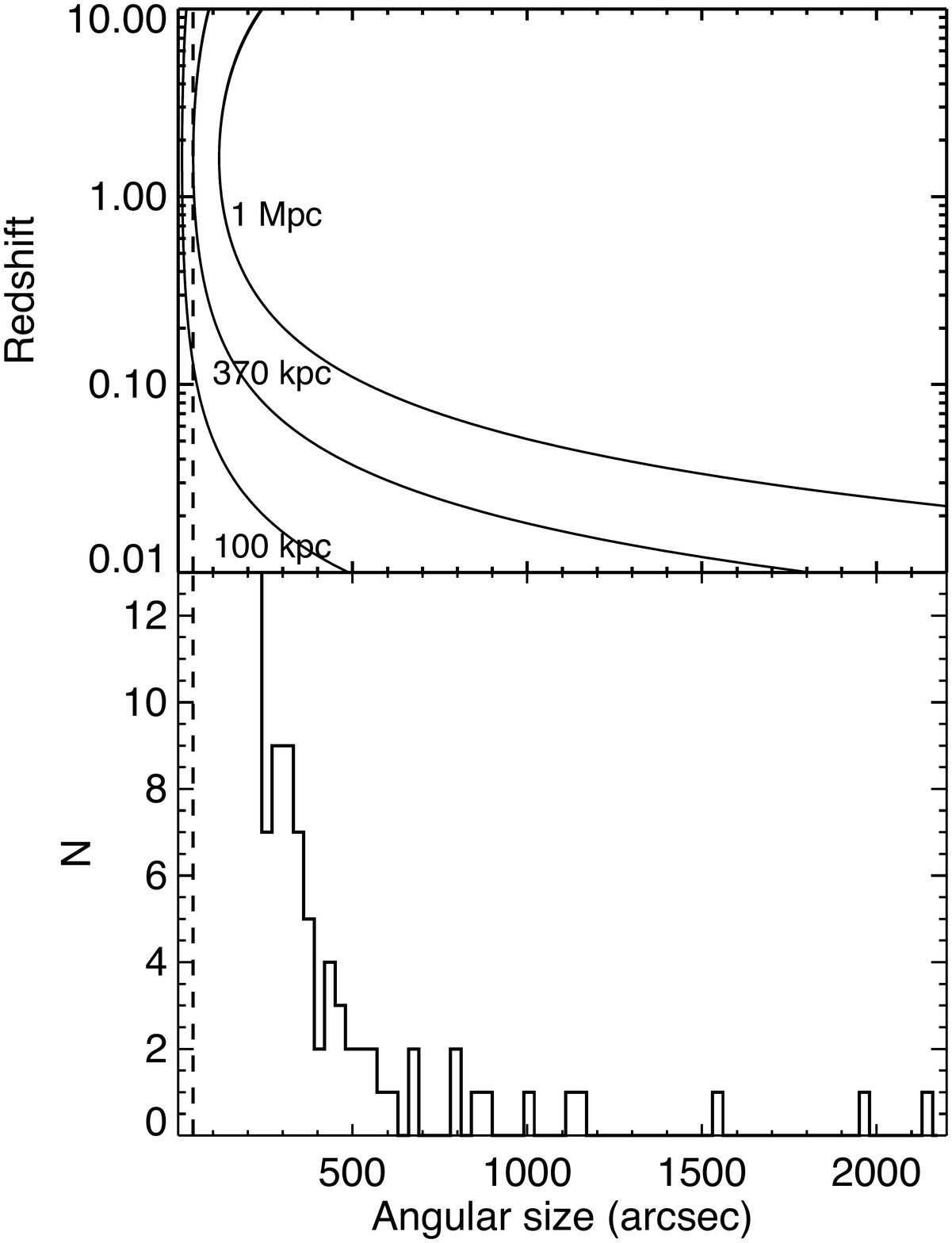} 
	} 
\caption{The solid lines in the upper panels of this figure show how the observed angular-size varies with redshift for a source of fixed physical-size (100\,kpc, 370\,kpc, 1\,Mpc). These functions are calculated in accordance with the cosmology described at the end of Section~\ref{sec:intro}. The lower panels show the angular-size distribution for sources in the G4Jy Sample, with the median angular-size marked by a dashed, vertical line (Section~\ref{sec:sizeanalysis}). \label{Ivysplots}}
\end{figure*}

In the meantime, we can obtain sample-wide information by considering the distribution in the {\it angular} sizes of the \nsources sources, since we know how the angular-size scale (kpc/arcsec) varies with redshift. This distribution is presented in Figure~\ref{Ivysplots}, where the median angular-size of 43.8$''$ is marked by a vertical, dashed line. (Note that for this analysis we fix all angular sizes to their upper limits. This is applicable for 855 sources, with the largest upper-limit being 129.8$''$.) As shown in the left-hand panels of Figure~\ref{Ivysplots}, (at least) half of the G4Jy Sample is smaller than 370\,kpc, regardless of redshift. Furthermore, if a source is $\gtrsim$\,500$''$ in angular size, its physical size must be at least 100\,kpc, even at the very low redshift of 0.01 (Figure~\ref{Ivysplots}b). However, we remind the reader that the angular sizes are derived using NVSS or SUMSS (Section~\ref{sec:angsizesforcatalogue}), and so are limited by the 45$''$ resolution of these surveys. If we consider only the 657 sources that are {\it multi-component} in NVSS or SUMSS, we find that the median angular-size is 74.5$''$.

Next, particularly with unresolved G4Jy sources in mind, we explore the compactness of the radio emission by considering whether the source exhibits interplanetary scintillation \citep{Little1966,Morgan2018}. This is where radio sources appear to `twinkle' due to the turbulence of the intervening solar wind, allowing sub-arcsecond scales to be probed. \citet{Chhetri2018} demonstrated the power of this method over $30\times30$\,deg$^2$ (i.e. the field of view for a single MWA pointing) by characterising the scintillation properties of 2,550 sources. 131 of these sources are in the G4Jy Sample, with two being described as `strong' scintillators (normalised scintillation index, NSI~$\geq0.9$). This means that a single sub-arcsecond component is dominating the flux density at 162\,MHz (their observing frequency). Another 20 sources are `moderate' scintillators ($0.4\leq$~NSI~$<0.9$), which may include sources with multiple sub-arcsecond components, and sources where there is a single sub-arcsecond component but it is surrounded by more-extended low-frequency emission. The 22 moderate/strong scintillators cross-matched with the G4Jy Sample are listed in Table~\ref{tab:ips}. However, another 43 of the 131 cross-matched sources may also have compact emission on sub-arcsecond scales, but for them \citet{Chhetri2018} could only provide upper limits in NSI, due to the scintillation not reaching the detection threshold.

\begin{table}
\centering 
\caption{ 22 G4Jy sources previously identified by \citet{Chhetri2018} as showing moderate ($0.4\leq$\,NSI\,$<0.9$) or strong (NSI\,$\geq0.9$) interplanetary scintillation (Section~\ref{sec:sizeanalysis}). NSI = normalised scintillation index. }
\begin{tabular}{@{}rcc@{}}
\hline
Source  & GLEAM component  & NSI \\
\hline
  G4Jy~1  &  GLEAM~J000057$-$105435  & 0.630 $\pm$ 0.055 \\
  G4Jy~17  & GLEAM~J000829$-$055839  & 0.790 $\pm$ 0.063 \\
  G4Jy~30  & GLEAM~J001707$-$125625  & 0.400  $\pm$ 0.031 \\
  G4Jy~34  & GLEAM~J001859$-$102248  & 0.440 $\pm$ 0.040  \\
  G4Jy~41  & GLEAM~J002125$-$005540  & 0.600 $\pm$ 0.260  \\
  G4Jy~48  & GLEAM~J002549$-$260211  & 0.780 $\pm$ 0.056 \\
  G4Jy~49  & GLEAM~J002609$-$124749  & 0.440 $\pm$ 0.035 \\
  G4Jy~64  & GLEAM~J003508$-$200354  & 0.470 $\pm$ 0.034 \\
  G4Jy~82  & GLEAM~J004441$-$353029  & 0.410 $\pm$ 0.041 \\
  G4Jy~98  & GLEAM~J005408$-$033354  & 0.750 $\pm$ 0.056 \\
  G4Jy~108  & GLEAM~J005906$-$170033  & 0.490 $\pm$ 0.035 \\
  G4Jy~109  & GLEAM~J010010$-$174841  & 0.480 $\pm$ 0.035 \\
  G4Jy~127  & GLEAM~J010925$-$344712  & 0.540 $\pm$ 0.062 \\
  G4Jy~134  & GLEAM~J011612$-$113610  & 0.490 $\pm$ 0.037 \\
  G4Jy~136  & GLEAM~J011651$-$205202  & 0.900 $\pm$ 0.065 \\ 
  G4Jy~148  & GLEAM~J012227$-$042123  & 0.800 $\pm$ 0.070  \\
  G4Jy~178  & GLEAM~J014013$-$095654  & 0.960 $\pm$ 0.077 \\  
  G4Jy~180  & GLEAM~J014127$-$270606  & 0.650 $\pm$ 0.048 \\
  G4Jy~187  & GLEAM~J014645$-$053758  & 0.540 $\pm$ 0.081 \\
  G4Jy~199  & GLEAM~J015323$-$033359  & 0.760 $\pm$ 0.093 \\
  G4Jy~211  & GLEAM~J015843$-$141308  & 0.870 $\pm$ 0.085 \\
  G4Jy~216  & GLEAM~J020157$-$113234  & 0.670 $\pm$ 0.074 \\
\hline
\label{tab:ips}
\end{tabular}
\end{table}

\subsection{Spectral information}
\label{sec:indexanalysis}

\begin{figure*}
\vspace{-0.7cm}
\centering
\subfigure[Spectral indices in the G4Jy catalogue]{
	\includegraphics[scale=0.4]{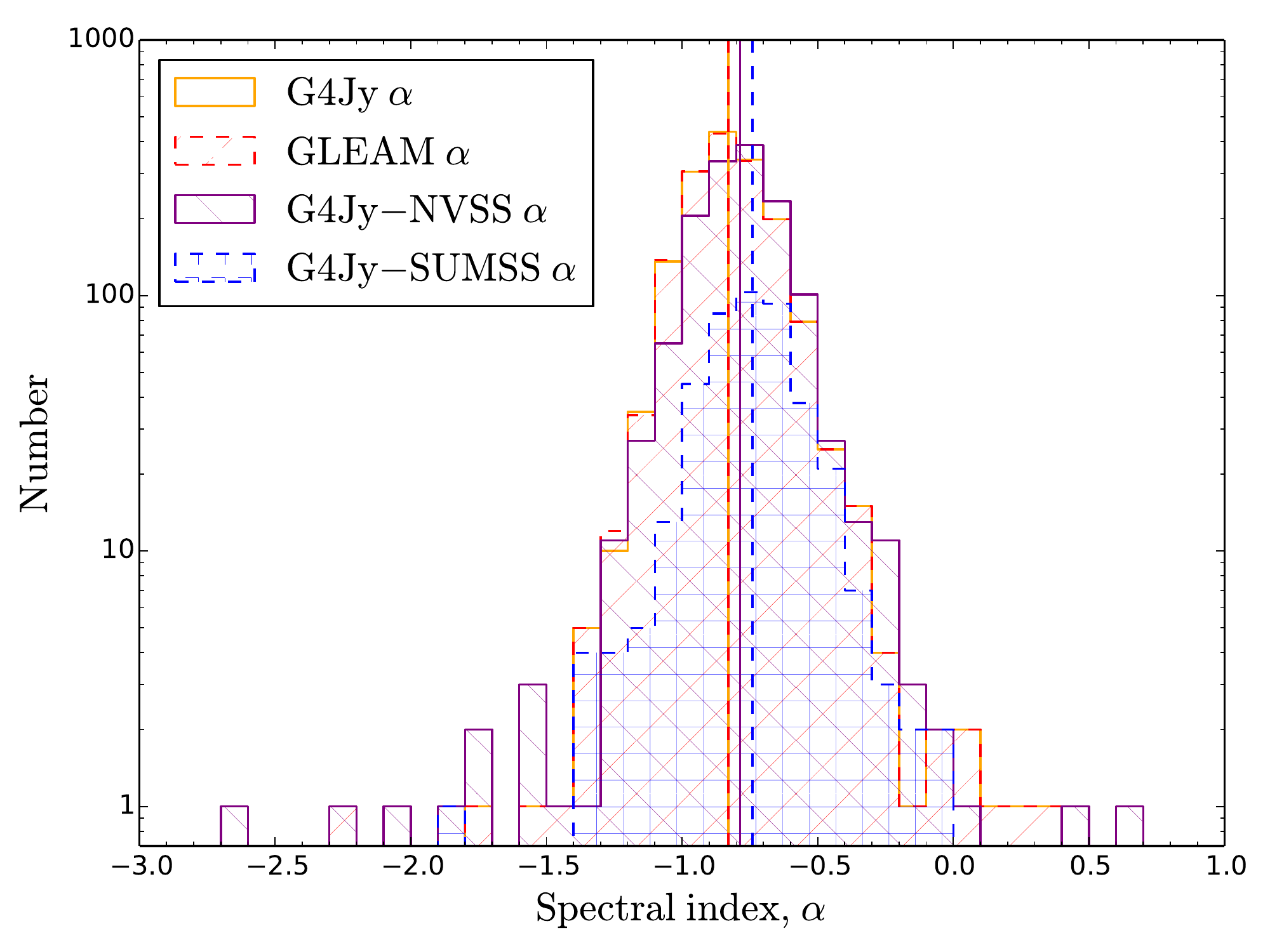} 
	} 
\subfigure[G4Jy~$\alpha$ by morphology]{
	\includegraphics[scale=0.4]{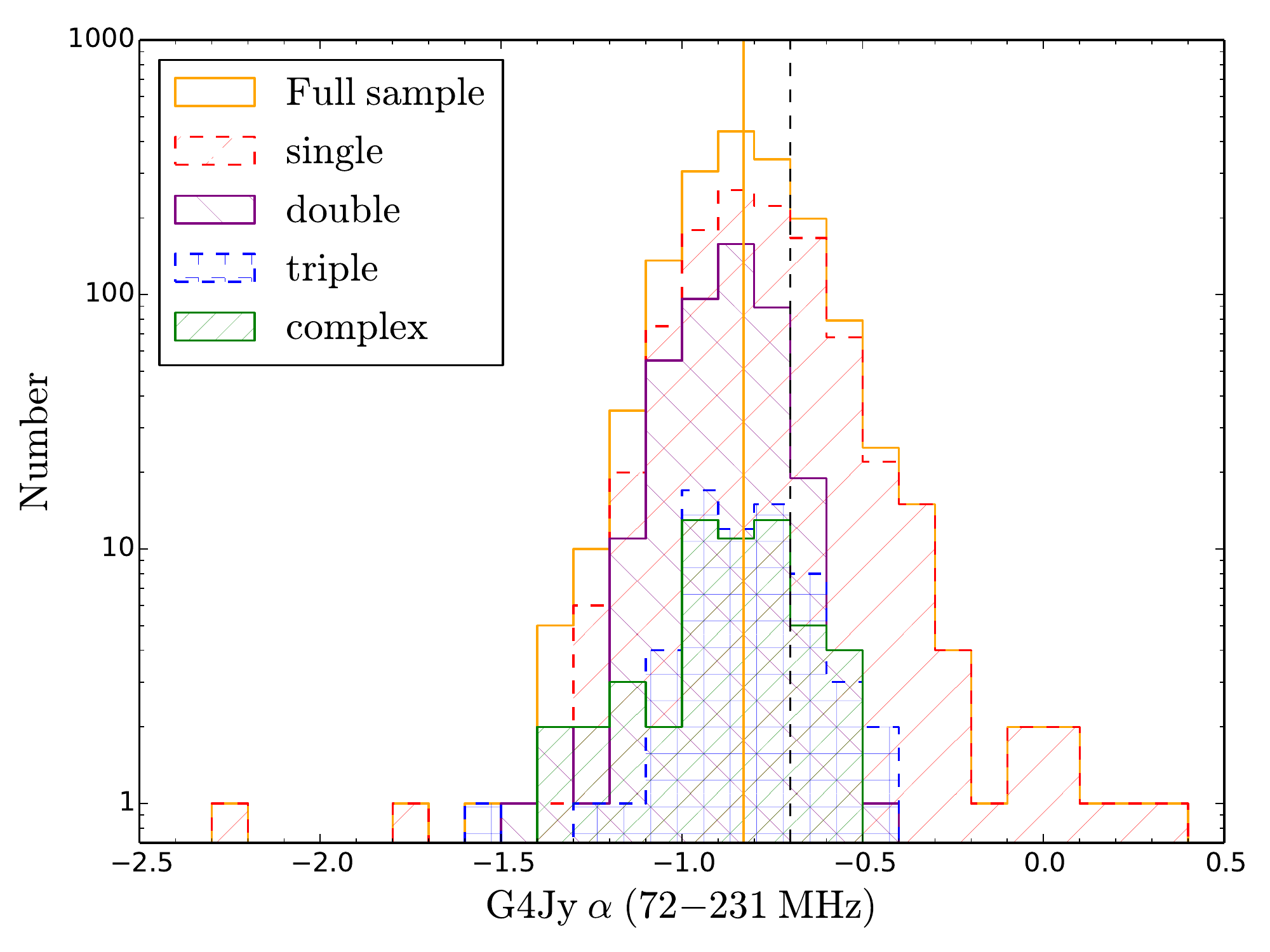}
	} 
\caption{ (a) The distributions for the four sets of spectral index provided in the G4Jy catalogue: G4Jy~$\alpha$, GLEAM~$\alpha$, G4Jy--NVSS~$\alpha$, and G4Jy--SUMSS~$\alpha$ (see Section~\ref{sec:fouralphas}). The median values for each spectral index are indicated by vertical lines (using the same colour and linestyle as for the corresponding histogram; see legend). (b) The distribution in G4Jy~$\alpha$ for the full sample, and for sources with `single', `double', `triple', and `complex' morphology in NVSS/SUMSS/TGSS (Sections~\ref{sec:morphology} and \ref{sec:indexanalysis}). The black, dashed, vertical line is where $\alpha = -0.7$, which is the canonical spectral-index that we use for extrapolation of flux densities (assuming $S_{\nu} \propto \nu^{\alpha}$). For comparison, we also plot the median G4Jy~$\alpha$ value for the full sample (orange, solid, vertical line).  }\label{alphadistributions}
\end{figure*}

In Figure~\ref{alphadistributions}a we present the distributions for each of the four spectral indices that we provide in the G4Jy catalogue (see Section~\ref{sec:fouralphas}). Although these refer to the `full sample', we remind the reader that a different number of G4Jy sources (or GLEAM components, as the case may be) is used for each distribution, as indicated in Table~\ref{tab:alphastatistics}. In the following analysis, $\alpha^{\mathrm{231\,MHz}}_{\mathrm{72\,MHz}}$ = G4Jy~$\alpha$, $\alpha^{\mathrm{1400\,MHz}}_{\mathrm{151\,MHz}}$ = G4Jy--NVSS~$\alpha$, and $\alpha^{\mathrm{843\,MHz}}_{\mathrm{151\,MHz}}$ = G4Jy--SUMSS~$\alpha$ (with each assuming radio emission of the spectral form, $S_{\nu} \propto \nu^{\alpha}$).

Note that for the majority of the sample (i.e. 1603/1863 = 86\% of the sources), a power-law spectrum is an accurate description of the total radio emission between 72 and 231\,MHz. If this were not the case, the reduced-$\chi^2$ value corresponding to $\alpha^{\mathrm{231\,MHz}}_{\mathrm{72\,MHz}}$ (the `G4Jy\_alpha' column in the G4Jy catalogue) would be $>1.93$ and we would mask the spectral index for the catalogue. Therefore, for the remaining 14\% of sources, the radio emission shows evidence of spectral curvature within the GLEAM band. Further evidence of curvature is apparent from the median value for the low-frequency spectral-index ($\alpha^{\mathrm{231\,MHz}}_{\mathrm{72\,MHz}}$) being steeper than the median value for the spectral-index calculated between 151\,MHz and 843\,MHz/1400\,MHz ($\alpha^{\mathrm{843\,MHz}}_{\mathrm{151\,MHz}}$/$\alpha^{\mathrm{1400\,MHz}}_{\mathrm{151\,MHz}}$, respectively) -- see Table~\ref{tab:alphastatistics} and the vertical lines in Figure~\ref{alphadistributions}a. The reasons for this will be discussed further in Papers~III and IV, following additional analysis (White et al., in preparation).

Conversely, a source with a flatter spectrum within the GLEAM band than towards higher frequencies (i.e. $\alpha^{\mathrm{231\,MHz}}_{\mathrm{72\,MHz}} > \alpha^{\mathrm{843\,MHz}}_{\mathrm{151\,MHz}}$ or $\alpha^{\mathrm{1400\,MHz}}_{\mathrm{151\,MHz}}$) is likely to be turning over due to free-free absorption or synchrotron self-absorption \citep{Lacki2013}. Such sources have previously been identified in the EGC by \citet{Callingham2017}, and in cross-matching the G4Jy Sample with their catalogues, we find an overlap of: one GHz-peaked spectrum (GPS) source (G4Jy 1533; GLEAM~J192451$-$291426), 67 sources with a spectral peak between 72 and 1400\,MHz (listed in Table~\ref{tab:pkfreq}), and 19 sources with a spectral peak below 72\,MHz (listed in Table~\ref{tab:pk72}). Each of these sources are `single' in morphology, with the exception of G4Jy~1233 (GLEAM~J151340+260718), which we label as `complex'. This is due to its X-shaped morphology, which is shown in figure~4d of Paper~II. Furthermore, G4Jy~352, G4Jy~420, G4Jy~819, G4Jy~965, G4Jy~1597, G4Jy~1772, and G4Jy~1801 are 7 of 15 sources found to have low-frequency variability by \citet{Bell2019}, who use a preliminary version of the G4Jy Sample for their study.  

\begin{table*}
\centering 
\caption{67 G4Jy sources previously identified by \citet{Callingham2017} as having a spectral peak at a frequency ($\nu_{\mathrm{peak}}$) between 72 and 1400\,MHz (Section~\ref{sec:indexanalysis}). G4Jy~136 and G4Jy~178 are the strong scintillators mentioned in Section~\ref{sec:sizeanalysis}. }
\begin{tabular}{@{}rccrcc@{}}
 \hline
Source & GLEAM component & $\nu_{\mathrm{peak}}$ / MHz  & Source & GLEAM component  & $\nu_{\mathrm{peak}}$ / MHz \\
 \hline
  G4Jy~48  &  GLEAM~J002549$-$260211  &  $145\pm19$    &      G4Jy~1023  &  GLEAM~J124823$-$195915  &  $728\pm254$           \\
  G4Jy~124  &  GLEAM~J010837$-$285124  &  $132\pm11$      &        G4Jy~1055  &  GLEAM~J131139$-$221640  &  $86\pm11$            \\
  G4Jy~136  &  GLEAM~J011651$-$205202  &  $120\pm10$      &        G4Jy~1082  &  GLEAM~J133737$-$181138  &  $138\pm8$             \\
  G4Jy~137  &  GLEAM~J011815$-$012037  &  $98\pm14$      &        G4Jy~1096  &  GLEAM~J135050$-$131210  &  $89\pm10$           \\
  G4Jy~166  &  GLEAM~J013119+041532  &  $152\pm29$      &      G4Jy~1100  &  GLEAM~J135210$-$264927  &  $86\pm20$           \\
  G4Jy~169  &  GLEAM~J013212$-$065232  &  $85\pm12$      &      G4Jy~1128  &  GLEAM~J141335$-$202037  &  $103\pm34$            \\
  G4Jy~178  &  GLEAM~J014013$-$095654  &  $124\pm18$      &        G4Jy~1181  &  GLEAM~J143810+282136  &  $96\pm34$            \\
  G4Jy~212  &  GLEAM~J015951$-$743054  &  $142\pm9$       &        G4Jy~1217  &  GLEAM~J150506+034709  &  $82\pm33$            \\
  G4Jy~303  &  GLEAM~J025216+082612  &  $75\pm9$       &     G4Jy~1240  &  GLEAM~J151656+183021  &  $95\pm21$            \\
  G4Jy~304  &  GLEAM~J025245$-$710434  &  $244\pm36$      &      G4Jy~1244  &  GLEAM~J152005+201602  &  $81\pm37$            \\
  G4Jy~308  &  GLEAM~J025516$-$665653  &  $89\pm38$      &     G4Jy~1256  &  GLEAM~J152512$-$190250  &  $107\pm8$             \\
  G4Jy~339  &  GLEAM~J031610$-$682104  &  $121\pm9$       &       G4Jy~1258  &  GLEAM~J152548+030825  &  $112\pm30$            \\
  G4Jy~352  &  GLEAM~J032320+053413  &  $944\pm330$     &      G4Jy~1295  &  GLEAM~J160127$-$242956  &  $81\pm23$           \\
  G4Jy~370  &  GLEAM~J033626+130219  &  $122\pm13$      &      G4Jy~1344  &  GLEAM~J163145+115601  &  $72\pm26$            \\
  G4Jy~416  &  GLEAM~J040820$-$654458  &  $225\pm14$      &        G4Jy~1346  &  GLEAM~J163449$-$222208  &  $124\pm19$            \\
  G4Jy~419  &  GLEAM~J040906$-$175708  &  $124\pm12$      &      G4Jy~1356  &  GLEAM~J164048+122000  &  $112\pm26$            \\
  G4Jy~420  &  GLEAM~J041022$-$523247  &  $183\pm55$      &       G4Jy~1416  &  GLEAM~J173250+203813  &  $87\pm35$            \\
  G4Jy~588  &  GLEAM~J060308$-$793446  &  $195\pm22$      &        G4Jy~1472  &  GLEAM~J181935$-$634546  &  $153\pm22$            \\
  G4Jy~603  &  GLEAM~J061732$-$363412  &  $73\pm29$      &       G4Jy~1487  &  GLEAM~J183059$-$360229  &  $166\pm25$            \\
  G4Jy~699  &  GLEAM~J082143+174813  &  $76\pm33$      &      G4Jy~1528  &  GLEAM~J192103$-$621726  &  $76\pm27$            \\
  G4Jy~717  &  GLEAM~J083710$-$195152  &  $450\pm70$      &       G4Jy~1545  &  GLEAM~J192908$-$373251  &  $89\pm37$            \\
  G4Jy~752  &  GLEAM~J090652$-$682940  &  $82\pm27$      &      G4Jy~1597  &  GLEAM~J200608$-$022332  &  $103\pm27$            \\
  G4Jy~765  &  GLEAM~J092011+175322  &  $109\pm43$      &        G4Jy~1691  &  GLEAM~J212252$-$100316  &  $72\pm8$             \\
  G4Jy~819  &  GLEAM~J100557$-$414849  &  $240\pm15$      &       G4Jy~1699  &  GLEAM~J213121+091232  &  $85\pm32$            \\
  G4Jy~857  &  GLEAM~J103338$-$193406  &  $84\pm12$      &        G4Jy~1709  &  GLEAM~J213750$-$204234  &  $117\pm17$            \\
  G4Jy~862  &  GLEAM~J103828$-$700310  &  $257\pm18$      &       G4Jy~1728  &  GLEAM~J215023+144939  &  $190\pm36$            \\
  G4Jy~863  &  GLEAM~J103848$-$043115  &  $141\pm11$      &        G4Jy~1769  &  GLEAM~J221655$-$452144  &  $164\pm35$            \\
  G4Jy~897  &  GLEAM~J110735$-$442053  &  $91\pm9$       &       G4Jy~1772  &  GLEAM~J221942$-$275626  &  $134\pm20$            \\
  G4Jy~913  &  GLEAM~J111925$-$030255  &  $92\pm12$      &       G4Jy~1782  &  GLEAM~J222946$-$382358  &  $125\pm8$             \\
  G4Jy~925  &  GLEAM~J113331$-$271521  &  $146\pm21$      &        G4Jy~1811  &  GLEAM~J231109$-$562439  &  $73\pm37$            \\
  G4Jy~965  &  GLEAM~J115421$-$350525  &  $274\pm54$      &      G4Jy~1826  &  GLEAM~J232503$-$405129  &  $88\pm28$           \\
  G4Jy~971  &  GLEAM~J120232$-$024005  &  $99\pm12$      &      G4Jy~1839  &  GLEAM~J233343$-$305753  &  $112\pm16$         \\
  G4Jy~978  &  GLEAM~J121256+203237  &  $73\pm26$      &        G4Jy~1853  &  GLEAM~J235025$-$022442  &  $93\pm14$         \\
  G4Jy~986  &  GLEAM~J121806$-$460030  &  $75\pm25$      &     &     &   \\
\hline
\label{tab:pkfreq}
\end{tabular}
\end{table*}

\begin{table}
\centering 
\caption{19 G4Jy sources previously identified by \citet{Callingham2017} as having a spectral peak below 72\,MHz (Section~\ref{sec:indexanalysis}). }
\begin{tabular}{@{}rc@{}}
\hline
Source  & GLEAM component   \\
\hline
    G4Jy~54  &  GLEAM~J002853$-$552328 \\
    G4Jy~97  &  GLEAM~J005351$-$491402 \\
    G4Jy~148  &  GLEAM~J012227$-$042123 \\
    G4Jy~179  &  GLEAM~J014109+135319 \\
    G4Jy~389  &  GLEAM~J035154+062757 \\
    G4Jy~411  &  GLEAM~J040534$-$130813 \\ 
    G4Jy~497  &  GLEAM~J044642$-$354217 \\
    G4Jy~538  &  GLEAM~J052139$-$204737 \\
    G4Jy~777  &  GLEAM~J093631+042207 \\
    G4Jy~835  &  GLEAM~J101809$-$314411 \\ 
    G4Jy~840  &  GLEAM~J102154+215930 \\
    G4Jy~947  &  GLEAM~J114426$-$174119 \\ 
    G4Jy~1121  &  GLEAM~J140635$-$273613 \\
    G4Jy~1134  &  GLEAM~J141510$-$105750 \\
    G4Jy~1144  &  GLEAM~J141908+062832 \\
    G4Jy~1233  &  GLEAM~J151340+260718 \\
    G4Jy~1471  &  GLEAM~J181806$-$515801 \\ 
    G4Jy~1476  &  GLEAM~J182159$-$745745 \\
    G4Jy~1801  &  GLEAM~J230223$-$371805 \\
\hline
\label{tab:pk72}
\end{tabular}
\end{table}

Returning to the distribution in $\alpha^{\mathrm{231\,MHz}}_{\mathrm{72\,MHz}}$ for the full sample, we also present this spectral index for each of the categories in morphology (Figure~\ref{alphadistributions}b). The `doubles' and `triples' have (mostly) steep spectral-indices that span a range of $-1.6$ to $-0.4$, which is as expected if the lobes are dominating the radio emission. Meanwhile, for `single' sources, the range in $\alpha^{\mathrm{231\,MHz}}_{\mathrm{72\,MHz}}$ is considerably wider, from $-2.3$ to $0.4$. This subset encompasses ultra-steep sources at high redshift, and flat-spectrum sources (where the core is believed to be dominating the radio emission)\footnote{We consider flat-spectrum sources as having $-0.5 < \alpha < 0.5$, and so $\alpha > 0.5$ as signifying an inverted spectrum. However, note that the radio spectrum of the Flame Nebula (G4Jy~571) is clearly inverted (figure~2 of Paper~II) but its spectral curvature within the GLEAM band is such that G4Jy~$\alpha$ is masked for the catalogue. On the other hand, its spectral index between 151 and 1400\,MHz is provided ($-0.67\pm0.01$) and appears in the G4Jy--NVSS~$\alpha$ distribution in Figure~\ref{alphadistributions}a.}. However, this is complicated by the size of the MWA beam, which leads to G4Jy/GLEAM flux-densities being affected by confusion. As a result, the fitted spectral-index of the G4Jy source differs from its true value.

\section{Summary} \label{sec:finalsummary}

Due to radio sources exhibiting a great variety of morphologies, identifying the correct host-galaxy (if appropriate) is a difficult task. We have invested considerable effort into this for the brightest radio-sources in the GLEAM extragalactic catalogue (EGC), including repeated visual inspection and thorough, time-intensive consultation of the literature (see Paper~II for details on individual sources). Here we summarise the work done in defining the GLEAM 4-Jy (G4Jy) Sample (i.e. sources with $S_{\mathrm{151\,MHz}}>4$\,Jy) and preparing the G4Jy catalogue:

\begin{enumerate}

\item{We confirm that `A-team' sources, the Magellanic Clouds, and the Orion Nebula all have $S_{\mathrm{151\,MHz}}>4$\,Jy but are {\it not} in the G4Jy Sample, following their absence from the parent catalogue, EGC. However, we provide an estimate of the integrated flux-density for these sources for reference.}

\item{Since the MWA provides $\sim$2$'$ resolution, we use the better spatial-resolution of TGSS, NVSS and SUMSS to interpret the morphology of the brightest radio-sources in the southern sky. This allows us to determine what radio emission is associated together, and which G4Jy sources are affected by confusion. We also use NVSS and SUMSS to calculate the summed flux-density and angular size at $\sim$1\,GHz.}

\item{In addition, we use the NVSS/SUMSS components to calculate brightness-weighted centroid positions, which aid our identification of the host galaxy in the mid-infrared. The latter is done through visual inspection, and all images and overlays are made available online (see https://github.com/svw26/G4Jy for details).}

\item{When inspecting the overlays, we remind the user to be aware of artefacts in NVSS, SUMSS, and TGSS. We characterise the TGSS artefacts in terms of position angle and angular separation, with respect to the G4Jy source, as this has not been done before.}

\item{The G4Jy catalogue contains 10 GLEAM components (corresponding to 6 G4Jy sources) that have been re-fitted using {\sc Aegean}, which is the source-finding software used to create the EGC. As such, the GLEAM flux-densities for these components do not appear in the parent catalogue.}

\item{Also within our value-added catalogue are AllWISE host-galaxy positions (and magnitudes) for 1,606 of the 1,863 sources in the G4Jy Sample. In the case of a cluster relic and a nebula, identifying a `host' is inappropriate, so these fields are left blank. The remaining sources either have a host galaxy that is {\it not} in the AllWISE catalogue, or there is uncertainty as to the correct identification. The latter are being followed up using MeerKAT Open Time (PI: White), to confirm the position of the radio core.}

\item{78 G4Jy sources are resolved into multiple GLEAM components by the MWA Phase-I beam. Therefore, we provide integrated flux-densities summed over these components, per source, and indicate their association via our source-naming scheme. Whilst the GLEAM-band spectral-index is inherited from the EGC {\it per GLEAM component} (for all but the re-fitted components), we use the summed, integrated flux-densities to re-calculate this spectral index {\it per G4Jy source}. In addition, we use the total $S_{\mathrm{151\,MHz}}$ to determine the spectral index between 151\,MHz and $\sim$1\,GHz (assuming a power-law description, $S_{\nu} \propto \nu^{\alpha}$).}

\item{In order to improve the completeness of the G4Jy Sample, we perform cross-checks against existing radio-source samples, and use internal matching of the EGC to identify extended sources that would otherwise have been missed. Following this, we estimate that the sample is 99.50--99.95\% complete to $S_{\mathrm{151\,MHz}}=4$\,Jy on the GLEAM flux-density scale, which corresponds to a completeness of 95.5\% on the flux-density scale of \citet{Perley2017}. }

\item{Note that the above estimates of sample completeness are relevant over the footprint of the EGC minus the region defined by $-18.3^{\circ} < b < -10.0^{\circ}$, $65.4^{\circ} < l < 81.1^{\circ}$, Dec.\,$<30.0^{\circ}$. The reason for this subtraction is that GLEAM flux-densities are not well-characterised over this area of the sky, due to the influence of Cygnus A.}

\item{Of the 173 radio galaxies belonging to the well-studied 3CRR sample, 67 of these overlap with the G4Jy Sample. We compare the GLEAM flux-density at 178\,MHz to that provided in the 3CRR catalogue, and find that the GLEAM value is systematically lower. However, we note that this may be due to several factors: the larger `3CRR' beams detecting unrelated emission, errors (and possible non-linearity) in the flux-density scales of the two catalogues, and that the GLEAM calibration-error is worse at these high declinations ($>10^{\circ}$).} 

\item{Preliminary analysis of the full sample shows that the median angular size (at 843/1400\,MHz) is 43.8$''$, and that the radio spectrum is more-often steeper at low frequencies (72--231\,MHz) than between 151 and 843/1400\,MHz. For the `doubles' and `triples' that make up 30\% of the sample, the 72--231\,MHz spectral index spans a range of $-1.6$ to $-0.4$, as expected for lobe-dominated radio sources. However, 21\% of the G4Jy Sample has low-frequency flux-densities that may be affected by confusion, and so these measurements and the derived spectral indices should be treated with caution.}

\end{enumerate}

The result of many iterations, between the G4Jy catalogue and accompanying overlays, is a firm base upon which this legacy dataset can be reliably cross-matched with information at other wavelengths. Such a large, complete, unbiased radio-source sample is required for investigating, for example, the production of powerful jets and their interaction with the environment. Furthermore, by exploiting the excellent spectral-coverage provided by the MWA, we can tightly constrain the spectral behaviour of the G4Jy sources, and determine the prevalence of `restarted' AGN activity. 


\section{Dedication}

Papers I and II are dedicated to the memory of Richard Hunstead, who was very helpful with the assessment of the sources presented in this work, and provided hitherto unpublished radio-images.

\begin{acknowledgements}

We thank the anonymous referee for their time in reviewing our manuscript. SVW would like to thank Chris Jordan, for his help with installing and running the cross-matching software (MCVCM), as well as Dave Pallot and the ICRAR Data Intensive Astronomy team, for their help with fixing and updating the G4Jy Sample Server. In addition, SVW thanks Ron Ekers, Robert Laing, Elaine Sadler, Tom Mauch, and Huib Intema for useful discussions. We acknowledge the International Centre for Radio Astronomy Research (ICRAR), which is a joint venture between Curtin University and The University of Western Australia, funded by the Western Australian State government. We acknowledge the Pawsey Supercomputing Centre which is supported by the Western Australian and Australian Governments. The financial assistance of the South African Radio Astronomy Observatory (SARAO) towards this research is hereby acknowledged (www.ska.ac.za).

This scientific work makes use of the Murchison Radio-astronomy Observatory, operated by CSIRO. We acknowledge the Wajarri Yamatji people as the traditional owners of the Observatory site. Support for the operation of the MWA is provided by the Australian Government (NCRIS), under a contract to Curtin University administered by Astronomy Australia Limited. 

GMRT is run by the National Centre for Radio Astrophysics of the Tata Institute of Fundamental Research (TIFR). The Australia Telescope Compact Array (ATCA) is part of the Australia Telescope National Facility which is funded by the Australian Government for operation as a National Facility managed by CSIRO.

This publication made use of data products from the {\it Wide-field Infrared Survey Explorer}, which is a joint project of the University of California, Los Angeles, and the Jet Propulsion Laboratory/California Institute of Technology, funded by the National Aeronautics and Space Administration (NASA).

This research has made use of the NASA/IPAC Extragalactic Database (NED), which is operated by the Jet Propulsion Laboratory, California Institute of Technology, under contract with the National Aeronautics and Space Administration. We also acknowledge the use of NASA's SkyView facility [http://skyview.gsfc.nasa.gov] located at NASA Goddard Space Flight Center. This research has made use of the SIMBAD astronomical database \citep{Wenger2000},
operated at CDS, Strasbourg, France.

This research has made use of the VizieR catalogue access tool, CDS, Strasbourg, France (DOI: 10.26093/cds/vizier). The original description of the VizieR service was published in A\&AS 143, 23 \citep{Ochsenbein2000}.

This research made use of Montage. It is funded by the National Science Foundation under Grant Number ACI-1440620, and was previously funded by the National Aeronautics and Space Administration's Earth Science Technology Office, Computation Technologies Project, under Cooperative Agreement Number NCC5-626 between NASA and the California Institute of Technology.

Finally, the following open-source software was used for data visualisation and processing: {\sc topcat} \citep{Taylor2005}, SAOImage DS9 \citep{SAO2000, Joye2003}, NumPy \citep{Oliphant2006}, Astropy \citep{Astropy2013}, APLpy \citep{Robitaille2012}, Matplotlib \citep{Hunter2007}, and SciPy \citep{Jones2001}.

\end{acknowledgements}

\begin{appendix}

\section[]{The Orion Nebula}
\label{sec:orion}

The GLEAM images of Orion A resolve the nebula into a rough trapezoid about $30'\times25'$ in size and a $5'$-diameter circular source to the north-east. We use the \textsc{poly\_flux} script\footnote{https://github.com/nhurleywalker/polygon-flux} to determine integrated flux-density measurements covering the entire nebula. This software calculates a background level from a region surrounding the object of interest, excluding any selected regions, which in this case we set to obvious areas of unrelated emission. The flux-density measurements at 88, 118, 154, and 200\,MHz are, respectively, $22.8\pm1.8$, $46.3\pm3.7$, $65.7\pm5.3$, and $81.8\pm6.5$\,Jy. The measurement errors are dominated by the 8\,\% flux-density accuracy of GLEAM at this declination.

\cite{Terzian1970} used the Arecibo Observatory to measure the flux density of the Orion Nebula at 73.8, 111.5, and 196.5\,MHz, measuring flux densities of $32\pm15$, $62\pm7$, and $108\pm11$\,Jy, respectively. The resolution of Arecibo at these frequencies is $85'\times120'$, $54'\times77'$, and $33'\times43\farcm5$, respectively. The larger values measured by this instrument are due to its low-resolution beam confusing surrounding radio sources with emission from the nebula\footnote{It is unlikely to be due to the MWA resolving out extended structure, as this instrument is sensitive to radio emission at angular scales up to $\sim$600$'$ \citep{Wayth2015}, which is larger than the Arecibo beam-sizes.}. We correct for this confusion by measuring the flux density contained within ellipses of the beam size, centred on the Orion Nebula, and comparing it to GLEAM measurements, deriving correction factors at 154\,MHz. These correction factors are 54\,\%, 64\,\%, and 76\,\%, respectively, leading to corrected values of $17\pm8$, $40\pm4$, and $82\pm8$\,Jy, respectively.

We fit a curved spectrum to the data, of the form $S_{\nu}\propto\nu^\alpha \exp{q(\ln{\nu})^2}$. Using solely the GLEAM measurements, we obtain $S_\mathrm{151\,MHz}=67.3\pm0.1$\,Jy, while including the corrected data from \cite{Terzian1970} yields $S_\mathrm{151\,MHz}=66.6\pm0.1$\,Jy. This is remarkably close given the simplicity of the correction method. The shape of the spectrum is identical, with $\alpha=1.1\pm0.2$ and $q=-1.5\pm0.5$.

\section[]{GLEAM components removed from the G4Jy Sample} \label{sec:removedcomponents}

\begin{figure*}
\centering
\subfigure[GLEAM~J093918+015948]{
	\includegraphics[scale=1.0]{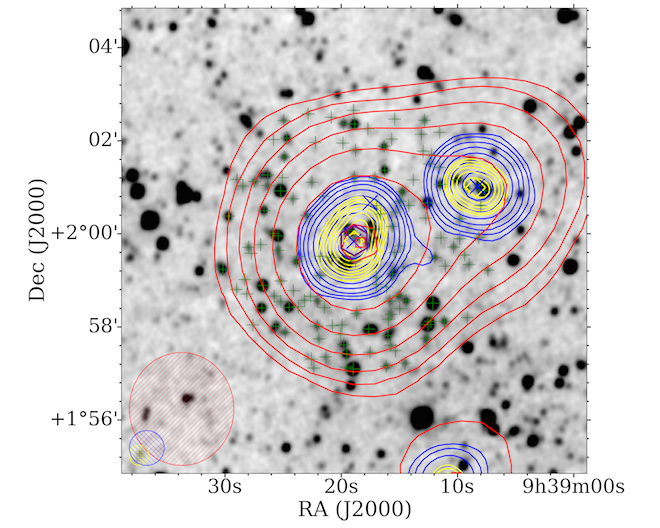} 
	} 
\subfigure[GLEAM~J101051$-$020137]{
	\includegraphics[scale=1.0]{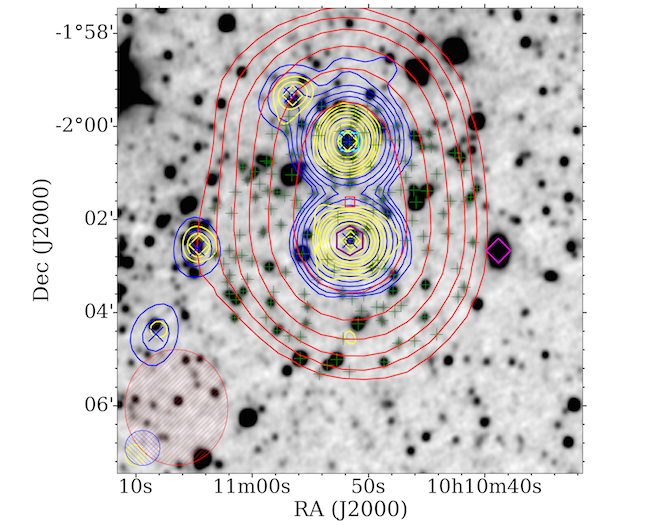} 
	} \\ 
\subfigure[GLEAM~J201707$-$310305]{
	\includegraphics[scale=1.0]{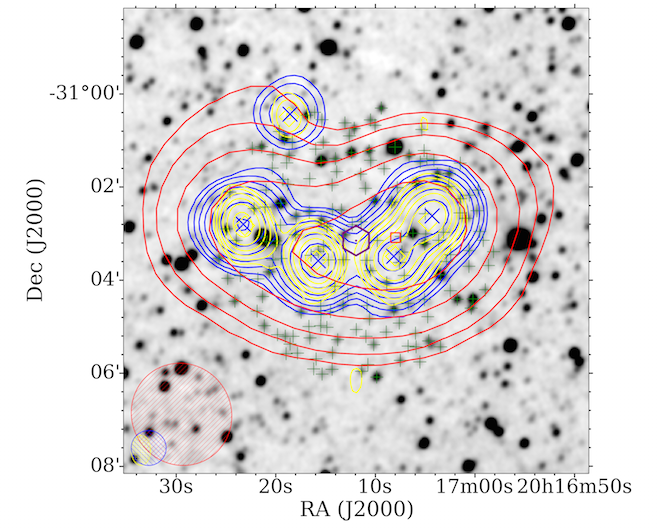} 
	} 
\subfigure[GLEAM~J202336$-$191144]{
	\includegraphics[scale=1.0]{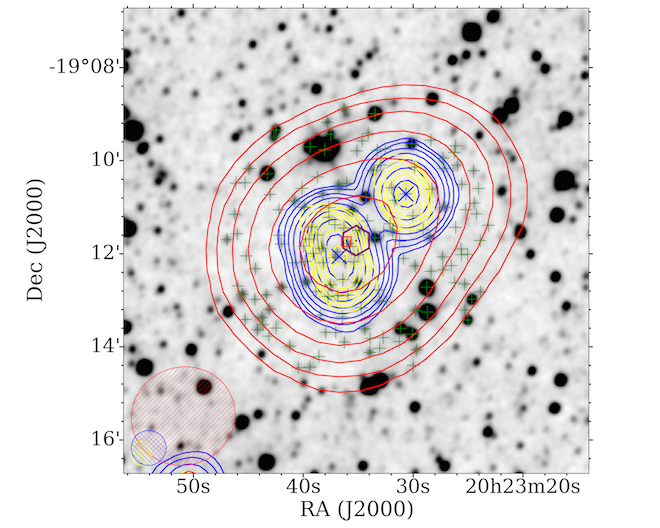} 
	} \\ 
\subfigure[GLEAM~J222751$-$303344]{
	\includegraphics[scale=1.0]{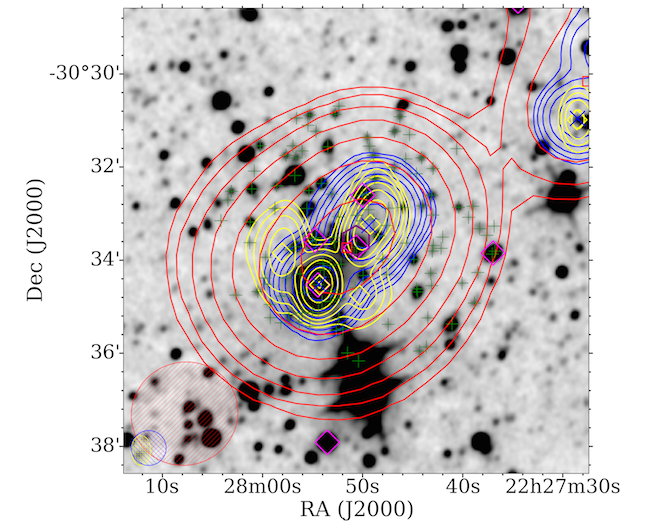} 
	} 
\caption{Overlays for five GLEAM components that were removed from the G4Jy Sample, following visual inspection and investigation of the relative flux-densities for the confused sources (Section~\ref{sec:morphology} and Appendix~\ref{sec:removedcomponents}). The contours, symbols, and beams are the same as for Figure~\ref{tiefighterOverlay}, with AllWISE positions (green plus signs) within 3$'$ of the centroid position (purple hexagon) also plotted. \label{figureofremoved}}
\end{figure*}

Five GLEAM components with $S_{\mathrm{151\,MHz}}>4$\,Jy were removed from the G4Jy Sample (Section~\ref{sec:morphology}), following the initial selection (Section~\ref{sec:sampleselection}). This is because multiple unrelated sources were identified through visual inspection of their overlays (Figure~\ref{figureofremoved}), suggesting that confused GLEAM emission was the reason for these components appearing above the selection threshold. This is supported by flux-density measurements at higher spatial-resolution from TGSS, which indicate that no {\it single source} is likely to have $S_{\mathrm{151\,MHz}}>4$\,Jy. Details are provided here for reference, including the relevant TGSS flux-densities ($S_{\mathrm{150\,MHz}}$). 

{\it GLEAM~J093918+015948:} $S_{\mathrm{151\,MHz}} = 4.55 \pm 0.03 $\,Jy. Two unrelated sources with $S_{\mathrm{150\,MHz}} = 4.40$, 1.08\,Jy. We note that a correction needs to be applied in order for the TGSS catalogue to have the same flux-scale as GLEAM \citep{HurleyWalker2017b}, but for this work we are only interested in the {\it relative} flux-densities of sources that are confused by the MWA beam. Here, the brighter source accounts for 80\% of the total TGSS emission. This corresponds to 3.66\,Jy at 151\,MHz in GLEAM.

{\it GLEAM~J101051$-$020137:} $S_{\mathrm{151\,MHz}} = 4.81 \pm 0.03 $\,Jy. This GLEAM component is dominated by two unrelated point-sources ($S_{\mathrm{150\,MHz}} = 3.31$, 2.18\,Jy), with two fainter sources ($S_{\mathrm{150\,MHz}} = 0.06$, 0.04\,Jy) nearby. None of the sources are above the 4-Jy threshold at 151\,MHz.

{\it GLEAM~J201707$-$310305:} $S_{\mathrm{151\,MHz}} = 4.60 \pm 0.06 $\,Jy. We agree with \citet{Jones1992}, regarding 2014$-$312, that the bulk of the radio emission is from a `double' ($S_{\mathrm{150\,MHz}} = 1.74 + 1.27 = 3.01$\,Jy) and two point sources ($S_{\mathrm{150\,MHz}} = 1.14$, 0.93\,Jy) to the east of this. The confused source ($S_{\mathrm{150\,MHz}} = 0.07$\,Jy) towards the north is also considered, but none of these sources have $S_{\mathrm{151\,MHz}} > 4$\,Jy. Based on the presence of a bright mid-infrared source between the two point sources, we also consider the possibility that they are associated and form another `double'. However, its summed flux-density ($S_{\mathrm{150\,MHz}} = 1.14 + 0.93 = 2.07$\,Jy) would still be too low to warrant retaining this GLEAM component for the G4Jy Sample. 

{\it GLEAM~J202336$-$191144:} $S_{\mathrm{151\,MHz}} = 4.54 \pm 0.05 $\,Jy. Two unrelated sources with $S_{\mathrm{150\,MHz}} = 3.47$, 1.08\,Jy. Neither source is above the 4-Jy threshold at 151\,MHz.

{\it GLEAM~J222751$-$303344:} $S_{\mathrm{151\,MHz}} = 4.40 \pm 0.02 $\,Jy. The MWA beam has blended emission from a `compact' source ($S_{\mathrm{150\,MHz}} = 1.23$\,Jy) and a head-tail galaxy to the north-west ($S_{\mathrm{150\,MHz}} = 1.83$\,Jy). This interpretation is based upon a VLA observation of 2225$-$308 by \citet{Ekers1989}. However, we draw attention to the unusual extension of the TGSS contours ($S_{\mathrm{150\,MHz}} = 0.53$, 0.17\,Jy) compared to the NVSS contours. The TGSS contours suggest that the compact radio-source may in fact be a `triple', with old, lobe emission evident only at low radio-frequencies. Even if the three southern-most TGSS components are associated, the integrated flux-density is still insufficient for the southern source to cross the 4-Jy threshold.

\section{Multi-component G4Jy sources}
\label{app:multiGLEAMsources}

\begin{figure*}
\centering
\subfigure[G4Jy~7]{
	\includegraphics[scale=1.1]{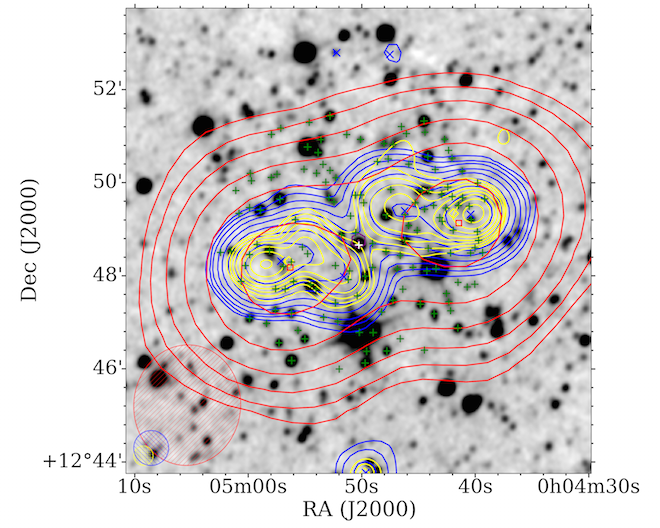}
	}
\subfigure[G4Jy~126]{
	\includegraphics[scale=1.1]{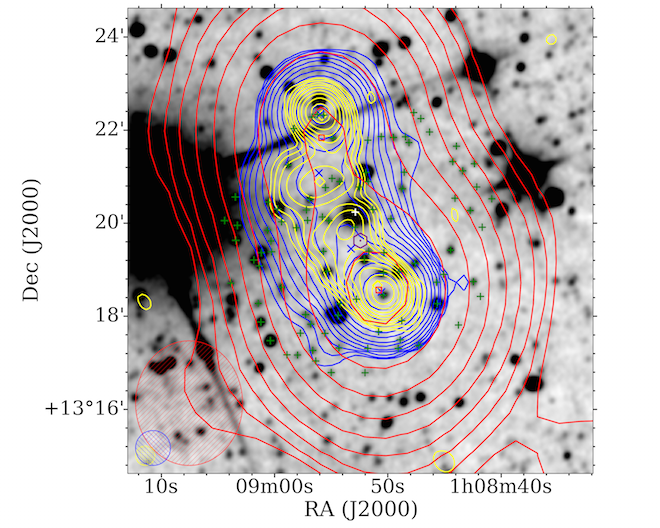}
	}
\subfigure[G4Jy~360]{
	\includegraphics[scale=1.1]{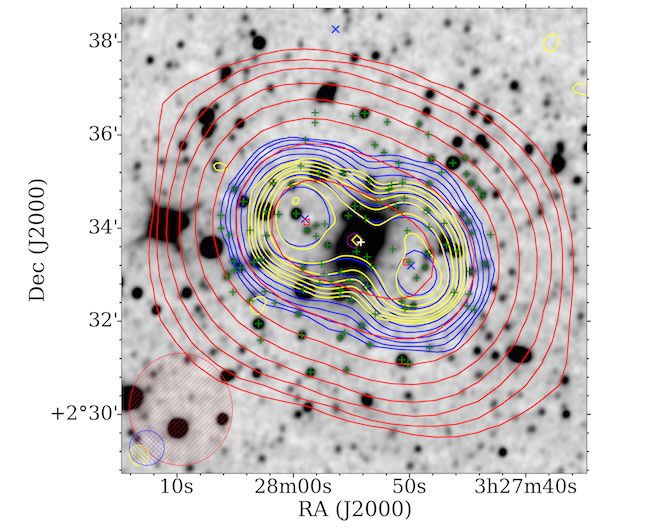}
	}
\subfigure[G4Jy~366]{
	\includegraphics[scale=1.1]{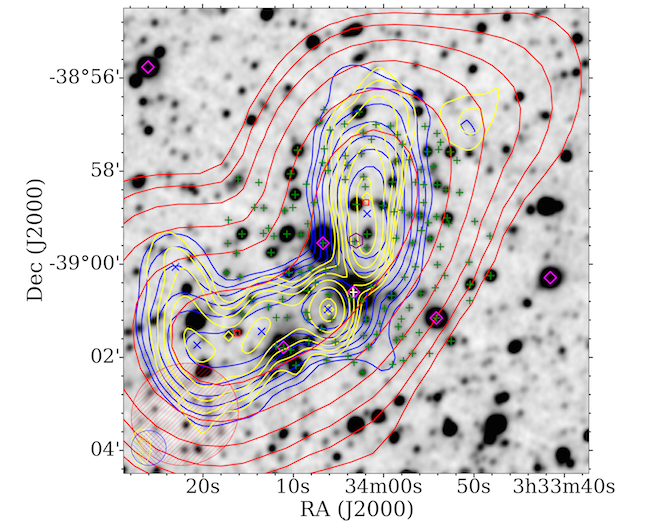}
	}
\subfigure[G4Jy~386]{
	\includegraphics[scale=1.1]{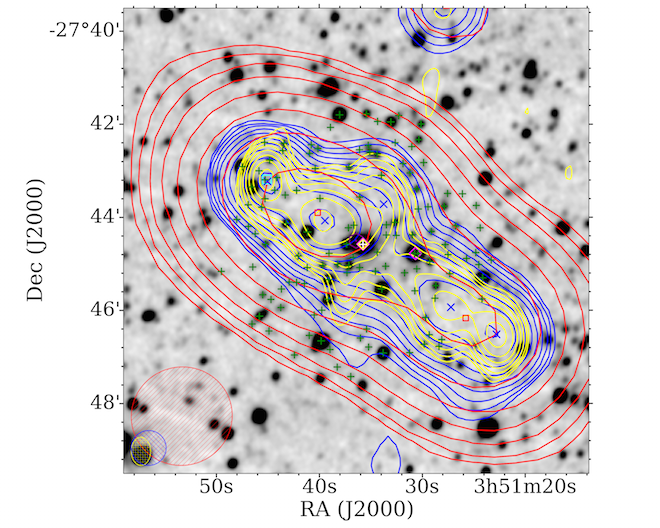}
	}
\subfigure[G4Jy~400]{
	\includegraphics[scale=1.1]{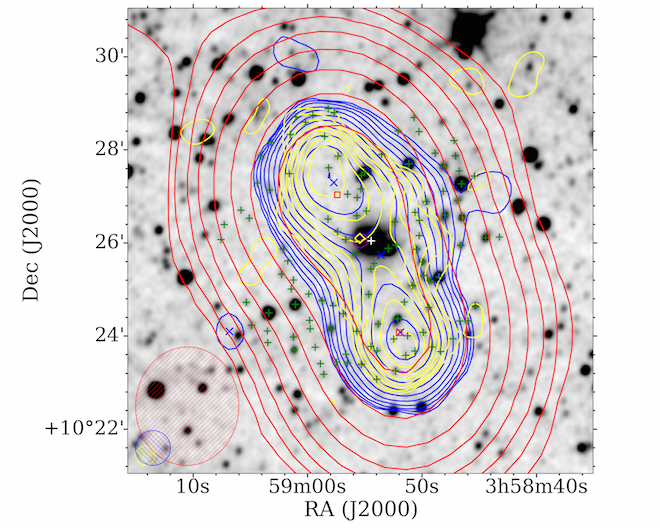}
	}
\caption{Overlays for G4Jy sources that have multiple GLEAM components (Table~\ref{multiGLEAMsources}, Appendix~\ref{app:multiGLEAMsources}). The datasets, contours, symbols, and beams are the same as those used for Figure~\ref{tiefighterOverlay}, but where blue contours, crosses, and ellipses correspond to NVSS {\it or} SUMSS. In addition, positions from AllWISE are indicated by green plus signs, with cross-identified host-galaxies highlighted in white. }
\label{remaining1}
\end{figure*}

\begin{figure*}
\centering
\subfigure[G4Jy~414]{
	\includegraphics[scale=1.1]{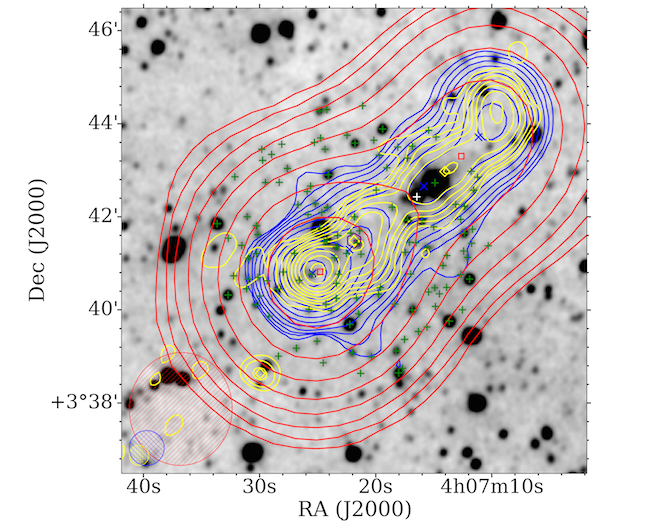}
	}
\subfigure[G4Jy~462]{
	\includegraphics[scale=1.1]{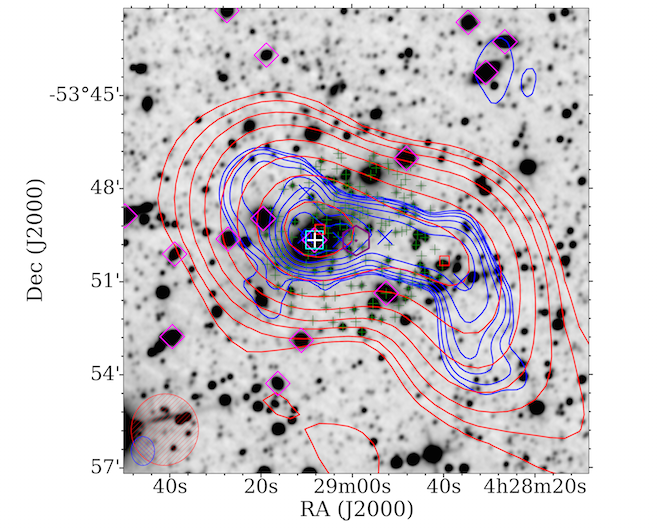}
	}
\subfigure[G4Jy~531]{
	\includegraphics[scale=1.1]{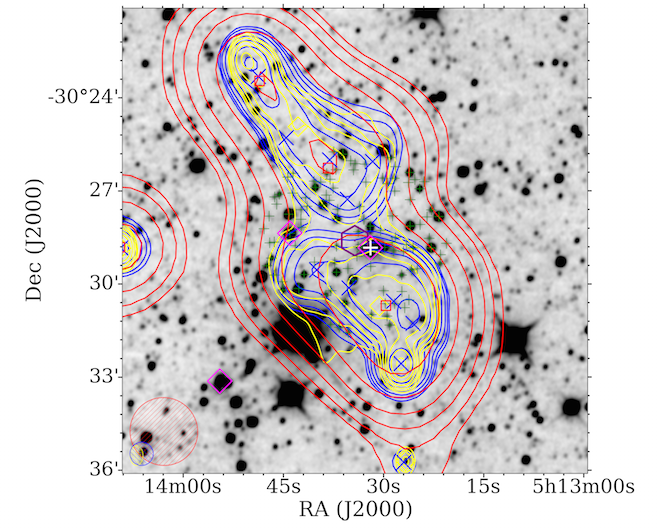}
	}
\subfigure[G4Jy~619]{
	\includegraphics[scale=1.1]{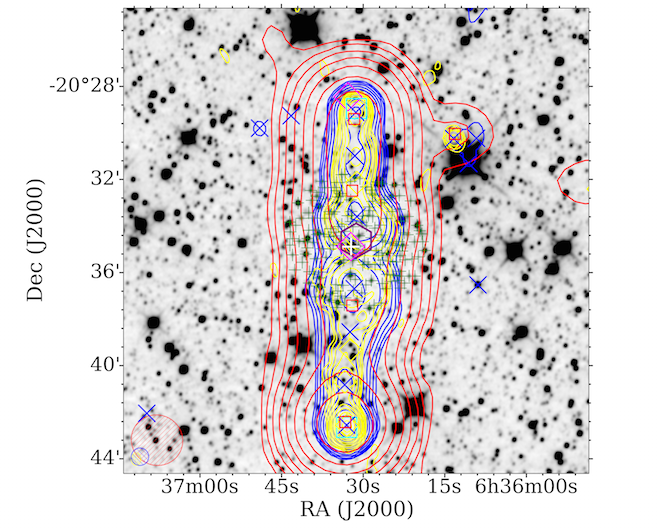}
	}
\subfigure[G4Jy~644]{
	\includegraphics[scale=1.1]{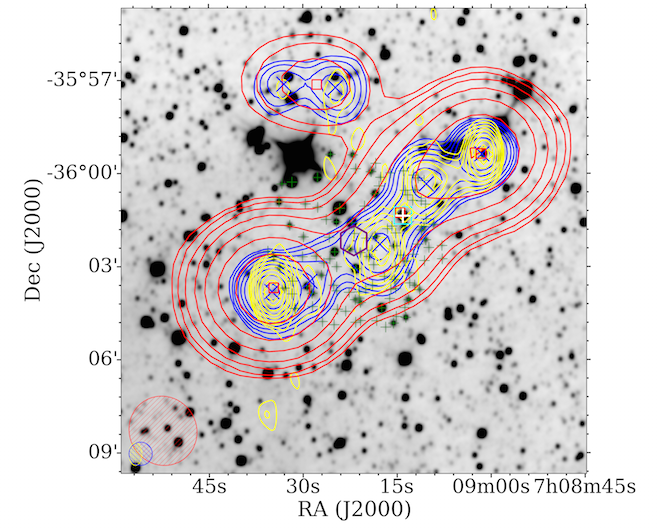}
	}
\subfigure[G4Jy~659]{
	\includegraphics[scale=1.1]{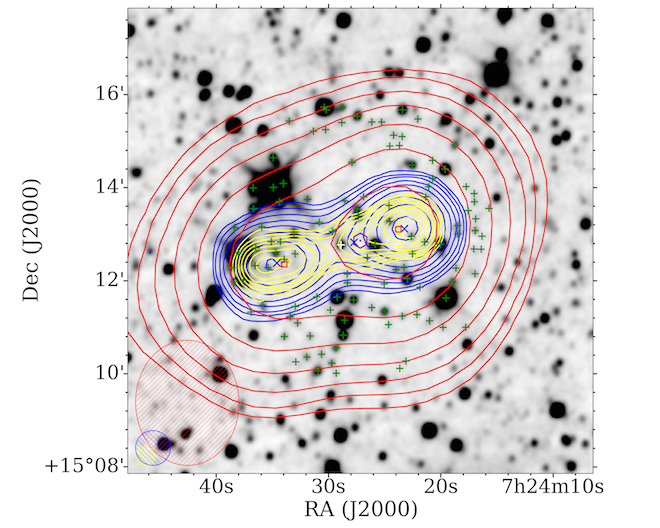}
	}
\caption{Overlays for G4Jy sources that have multiple GLEAM components (Table~\ref{multiGLEAMsources}, Appendix~\ref{app:multiGLEAMsources}). The datasets, contours, symbols, and beams are the same as those used for Figure~\ref{remaining1}. }
\label{remaining2}
\end{figure*}

\begin{figure*}
\centering
\subfigure[G4Jy~923]{
	\includegraphics[scale=1.1]{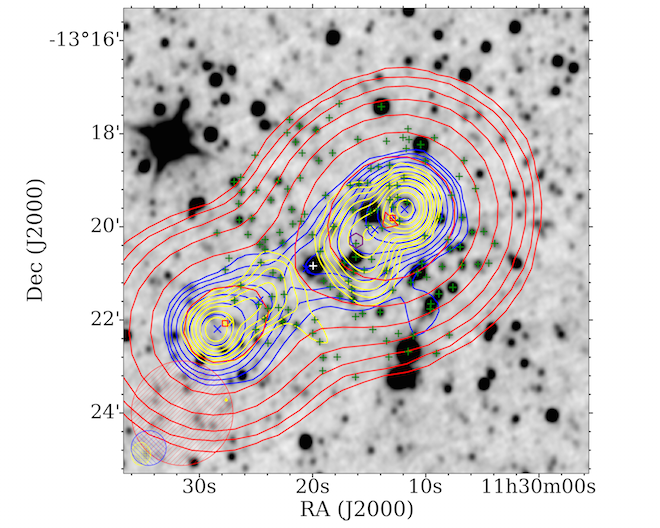}
	}
\subfigure[G4Jy~957]{
	\includegraphics[scale=1.1]{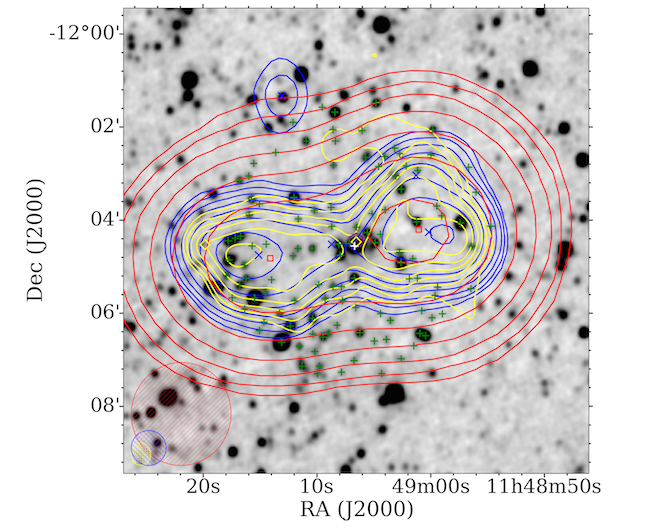}
	}
\subfigure[G4Jy~987]{
	\includegraphics[scale=1.1]{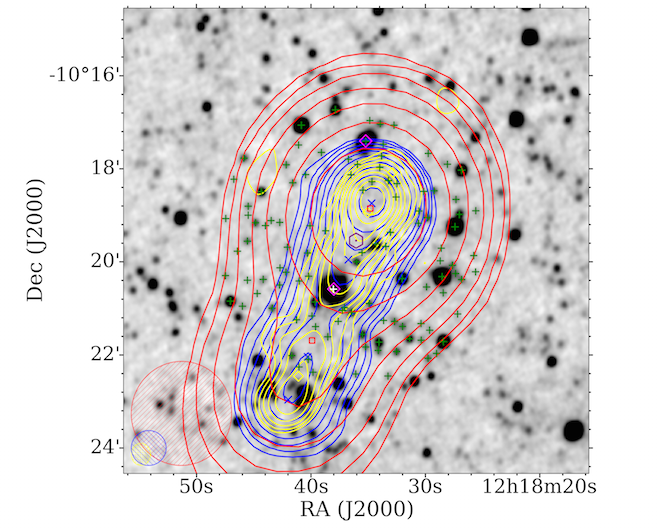}
	}
\subfigure[G4Jy~1048]{
	\includegraphics[scale=1.1]{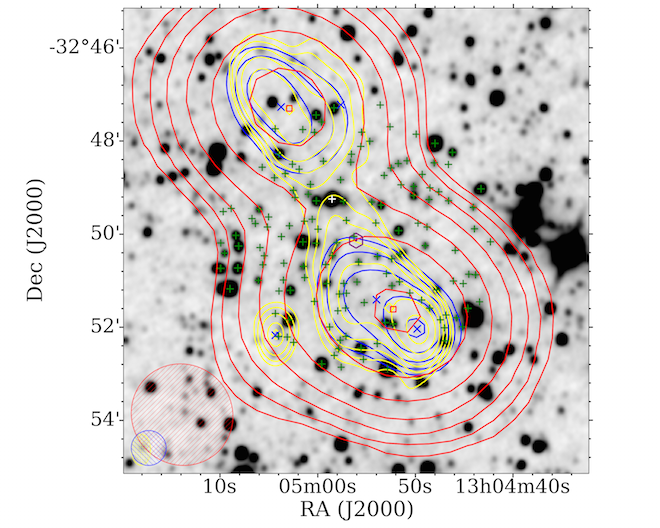}
	}
\subfigure[G4Jy~1197]{
	\includegraphics[scale=1.1]{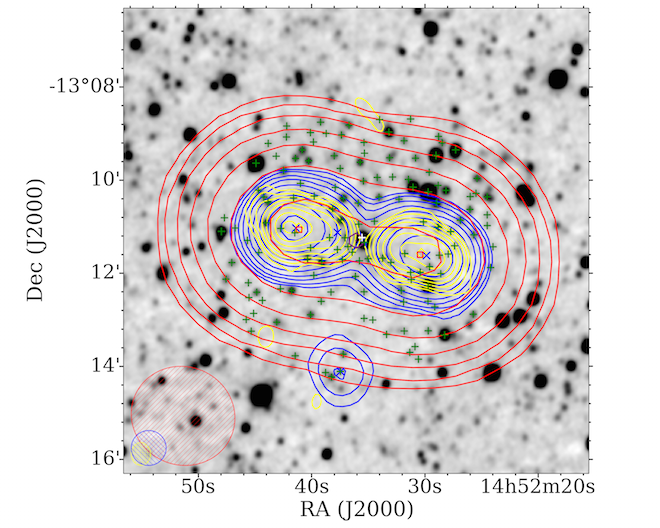}
	}
\subfigure[G4Jy~1200]{
	\includegraphics[scale=1.1]{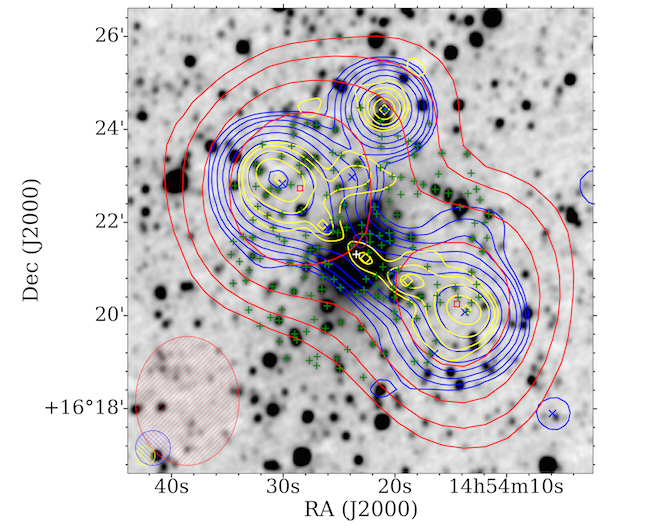}
	}
\caption{Overlays for G4Jy sources that have multiple GLEAM components (Table~\ref{multiGLEAMsources}, Appendix~\ref{app:multiGLEAMsources}). The datasets, contours, symbols, and beams are the same as those used for Figure~\ref{remaining1}. }
\label{remaining3}
\end{figure*}

\begin{figure*}
\centering
\subfigure[G4Jy~1238]{
	\includegraphics[scale=1.1]{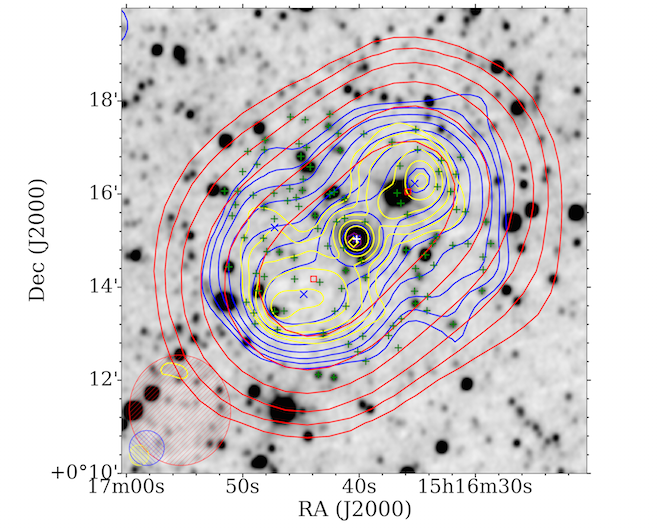}
	}
\subfigure[G4Jy~1289]{
	\includegraphics[scale=1.1]{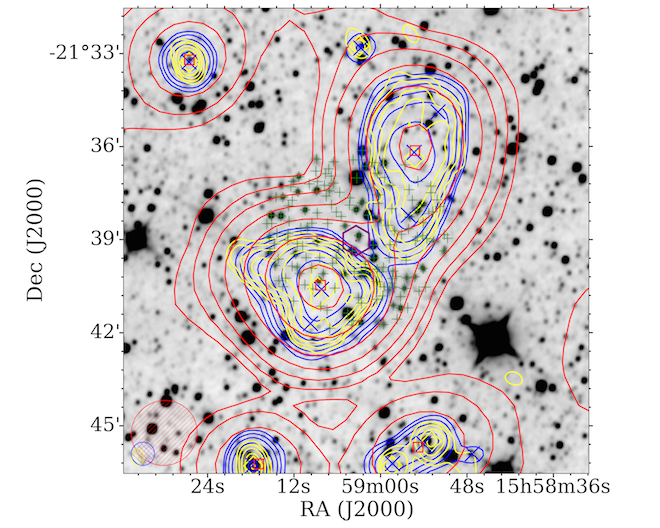}
	}
\subfigure[G4Jy~1296]{
	\includegraphics[scale=1.1]{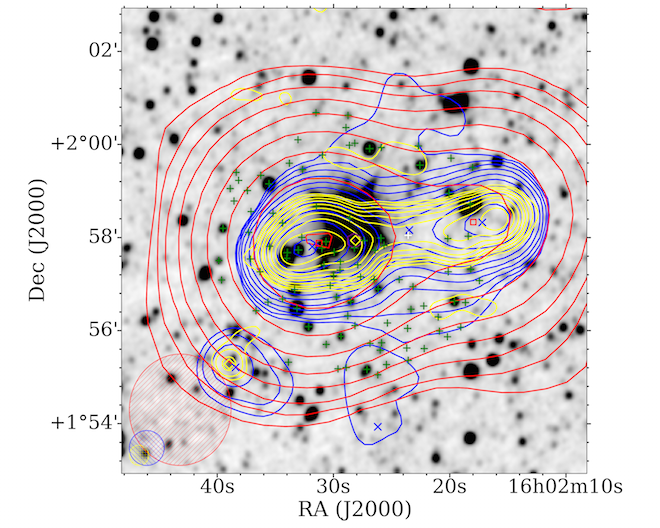}
	}
\subfigure[G4Jy~1303]{
	\includegraphics[scale=1.1]{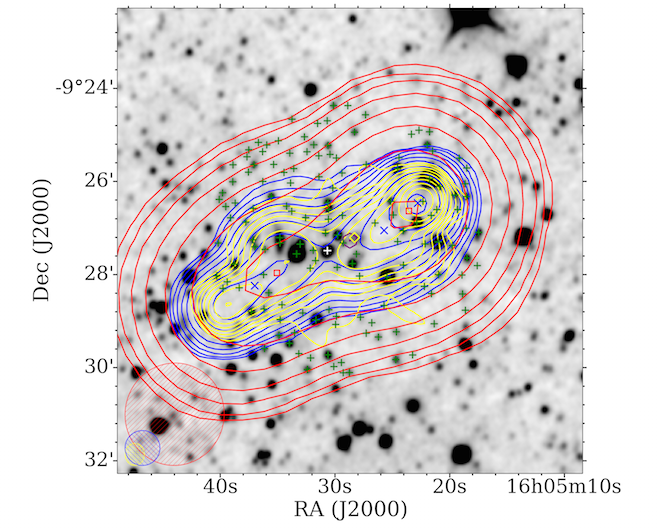}
	}
\subfigure[G4Jy~1423]{
	\includegraphics[scale=1.1]{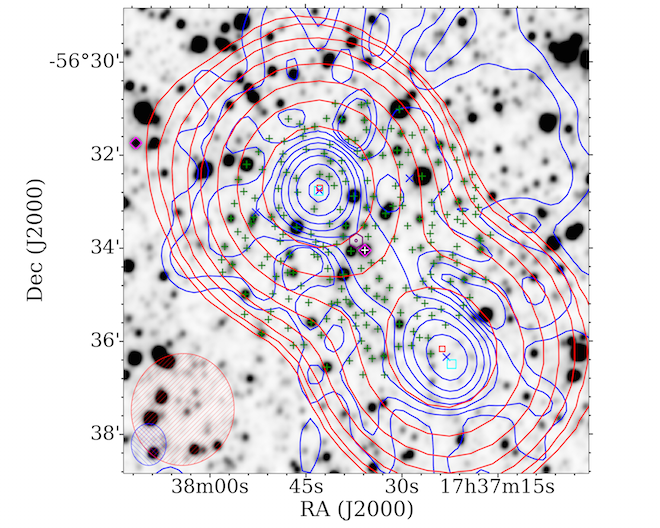}
	}
\subfigure[G4Jy~1484]{
	\includegraphics[scale=1.1]{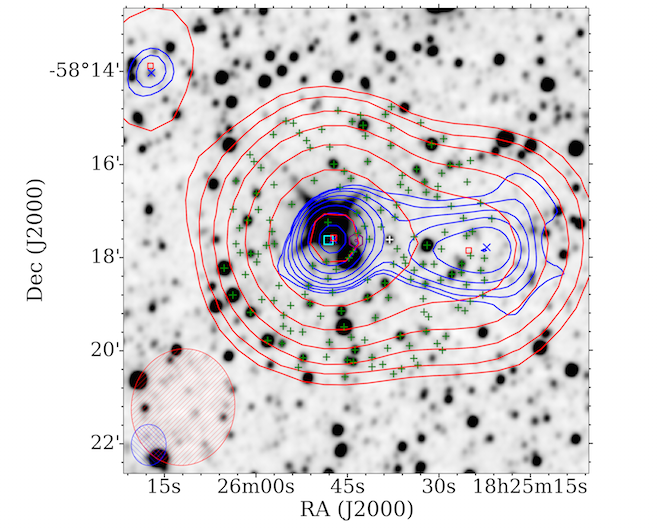}
	}
\caption{Overlays for G4Jy sources that have multiple GLEAM components (Table~\ref{multiGLEAMsources}, Appendix~\ref{app:multiGLEAMsources}). The datasets, contours, symbols, and beams are the same as those used for Figure~\ref{remaining1}. }
\label{remaining4}
\end{figure*}

\begin{figure*}
\centering
\subfigure[G4Jy~1569]{
	\includegraphics[scale=1.1]{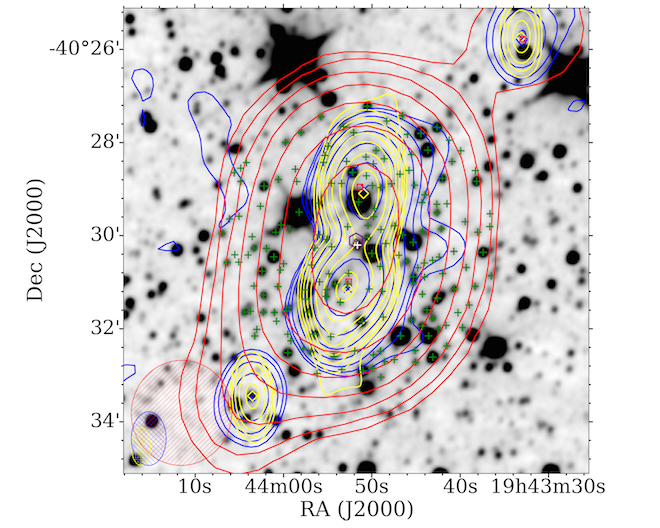}
	}
\subfigure[G4Jy~1617]{
	\includegraphics[scale=1.1]{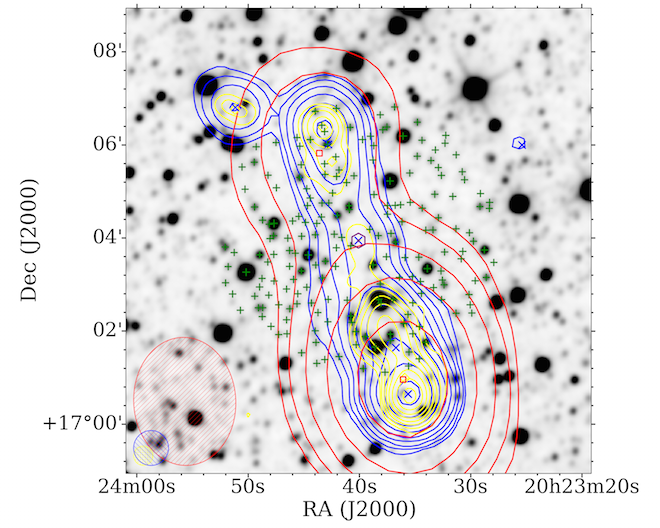}
	}
\subfigure[G4Jy~1643]{
	\includegraphics[scale=1.1]{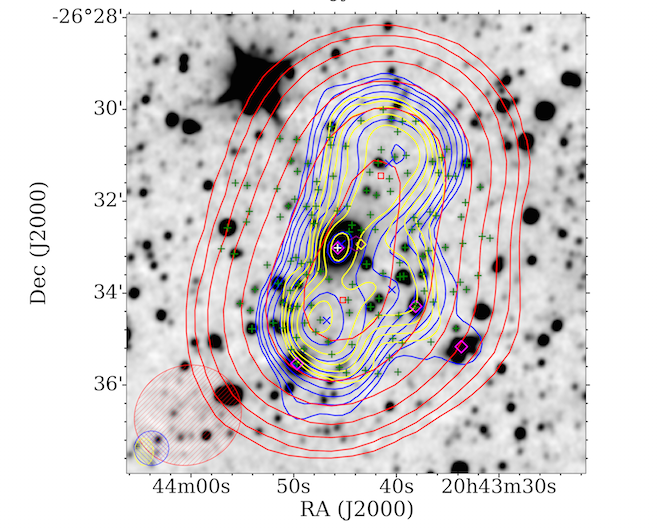}
	}
\subfigure[G4Jy~1671]{
	\includegraphics[scale=1.1]{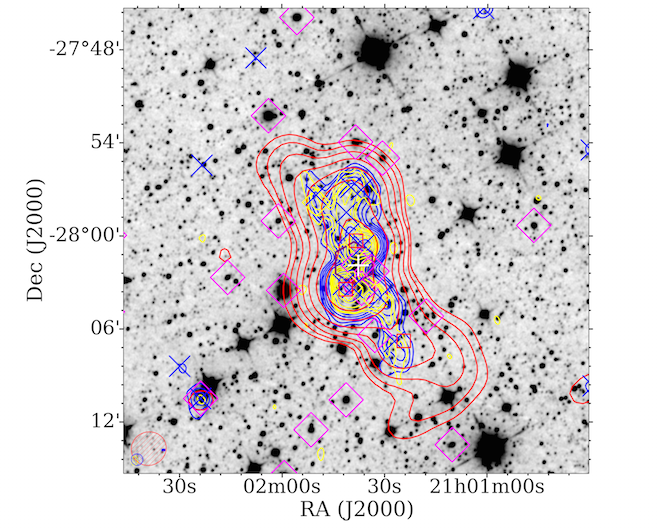}
	}
\subfigure[G4Jy~1775]{
	\includegraphics[scale=1.1]{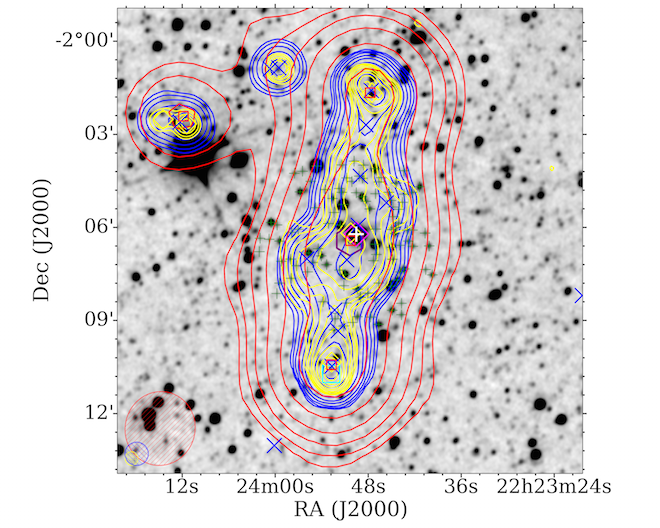}
	}
\subfigure[G4Jy~1863]{
	\includegraphics[scale=1.1]{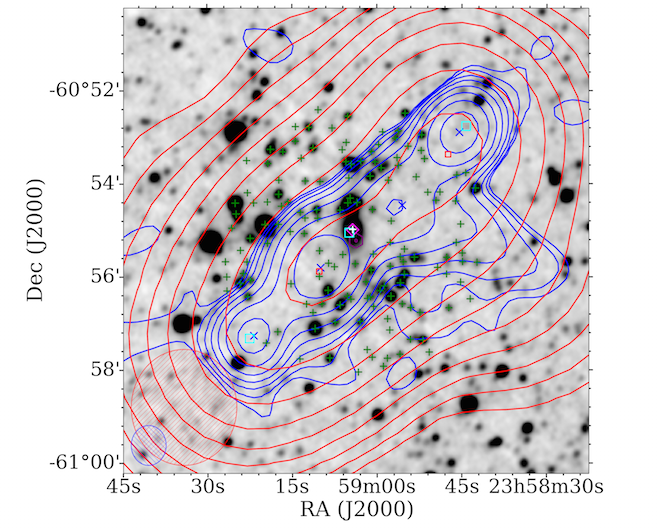}
	}
\caption{Overlays for G4Jy sources that have multiple GLEAM components (Table~\ref{multiGLEAMsources}, Appendix~\ref{app:multiGLEAMsources}). The datasets, contours, symbols, and beams are the same as those used for Figure~\ref{remaining1}.}
\label{remaining5}
\end{figure*}

Table~\ref{multiGLEAMsources} lists all of the G4Jy sources that have multiple GLEAM components. We also present overlays for 30 of these sources (Figures~\ref{remaining1}--\ref{remaining5}) that are not shown elsewhere in Paper~I (Figures~1, 3--9, \ref{galaxypair}, \ref{worrallgalaxies}) nor in Paper~II (figures 3--4, 6, 8, 12, 16--17, 19, 21, 23).  

\begin{table*}
\centering 
\caption{The 78 G4Jy sources that have multiple GLEAM components. (Their radio/mid-infrared overlays are shown in numerous figures throughout the paper, with 30 of them presented in Appendix~\ref{app:multiGLEAMsources}.) For 54~sources, just one component needed to have $S_{\mathrm{151\,MHz}}>4$\,Jy in order to be selected for the sample (Section~\ref{sec:sampleselection}). Other components were then identified via visual inspection (Section~\ref{sec:visualinspection}), and become part of the G4Jy Sample by association. Note that the GLEAM components associated with G4Jy~1080, 1677, 1678, 1704 and 1705 do not appear in the GLEAM extragalactic catalogue \citep{HurleyWalker2017}, as these are the result of re-fitting with \textsc{Aegean} (Appendix~D). In addition, for 24 multi-component sources, none of the GLEAM components are in the original selection, but further work (Section~\ref{sec:completeness}) leads to their inclusion in the G4Jy Sample. In this table we remove the `GLEAM' prefix of the GLEAM-component name, for space considerations.}
\begin{tabular}{@{}rccccc@{}}
 \hline
Source & \multicolumn{5}{c}{Corresponding GLEAM components}    \\
 \hline
G4Jy 7 &  J000441+124907 &  J000456+124810 &  & & \\ 
G4Jy 126 &  J010850+131833 &  J010855+132150 &  & & \\ 
G4Jy 131 &  J011257+153048 &  J011303+152654 &  &  &\\ 
G4Jy 133 &  J011609$-$471816 &  J011630$-$472542 &  &&  \\ 
G4Jy 151 &  J012603$-$012356 &  J012604$-$011802 &  &&  \\ 
G4Jy 171 &  J013332$-$362850 &  J013411$-$362913 &  & & \\ 
G4Jy 189 &  J014848+062147 &  J014850+062403 &  & & \\ 
G4Jy 234 &  J021305$-$474112 &  J021311$-$474615 &  &&  \\ 
G4Jy 270 &  J023132$-$203948 &  J023139$-$204104 &  & & \\ 
G4Jy 285 &  J024103+084523 &  J024107+084452 &  & & \\ 
G4Jy 315 &  J025738+060352 &  J025748+060201 &  &  &\\ 
G4Jy 318 &  J030115$-$250538 &  J030127$-$250354 &  &&  \\ 
G4Jy 347 &  J031939$-$452649 &  J032123$-$451021 &  &&  \\ 
G4Jy 360 &  J032750+023316 &  J032758+023407 &  & & \\ 
G4Jy 366 &  J033401$-$385840 &  J033416$-$390129 & & &  \\ 
G4Jy 386 &  J035125$-$274610 &  J035140$-$274354 &  &&  \\ 
G4Jy 400 &  J035852+102404 &  J035857+102702 &  &  &\\ 
G4Jy 414 &  J040712+034318 &  J040724+034049 &  &&  \\ 
G4Jy 447 &  J042220+140742 &  J042233+140733 &  & & \\ 
G4Jy 462 &  J042839$-$535020 &  J042907$-$534919 & & &  \\ 
G4Jy 475 &  J043409$-$132250 &  J043415$-$132717 &  &&  \\ 
G4Jy 517 &  J050535$-$285648 &  J050539$-$282627 &  J050544$-$282236 &&  \\ 
G4Jy 531 &  J051329$-$303042 &  J051338$-$302616 &  J051348$-$302327 & & \\ 
G4Jy 543 &  J052522$-$324121 &  J052531$-$324357 &  &&  \\ 
G4Jy 579 &  J054825$-$330128 &  J054836$-$325458 &  & & \\ 
G4Jy 604 &  J061740$-$485426 &  J061803$-$484610 &  J061812$-$484257 & & \\ 
G4Jy 619 &  J063631$-$202924 &  J063631$-$203725 &  J063632$-$203229 &  J063633$-$204225 & \\ 
G4Jy 641 &  J070525$-$451328 &  J070546$-$451158 &  & & \\
G4Jy 644 &  J070901$-$355921 &  J070914$-$360115 &  J070934$-$360341 & & \\
G4Jy 659 &  J072423+151306 &  J072433+151221 &  & & \\ 
G4Jy 680 &  J080225$-$095823 &  J080253$-$095822 &  & & \\ 
G4Jy 729 &  J084832+055502 &  J084841+055532 &  & & \\ 
G4Jy 886 &  J105846$-$361754 &  J105854$-$362051 &  & & \\ 
G4Jy 923 &  J113012$-$131948 &  J113027$-$132204 &  &  &\\ 
G4Jy 935 &  J113943$-$464032 &  J113956$-$463743 &  & & \\ 
G4Jy 957 &  J114901$-$120412 &  J114914$-$120449 &  &  &\\ 
G4Jy 987 &  J121834$-$101851 &  J121839$-$102141 &  & & \\ 
G4Jy 990 &  J121915+054929 &  J121933+054944 &  & & \\ 
G4Jy 1021 &  J124602+255359 &  J124612+255337 &  & & \\ 
G4Jy 1048 &  J130452$-$325137 &  J130502$-$324718 & & &  \\ 
G4Jy 1067 &  J132606$-$272641 &  J132616$-$272632 &  &&  \\ 
G4Jy 1079 &  J133351$-$100740 &  J133419$-$100937 &  J133442$-$101114 & & \\ 
G4Jy 1080 &  J133548$-$335240 &  J133630$-$335656 &  J133641$-$335829 &  J133739$-$340904 &\\ 
G4Jy 1110 &  J140134$-$113504 &  J140150$-$113801 &  & & \\ 
G4Jy 1173 &  J142955+072134 &  J143002+071505 &  & & \\ 
G4Jy 1197 &  J145230$-$131136 &  J145241$-$131104 & & &  \\ 
\hline
\label{multiGLEAMsources}
\end{tabular}
\end{table*}

\setcounter{table}{12} 

\begin{table*} 
\centering 
\caption{{\it Continued} -- G4Jy sources with multiple GLEAM components.} 
\begin{tabular}{@{}rccccc@{}} 
\hline 
Source & \multicolumn{5}{c}{Corresponding GLEAM components}  \\ 
\hline 
G4Jy 1200 &  J145414+162015 &  J145428+162244 &  & & \\ 
G4Jy 1238 &  J151635+001603 &  J151643+001410 &  &  &\\ 
G4Jy 1265 &  J153137+240542 &  J153150+240244 &  &&  \\ 
G4Jy 1279 &  J154851$-$321431 &  J154902$-$321811 &  & & \\ 
G4Jy 1282 &  J155120+200312 &  J155147+200424 &  J155226+200556 & & \\ 
G4Jy 1289 &  J155855$-$213608 &  J155908$-$214028 &  &&  \\ 
G4Jy 1296 &  J160217+015819 &  J160231+015752 &  & & \\ 
G4Jy 1303 &  J160523$-$092638 &  J160535$-$092757 &  &  &\\ 
G4Jy 1423 &  J173723$-$563610 &  J173742$-$563242 &  & & \\ 
G4Jy 1428 &  J174120$-$052240 &  J174132$-$052440 &  J174145$-$052554 & & \\ 
G4Jy 1480 &  J182239+121429 &  J182243+121754 &  & & \\ 
G4Jy 1484 &  J182525$-$581751 &  J182547$-$581735 & & &  \\ 
G4Jy 1496 &  J183626+193946 &  J183640+194318 &  J183649+194105 & & \\ 
G4Jy 1525 &  J191905$-$795737 &  J191931$-$800128 &  & & \\ 
G4Jy 1569 &  J194351$-$402857 &  J194352$-$403059 &  &  &\\ 
G4Jy 1582 &  J195222$-$011550 &  J195232$-$011729 &  & & \\ 
G4Jy 1605 &  J201021$-$562915 &  J201034$-$562417 &  J201110$-$562635 &  J201143$-$561904 &  J201215$-$562240 \\ 
G4Jy 1613 &  J201739$-$553242 &  J201749$-$553800 &  J201801$-$553938 &  J201814$-$554145 & \\ 
G4Jy 1617 &  J202336+170057 &  J202343+170549 &  & & \\ 
G4Jy 1628 &  J202932$-$411755 &  J202940$-$412011 & & &  \\ 
G4Jy 1643 &  J204341$-$263126 &  J204345$-$263409 &  &&  \\ 
G4Jy 1670 &  J210135$-$131754 &  J210154$-$131850 &  &&  \\ 
G4Jy 1671 &  J210138$-$280019 &  J210141$-$280327 &  & & \\ 
G4Jy 1677 &  J210716$-$252733 &  J210724$-$252953 & & &  \\ 
G4Jy 1678 &  J210722$-$252615 &  J210724$-$252514 &  &&  \\ 
G4Jy 1704 &  J213356$-$533509 &  J213418$-$533514 &  &&  \\ 
G4Jy 1705 &  J213415$-$533736 &  J213422$-$533756 &  & & \\ 
G4Jy 1718 &  J214352$-$563839 &  J214406$-$563558 & & &  \\ 
G4Jy 1732 &  J215122$-$552139 &  J215123$-$552604 &  J215133$-$551636 & & \\ 
G4Jy 1741 &  J215415$-$455319 &  J215435$-$454954 &  & & \\ 
G4Jy 1775 &  J222347$-$020139 &  J222350$-$020625 &  J222352$-$021025 & & \\ 
G4Jy 1863 &  J235847$-$605322 &  J235910$-$605553 &  & & \\ 
\hline
\end{tabular}
\end{table*}

\newpage 

\section[]{Details of Aegean re-fitting}
\label{app:refitting}

This appendix provides details as to GLEAM components -- identified through visual inspection -- that require re-fitting (Section~\ref{sec:refitting}). The source-finding software, {\sc Aegean} \citep{Hancock2012,Hancock2018}, is used in different modes for this procedure, as described in the following subsections. The resulting re-fitted components follow the same naming system as for the EGC \citep{HurleyWalker2017}, and are used for the G4Jy catalogue (Section~\ref{sec:finalcatalogue}; Appendix~E) and overlays.

\subsection{Unconstrained re-fitting}
\label{sec:unconstrained}

The value of inspecting large images is most-apparent for radio sources with very extended emission. Based on the original 20$'$ overlays, it was thought that GLEAM~J133549$-$335247 and GLEAM~J133637$-$335724 were associated, but `zooming out' revealed that these components accounted for only one of the lobes belonging to an extended radio-galaxy (Figure~\ref{tiefighterOverlay}). The reason that this was not appreciated earlier is that the second lobe is within the masked region (a circle of radius = 9\arcdeg) used to exclude Centaurus~A from the GLEAM catalogue \citep{HurleyWalker2017}. Consequently, we re-fit this `bisected' source using {\sc Aegean}. Without the constraints imposed by the previously-applied mask, we find that four GLEAM components describe the low-frequency emission of {\bf G4Jy~1080}: GLEAM~J133548$-$335240, GLEAM~J133630$-$335656, GLEAM~J133641$-$335829, and GLEAM~J133739$-$340904. We add these components to the GLEAM-component list, and remove GLEAM~J133549$-$335247 and GLEAM~J133637$-$335724.

Also known as IC~4296, this re-fitted source is the brightest cluster-galaxy (BCG) of Abell~3565, which is part of the Hydra--Centaurus Supercluster. Being at $z=0.012$ \citep{Mahony2011}, the galaxy's lobe-to-lobe extent of 33$'$ corresponds to a physical scale of 487\,kpc \citep{Wright2006}. In addition, we note that the backflow of plasma in the southern lobe is only indicated by the MWA contours (Figure~\ref{tiefighterOverlay}), thanks to the instrument's sensitivity to diffuse emission.

\subsection{Peeled sources}
\label{app:peeledsources}

Due to the regular arrangement of dipoles in its component tiles, the primary beam of the MWA has regularly-spaced sidelobes, which can have high sensitivity, typically $\approx10$\,\% at low frequencies and high elevations, and up to $100$\,\% at high frequencies and low elevations. For some pointings of GLEAM, bright sources appeared in these sidelobes and needed to be removed from the visibilities before self-calibration could be performed. \cite{HurleyWalker2017} used a `peeling' technique, in which the visibilities are phase-rotated to the source, a calibration solution is formed, and the solution applied to the model of the source, then subtracted from the visibilities, which are then phase-rotated back to the original pointing direction for self-calibration and imaging. However, to perfectly predict which observations need peeling and which do not, the model of the primary beam and the flux densities of the sources must be perfectly known, and this was not the case during the GLEAM data processing.

In the case of observations of PKS\,B1932$-$46 ({\bf G4Jy~1558}), Cygnus~A appeared in the far northern primary-beam sidelobe, with an apparent flux-density that varied as a complex function of frequency. The automatic peeling algorithm attempted to remove Cygnus~A, but at some sub-bands, the fitting converged on PKS\,B1932$-$46, removing it from the images. When the images were mosaicked, PKS\,B1932-46 therefore had incorrect flux-density measurements across the GLEAM band.

To correctly measure the SED, we re-imaged five GLEAM observations covering PKS\,B1932$-$46, each covering 30.72\,MHz of the GLEAM band in $4\times7.68$\,MHz sub-bands, without applying any peeling. The lowest band (72--103\,MHz) was found to be contaminated with RFI and was discarded. For the remaining 16\,sub-bands, 10 were not affected by the presence of Cygnus~A in the sidelobes. From these, we were able to measure the flux density of PKS\,B1932$-$46, flux-calibrate it to the 10 closest bright, isolated GLEAM sources, and fit an SED. Based on the fitted SED -- a power-law function, $S_{\nu} \propto \nu^{\alpha}$, with $\alpha = -1.0 \pm 0.1$ -- we estimate the integrated flux-density for the intervening, missing sub-band measurements. The G4Jy catalogue is updated to use these flux densities, which are provided in Table~\ref{tab:PKS1932fluxdensities} for reference. \\ 

\begin{table*} 
\centering 
\caption{Re-measured and fitted integrated flux-densities for G4Jy~1558, which is PKS~B1932$-$46 (Appendix~D.2). To avoid extrapolating the SED-fit beyond the frequency range for which it is valid, we blank existing measurements in the G4Jy catalogue (Section~\ref{sec:finalcatalogue}) for the following sub-bands: 76, 84, 92, 99, and 227\,MHz. The 200-MHz flux-density and associated error correspond to the wide-band (170--231\,MHz) measurement (replacing the original `int\_flux\_wide' and `err\_int\_flux\_wide' values; see Appendix~D). `Estimated' refers to flux densities obtained via a fitted SED, rather than the application of a corrective flux-scale factor (following re-imaging).} 
\begin{tabular}{@{}rccc@{}} 
\hline 
Central frequency & Method & Corrective    & Corrected (or estimated)   \\ 
 / MHz &  &  flux-scale factor  & integrated flux-density / Jy  \\ 
\hline 
107 & Re-measured & 1.6 $\pm$ 0.3 & 105.4 $\pm$ 22.3 \\ 
115 & Re-measured & 1.8 $\pm$ 0.3 & 107.5 $\pm$ 20.4 \\ 
122 & Re-measured & 1.8 $\pm$ 0.4 & 99.1 $\pm$ 22.5 \\ 
130 & Re-measured & 1.8 $\pm$ 0.4 & 94.7 $\pm$ 20.4 \\ 
143 & Fitted SED     & -- & 84.4 $\pm$ 6.7 \\ 
151 & Fitted SED      & -- & 80.0 $\pm$ 6.4 \\ 
158 & Fitted SED     & -- & 76.5 $\pm$ 6.1 \\ 
166 & Re-measured & 1.5 $\pm$ 0.5 & 74.0 $\pm$ 26.6 \\ 
174 & Re-measured & 1.5 $\pm$ 0.7 & 72.8 $\pm$ 31.7 \\ 
181 & Re-measured & 1.4 $\pm$ 0.6 & 66.9 $\pm$ 28.0 \\ 
189 & Fitted SED     & -- & 64.0 $\pm$ 4.9 \\ 
197 & Fitted SED     & -- & 61.4 $\pm$ 4.6 \\ 
200 & Fitted SED     & -- & 60.5 $\pm$ 4.5 \\ 
204 & Re-measured & 1.4 $\pm$ 0.1 & 57.6 $\pm$ 4.2 \\ 
212 & Re-measured & 1.3 $\pm$ 0.1 & 54.9 $\pm$ 4.8 \\ 
220 & Re-measured & 1.0 $\pm$ 0.1 & 55.0 $\pm$ 6.3 \\
\hline 
\label{tab:PKS1932fluxdensities} 
\end{tabular} 
\end{table*}

\subsection{Priorised re-fitting}
\label{app:priorised}

\begin{figure*}
\centering
\subfigure[The 10$'$ G4Jy overlay (after re-fitting)]{
	\includegraphics[scale=1.1]{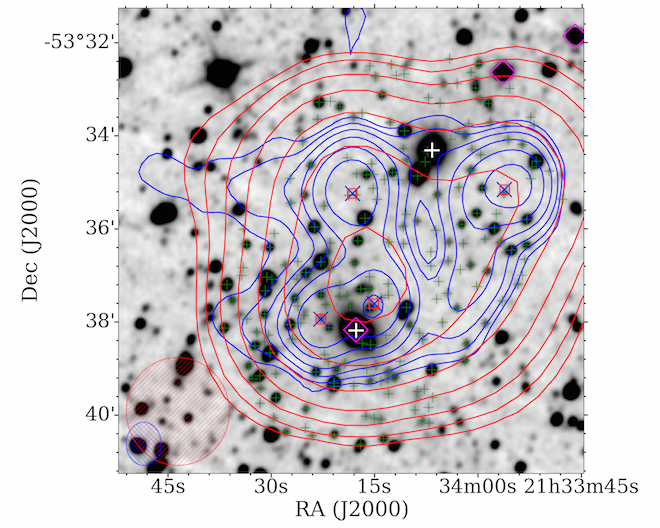}
	}
\subfigure[1.3\,GHz/optical overlay]{
	\includegraphics[scale=1.1]{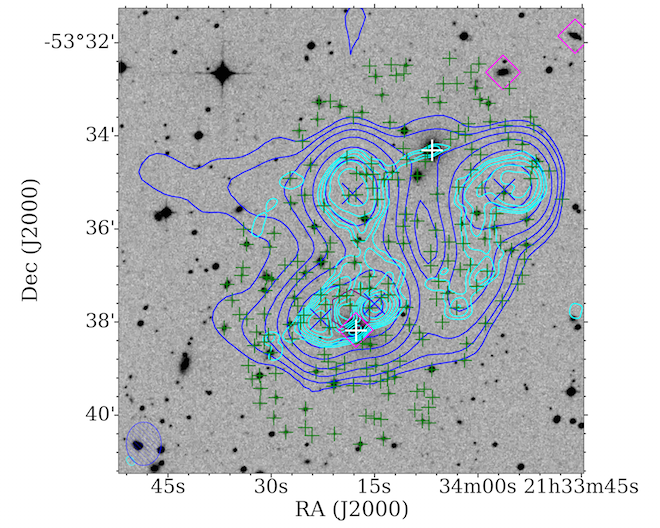}
	}
\caption{Two overlays for the radio galaxies (G4Jy~1704 and G4Jy~1705) in cluster Abell 3785. These sources required re-fitting with {\sc Aegean} (Appendix~D.3), with the new GLEAM positions (red squares) set to those of the SUMSS-catalogue positions (blue crosses), as shown in the first overlay, (a). Radio contours from GLEAM (170--231\,MHz; red) and SUMSS (843\,MHz; blue) are overlaid on a mid-infrared image from {\it WISE} (3.4\,$\mu$m; inverted greyscale). White plus signs indicate the host galaxies for the two G4Jy sources, whilst magenta diamonds represent 6dFGS positions. The second overlay, (b), uses an ATCA image at 1.3\,GHz (cyan contours) from \citet{Haigh2000}, which was provided courtesy of Richard Hunstead. This image was obtained using a combination of 6A and 6C configurations, with a restoring beam of $14.8'' \times 9.6''$ at position angle = $-$59\arcdeg (cyan ellipse in the bottom left-hand corner). The cyan contours are overlaid on an optical image (inverted greyscale) from SuperCOSMOS \citep{Hambly2001}, and SUMSS contours are again plotted in blue for reference. For each set of contours in this figure, the lowest contour is at the 3\,$\sigma$ level (where $\sigma$ is the local rms), with the number of $\sigma$ doubling with each subsequent contour (i.e. 3, 6, 12\,$\sigma$, etc.).}
\label{galaxypair}
\end{figure*}

The morphology of the blended components GLEAM~J213416$-$533648 and GLEAM~J213356$-$533524 is unclear (with the morphology of J21341775$-$5338101 listed as `unknown' by \citealt{vanVelzen2012}). Using earlier Molonglo Observatory Synthesis Telescope observations at higher resolution, \citet{Jones1992} interpreted the combined radio emission as arising from two double-lobed radio galaxies in the cluster A3785, confirming the optical identifications of \citet{Ekers1970}. Follow-up ATCA observations at 1.34\,GHz by \citet{Haigh2000} -- published here for the first time -- show clear evidence for relative motion, in opposite directions, between the host galaxies and the surrounding cluster gas (Figure~\ref{galaxypair}).  The radio structure of the northern galaxy, in particular, is not typical of a cluster radio-source, as the jet between the core and the eastern lobe is well-collimated. This suggests that we might be witnessing the early(?) stages of a cluster merger, before the jet becomes disrupted.

In an effort to de-blend the low-frequency emission from the two radio-galaxies, we re-fit them using the SUMSS detections as priorised positions. Therefore, we replace the original two GLEAM components with GLEAM~J213356$-$533509 and GLEAM~J213418$-$533514 (for the northern `double'), in addition to GLEAM~J213415$-$533736 and GLEAM~J213422$-$533756 (for the southern `double'). However, as a consequence of priorised fitting, the summation (per sub-band) of the resulting integrated flux-densities ($\Sigma\,S_{\mathrm{re-fitted}}$) is systematically lower than the summation calculated for the original GLEAM components ($\Sigma\,S_{\mathrm{original}}$). The ratio between these two sums is presented in Table~\ref{refittedratios} (under {\bf G4Jy~1704/G4Jy~1705}), and shows that the proportion of low-frequency emission that is `recovered' during the re-fitting is $> 79$\%. As we wish to provide the best-estimate of {\it integrated} flux-densities for these radio galaxies, we use the ratios to proportionally distribute $\Sigma\,S_{\mathrm{original}}$ across the re-fitted components. For example,
\begin{equation}
\begin{aligned}
S_{\mathrm{rescaled}}^{1} &= S_{\mathrm{refitted}}^{1}\,\Sigma\,S_{\mathrm{original}}\ /\ \Sigma\,S_{\mathrm{refitted}} \label{eqn:1}\\
&= \frac{S_{\mathrm{refitted}}^{1}\,(S_{\mathrm{original}}^{A} + S_{\mathrm{original}}^{B})}{S_{\mathrm{refitted}}^{1} + S_{\mathrm{refitted}}^{2} + S_{\mathrm{refitted}}^{3} + S_{\mathrm{refitted}}^{4} }, 
\end{aligned}
\end{equation}
where superscripts are used to denote individual GLEAM components. We apply the same re-scaling to the errors on the re-fitted, integrated flux-densities.

Following this re-scaling, we consider the summed flux-density at 151\,MHz for each of the radio galaxies. For the northern `double' this is $4.44 \pm 0.05$\,Jy, and for the southern `double' this is $4.19 \pm 0.05$\,Jy. As both radio galaxies cross the 4-Jy threshold, they are included in the G4Jy Sample (as G4Jy~1704 and G4Jy~1705, respectively). However, we emphasise that their integrated flux-densities are estimates (rather than direct measurements), and note that they will be superseded by new measurements using the recently-upgraded MWA \citep{Beardsley2019}. This will provide higher spatial-resolution (at $\sim1'$) than is currently available through GLEAM. For this reason, coupled with the involved process of re-fitting, we use the same method to de-blend only a few other confused sources in the sample. 

\begin{table*} 
\centering 
\caption{We present the ratio between the summed, integrated flux-density for the re-fitted GLEAM-components, and the summed, integrated flux-density for the original GLEAM-components (i.e. ratio $= \Sigma\,S_{\mathrm{refitted}}\ /\ \Sigma\,S_{\mathrm{original}}$). These ratios are calculated at each frequency for G4Jy~813, G4Jy~1410, and two sets of blended radio-galaxies: G4Jy~1677/G4Jy~1678 and G4Jy~1704/G4Jy~1705. The ratios at 200\,MHz correspond to the wide-band image (170--231\,MHz). We use these ratios to correct integrated flux-densities, which have been under-/over-estimated as a result of priorised re-fitting (Appendix~D.3). } 
\begin{tabular}{@{}rcccc@{}} 
\hline 
Central frequency & \multicolumn{4}{c}{Integrated flux-density ratio for} \\ 
/ MHz & G4Jy~813 & G4Jy~1410 & G4Jy~1677/G4Jy~1678 &  G4Jy~1704/G4Jy~1705 \\ 
\hline 
76 & 0.978 & 0.988 & 1.012 & 0.946 \\ 
84 & 0.958 & 0.978 & 1.012 & 0.941 \\ 
92 & 0.934 & 0.962 & 1.003 & 0.938 \\ 
99 & 0.920 & 0.950 & 0.997 & 0.932 \\ 
107 & 0.907 & 0.968 & 0.998 & 0.932 \\ 
115 & 0.911 & 0.962 & 0.990 & 0.922 \\ 
122 & 0.896 & 0.956 & 0.979 & 0.915 \\ 
130 & 0.886 & 0.952 & 0.963 & 0.907 \\ 
143 & 0.860 & 0.950 & 0.941 & 0.893 \\ 
151 & 0.849 & 0.939 & 0.929 & 0.883 \\ 
158 & 0.852 & 0.938 & 0.909 & 0.875 \\ 
166 & 0.837 & 0.928 & 0.893 & 0.867 \\ 
174 & 0.842 & 0.938 & 0.882 & 0.861 \\ 
181 & 0.828 & 0.942 & 0.868 & 0.851 \\ 
189 & 0.813 & 0.935 & 0.856 & 0.838 \\ 
197 & 0.806 & 0.937 & 0.844 & 0.829 \\ 
200 & 0.866 & 0.937 & 0.898 & 0.856 \\ 
204 & 0.822 & 0.934 & 0.828 & 0.821 \\ 
212 & 0.805 & 0.915 & 0.817 & 0.809 \\ 
220 & 0.784 & 0.907 & 0.805 & 0.804 \\ 
227 & 0.793 & 0.904 & 0.787 & 0.790 \\ 
\hline 
\label{refittedratios} 
\end{tabular} 
\end{table*} 

\begin{figure*}
\centering
\subfigure[G4Jy~813]{
	\includegraphics[scale=1.1]{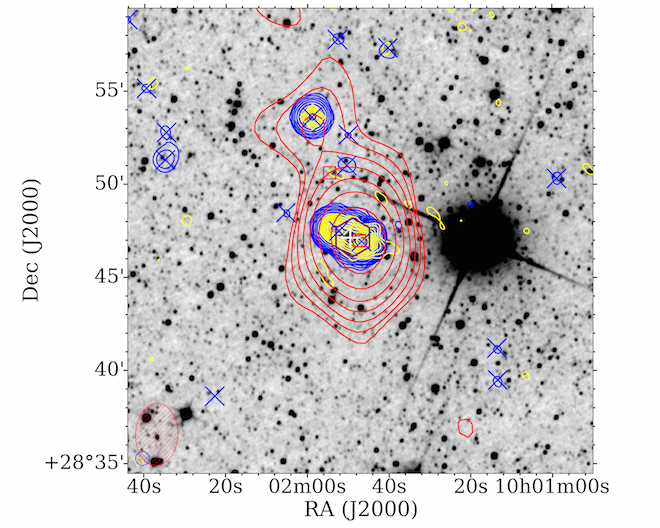}
	}
\subfigure[G4Jy~1410]{
	\includegraphics[scale=1.1]{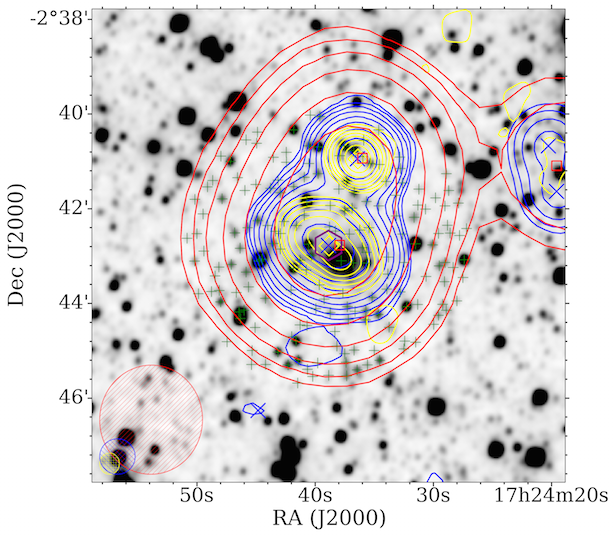}
	}
\caption{Overlays for (a) G4Jy~813 and (b) G4Jy~1410 (Appendix~D.3), with the same datasets, contours, symbols, and beams as used for Figure~\ref{tiefighterOverlay}. The red squares represent the original GLEAM positions for G4Jy~813 (shown for illustration), whilst for G4Jy~1410, they indicate the re-fitted GLEAM positions.}
\label{morerefitting}
\end{figure*}

Meanwhile, GLEAM~J100154+285037 ($S_{\mathrm{151\,MHz}}=4.99 \pm 0.11 $\,Jy) was found to be a poorly-fitted component (Figure~\ref{morerefitting}a) near to GLEAM~J100147+284659 ({\bf G4Jy~813}, $S_{\mathrm{151\,MHz}}=35.12 \pm 0.06$\,Jy). This region was therefore re-fitted, with the positions of associated NVSS sources acting as priors. Again, the re-fitted components do not recover all of the low-frequency emission, as characterised by the original run of {\sc Aegean}. Therefore we re-scale the re-fitted, integrated flux-densities (and their errors) using the ratios for G4Jy~813, given in Table~\ref{refittedratios}. The result is GLEAM~J100147+284659 now having $S_{\mathrm{151\,MHz}}=39.04 \pm 0.06$\,Jy, and GLEAM~J100154+285037 being replaced with GLEAM~J100159+285336 ($S_{\mathrm{151\,MHz}}=1.06 \pm 0.06 $\,Jy). Since this new component is not above the 4-Jy threshold, it is not retained for this work.

In the overlay for GLEAM~J172438$-$024205 ($S_{\mathrm{151\,MHz}}=9.54 \pm 0.07 $\,Jy) we see that two unrelated sources have been blended together (Figure~\ref{morerefitting}b). Using TGSS to determine whether either or both sources cross the 4-Jy threshold (see Section~\ref{sec:morphology}, and details for other GLEAM components in Appendix~\ref{sec:removedcomponents}), we find that the southern source ({\bf G4Jy~1410}) is bright enough to be included in the G4Jy Sample. We proceed by re-fitting the low-frequency emission, using the two NVSS positions as priors. This gives rise to the GLEAM components, GLEAM~J172436$-$024055 and GLEAM~J172437$-$024246. After applying the re-scaling ratios calculated for G4Jy~1410 (Table~\ref{refittedratios}), their integrated flux-densities are $S_{\mathrm{151\,MHz}}=3.79 \pm 0.07 $\,Jy and $S_{\mathrm{151\,MHz}}=5.74 \pm 0.07 $\,Jy, respectively. We therefore remove GLEAM~J172438$-$024205 from the G4Jy Sample, and replace it with GLEAM~J172437$-$024246. 

\begin{figure*}
\centering
\includegraphics[scale=2.0]{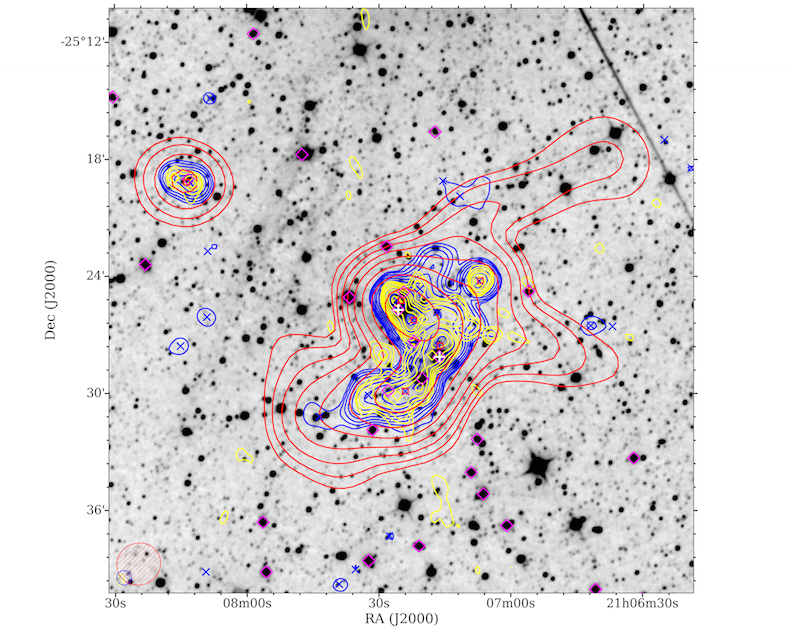}
\caption{An overlay for G4Jy~1677 and G4Jy~1678 (Appendix~D.3), with their host galaxies indicated by white plus signs (towards the west and east, respectively). Radio contours from TGSS (150\,MHz; yellow), GLEAM (170--231\,MHz; red) and NVSS (1.4\,GHz; blue) are overlaid on a mid-infrared image from AllWISE ($3.4\,\mu$m; inverted greyscale). For each set of contours, the lowest contour is at the 3\,$\sigma$ level (where $\sigma$ is the local rms), with the number of $\sigma$ doubling with each subsequent contour (i.e. 3, 6, 12\,$\sigma$, etc.). Also plotted, in the bottom left-hand corner, are ellipses to indicate the beam sizes for TGSS (yellow with `+' hatching), GLEAM (red with `/' hatching), and NVSS (blue with `\textbackslash' hatching). Magenta diamonds represent optical positions for sources in 6dFGS, yellow diamonds represent TGSS positions, and blue crosses represent NVSS positions. Of the six GLEAM positions shown in this overlay (red squares), the five furthest west are following re-fitting with {\sc Aegean}.}
\label{worrallgalaxies}
\end{figure*}

Together, GLEAM~J210722$-$252556 and GLEAM~J210724$-$252953 characterise the low-frequency emission of two `double' radio-galaxies and a single point-source (Figure~\ref{worrallgalaxies}). In order to determine integrated flux-densities for each source, separately, we again perform re-fitting. In this case, we set a total of five priorised positions: one for the point source, and one for each radio lobe belonging to the two doubles. Following re-scaling (Table~\ref{refittedratios}), the point source has a flux-density below 4\,Jy, and so is not considered any further. The southern double (GLEAM~J210716-252733 and GLEAM~J210724-252953) has total $S_{\mathrm{151\,MHz}}=18.56 \pm 0.05 $\,Jy, and becomes listed in the G4Jy Sample as {\bf G4Jy~1677}. The northern double (GLEAM~J210722$-$252615 and GLEAM~J210724$-$252514) has total $S_{\mathrm{151\,MHz}}=23.19 \pm 0.05$\,Jy, and becomes listed in the G4Jy Sample as {\bf G4Jy~1678}. Hence, the re-fitted, GLEAM components associated with these two double radio-galaxies replace GLEAM~J210722$-$252556 and GLEAM~J210724$-$252953 for the G4Jy catalogue.\\

\section[]{Column descriptions and first row of the G4Jy catalogue} \label{sec:cataloguecolumns}

In Table~\ref{tab:catalogueexample} we list columns that are newly-created for the G4Jy catalogue (Section~\ref{sec:finalcatalogue}), in addition to columns for wide-band (170--231\,MHz) and 151-MHz measurements, inherited from the EGC \citep{HurleyWalker2017}. Equivalent columns for the remaining GLEAM sub-bands (76, 84, 92, 99, 107, 115, 122, 130, 143, 158, 166, 174, 181, 189, 197, 204, 212, 220, and 227\,MHz) are listed in appendix~A of \citet{HurleyWalker2017}. Example entries, for the first row of the G4Jy catalogue, are also provided in Table~\ref{tab:catalogueexample}.\\

\section[]{BROADBAND RADIO SPECTRA FOR G4JY--3CRR SOURCES} \label{app:sedsbyJoe}

As part of our comparison with the 3CRR sample (Section~7.3), we plot summed, GLEAM integrated flux-densities alongside measurements obtained for {\it 3CR} sources, spanning 10\,MHz to 15\,GHz \citep{Laing1980}. We thank Robert Laing for providing a compilation of the latter. Note that this does not include data for the following sources: G4Jy~18 (4C~+12.03), G4Jy~432 (4C~+14.11), G4Jy~714 (4C~+14.27), G4Jy~1004
(1227+119), G4Jy~1282 (3C~326), G4Jy~1419 (4C~+16.49) and G4Jy~1456 (4C+~13.66). This is because these sources were added later, during the creation of the 3CRR sample \citep{Laing1983}, or -- in the case of 3C~326 -- the source was omitted by \citet{Laing1980} because its `integrated flux densities are not well known'. These spectra are available online as supplementary material (through PASA and at https://github.com/svw26/G4Jy).

\begin{landscape}
\begin{table}
\centering
\caption{Column numbers, names, units, descriptions, and first-row entries for 117 of the 383 columns in the G4Jy catalogue (Section~\ref{sec:finalcatalogue}, Appendix~E). Mid-infrared information is taken from the AllWISE catalogue \citep{Cutri2013}, and `PSF' stands for `point spread function'. See appendix~A of \citet{HurleyWalker2017} for the remaining 266 G4Jy-catalogue columns.}
\begin{tabular}{@{}ccccr@{}} 
\hline
Column no. & Column name & Unit & Description & First-row entry  \\
\hline
1 & G4Jy\_name & -- & Name of source belonging to the G4Jy Sample & G4Jy~1 \\
2 & G4Jy\_component & -- & Label for an individual GLEAM component  & -- \\
3 & ncmp\_GLEAM & -- & Number of GLEAM components for this G4Jy source   & 1 \\
4 & GLEAM\_name & --  & Name of the GLEAM component  & GLEAM~J000057$-$105435  \\
5 & refitted\_flag & -- & Flag for sources with re-fitted GLEAM measurements  & 0 \\
6 & centroid\_RAJ2000 & $\arcdeg$ & Right ascension for the centroid position (J2000) &  0.24 \\
7 & err\_centroid\_RAJ2000 & $''$ & Error on RA for the centroid position   &  0.44\\
8 & centroid\_DEJ2000 & $\arcdeg$ & Declination for the centroid position (J2000)  & $-$10.91 \\
9 & err\_centroid\_DEJ2000 & $''$ & Error on Dec for the centroid position  & 0.60 \\
10 & centroid\_flag & -- & Flag for sources with an updated centroid-position  & 0 \\
11 & morphology & -- & Morphology of the source in NVSS/SUMSS/TGSS  & single \\
12 & ncmp\_NVSSorSUMSS & -- & Number of NVSS/SUMSS components  & 1 \\
13 & confusion\_flag & -- & Flag for sources affected by confusion in GLEAM  & 0 \\
14 & angular\_size\_limit & -- & Indicates that the angular size is an upper limit  & $<$ \\
15 & angular\_size & $''$ & Angular size in NVSS/SUMSS  & 16.6 \\
16 & S\_NVSSorSUMSS & Jy & Total flux-density in NVSS/SUMSS  &  0.397 \\
17 & err\_S\_NVSSorSUMSS & Jy & Error on the total flux-density in NVSS/SUMSS  &  0.012  \\
18 & Freq & MHz & Indicates whether NVSS or SUMSS was used  & 1400 \\
19 & G4Jy\_NVSS\_alpha & -- & Spectral index between 151 and 1400\,MHz  & $-$1.13 \\
20 & err\_G4Jy\_NVSS\_alpha & -- & Error on spectral index between 151 and 1400\,MHz  & 0.01 \\
21 & G4Jy\_SUMSS\_alpha & -- & Spectral index between 151 and 843\,MHz  &  \\
22 & err\_G4Jy\_SUMSS\_alpha & -- & Error on spectral index between 151 and 843\,MHz  &  \\
23 & G4Jy\_alpha & -- & Fitted spectral-index using total GLEAM flux-densities  & $-$1.06  \\
24 & err\_G4Jy\_alpha & -- & Error on fitted, G4Jy spectral-index  & 0.01 \\
25 & reduced\_chi2\_G4Jy\_alpha & -- & Reduced-$\chi^2$ statistic for G4Jy spectral-index fit  & 1.18 \\
26 & host\_flag & -- & Flag for the type of host-galaxy identification  & i \\
27 & AllWISE\_name & -- & Name of the host galaxy in AllWISE  &  J000057.57$-$105432.1 \\
28 & AllWISE\_RAJ2000 & $\arcdeg$ & Right ascension for the host galaxy (J2000)  & 0.23990 \\
29 & AllWISE\_DEJ2000 & $\arcdeg$ & Declination for the host galaxy (J2000)  & $-$10.90893 \\
30 & W1mag & mag & W1 magnitude for the host galaxy  & 16.966 \\
31 & err\_W1mag & mag & Error on W1 magnitude for the host galaxy  & 0.134 \\
32 & W2mag & mag & W2 magnitude for the host galaxy  & 16.865 \\
33 & err\_W2mag & mag & Error on W2 magnitude for the host galaxy  & 0.415 \\
34 & W3mag & mag & W3 magnitude for the host galaxy  & 11.890 \\
35 & err\_W3mag & mag & Error on W3 magnitude for the host galaxy  &  \\
36 & W4mag & mag & W4 magnitude for the host galaxy  & 8.487 \\
37 & err\_W4mag & mag & Error on W4 magnitude for the host galaxy  &  \\
\hline
\label{tab:catalogueexample}
\end{tabular}
\end{table}
\end{landscape}

\setcounter{table}{15} 

\begin{landscape}
\begin{table}
\centering
\caption{{\it Continued} -- Column numbers, names, units, descriptions, and first-row entries for 117 of the 383 columns in the G4Jy catalogue (Section~\ref{sec:finalcatalogue}, Appendix~E).}
\begin{tabular}{@{}ccccr@{}} 
\hline
Column no. & Column name & Unit & Description & First-row entry  \\
\hline
38 & GLEAM\_ra\_str & h:m:s & Right ascension for the GLEAM component (J2000) &  00:00:57.39   \\
39 & GLEAM\_dec\_str & d:m:s & Declination for the GLEAM component (J2000) &   $-$10:54:35.00  \\
40 & GLEAM\_RAJ2000 & $\arcdeg$ & Right ascension for the GLEAM component (J2000) &  0.23910   \\
41 & err\_GLEAM\_RAJ2000 & $\arcdeg$ & Error on RA for the GLEAM component &  0.00003   \\
42 & GLEAM\_DEJ2000 & $\arcdeg$ & Declination for the GLEAM component (J2000) &  $-$10.90972   \\ 
43 & err\_GLEAM\_DEJ2000 & $\arcdeg$ & Error on Dec for the GLEAM component &  0.00006   \\ 
44 & background\_wide & Jy\,beam$^{-1}$ & Background level in wide-band image & -0.001    \\ 
45 & local\_rms\_wide & Jy\,beam$^{-1}$ & Local noise level in wide-band image &  0.010   \\ 
46 & peak\_flux\_wide & Jy\,beam$^{-1}$ & Peak flux-density in wide-band image &   3.734  \\ 
47 & err\_peak\_flux\_wide & Jy\,beam$^{-1}$ & Uncertainty in fit for peak flux-density in wide-band image &   0.010  \\ 
48 & int\_flux\_wide & Jy & Integrated flux-density in wide-band image &  3.580   \\ 
49 & err\_int\_flux\_wide & Jy & Uncertainty in fit for integrated flux-density in wide-band image &   0.011  \\ 
50 & a\_wide & $''$ & Fitted semi-major axis in wide-band image &  132.33   \\ 
51 & err\_a\_wide & $''$ & Uncertainty in fitted semi-major axis in wide-band image & 0.14    \\ 
52 & b\_wide & $''$ & Fitted semi-minor axis in wide-band image &  124.83   \\ 
53 & err\_b\_wide & $''$ & Uncertainty in fitted semi-minor axis in wide-band image &  0.15   \\ 
54 & pa\_wide & $\arcdeg$ & Fitted position angle in wide-band image &  $-$69.40   \\ 
55 & err\_pa\_wide & $\arcdeg$ & Uncertainty in fitted position angle in wide-band image &  0.01  \\ 
56 & residual\_mean\_wide & Jy\,beam$^{-1}$ & Mean value of data$-$model in wide-band image &  0.002   \\ 
57 & residual\_std\_wide & Jy\,beam$^{-1}$ & Standard deviation of data$-$model in wide-band image &   0.019  \\ 
58 & psf\_a\_wide & $''$ & Semi-major axis of PSF in wide-band image &  135.70   \\
59 & psf\_b\_wide & $''$ & Semi-minor axis of PSF in wide-band image &  126.98   \\
60 & psf\_pa\_wide & $\arcdeg$ & Position angle of PSF in wide-band image &  $-$71.31   \\
187 & background\_151 & Jy\,beam$^{-1}$ & Background level in 147--154\,MHz image &  $-$0.002   \\
188 & local\_rms\_151 & Jy\,beam$^{-1}$ & Local noise level in 147--154\,MHz image &   0.033  \\
189 & peak\_flux\_151 & Jy\,beam$^{-1}$ & Peak flux-density in 147--154\,MHz image &  5.078   \\
190 & err\_peak\_flux\_151 & Jy\,beam$^{-1}$ & Uncertainty in fit for peak flux-density in 147--154\,MHz image &  0.029   \\
191 & int\_flux\_151 & Jy & Integrated flux-density in 147--154\,MHz image &  4.920   \\
192 & err\_int\_flux\_151 & Jy & Uncertainty in fit for integrated flux-density in 147--154\,MHz image &   0.028  \\
193 & a\_151 & $''$ & Fitted semi-major axis in 147--154\,MHz image &  151.69  \\
194 & b\_151 & $''$ & Fitted semi-minor axis in 147--154\,MHz image &   147.28  \\
195 & pa\_151 & $\arcdeg$ & Fitted position angle in 147--154\,MHz image &  $-$69.40  \\
196 & residual\_mean\_151 & Jy\,beam$^{-1}$ & Mean value of data$-$model in 147--154\,MHz image &  0.007   \\
197 & residual\_std\_151 & Jy\,beam$^{-1}$ & Standard deviation of data$-$model in 147--154\,MHz image &  0.044  \\
198 & psf\_a\_151 & $''$ & Semi-major axis of PSF in 147--154\,MHz image &  154.64   \\
199 & psf\_b\_151 & $''$ & Semi-minor axis of PSF in 147--154\,MHz image &   149.11  \\
200 & psf\_pa\_151 & $\arcdeg$ & Position angle of PSF in 147--154\,MHz image &   $-$46.01  \\
341 & GLEAM\_alpha & -- & Fitted spectral-index using GLEAM-component flux-densities & $-$1.06   \\
342 & err\_GLEAM\_alpha & -- & Error on fitted, GLEAM spectral-index &   0.01  \\
343 & reduced\_chi2\_GLEAM\_alpha & -- & Reduced-$\chi^2$ statistic for GLEAM spectral-index fit &   1.18  \\
\hline
\label{tab:catalogueexample}
\end{tabular}
\end{table}
\end{landscape}

\setcounter{table}{15} 

\begin{landscape}
\begin{table}
\centering
\caption{{\it Continued} -- Column numbers, names, units, descriptions, and first-row entries for 117 of the 383 columns in the G4Jy catalogue (Section~\ref{sec:finalcatalogue}, Appendix~E).}
\begin{tabular}{@{}ccccr@{}} 
\hline
Column no. & Column name & Unit & Description & First-row entry  \\
\hline
344 & total\_int\_flux\_076 & Jy & Total, integrated flux-density in 72--80\,MHz image &   9.862  \\
345 & err\_total\_int\_flux\_076 & Jy & Error on total, integrated flux-density in 72--80\,MHz image &    0.092  \\
346 & total\_int\_flux\_084 & Jy & Total, integrated flux-density in 80--88\,MHz image &   8.910  \\
347 & err\_total\_int\_flux\_084 & Jy & Error on total, integrated flux-density in 80--88\,MHz image &   0.074  \\
348 & total\_int\_flux\_092 & Jy & Total, integrated flux-density in 88--95\,MHz image &   8.168  \\
349 & err\_total\_int\_flux\_092 & Jy & Error on total, integrated flux-density in 88--95\,MHz image &   0.062   \\
350 & total\_int\_flux\_099 & Jy & Total, integrated flux-density in 95--103\,MHz image &  7.488   \\
351 & err\_total\_int\_flux\_099 & Jy & Error on total, integrated flux-density in 95--103\,MHz image &  0.054   \\
352 & total\_int\_flux\_107 & Jy & Total, integrated flux-density in 103--111\,MHz image &  7.024   \\
353 & err\_total\_int\_flux\_107 & Jy & Error on total, integrated flux-density in 103--111\,MHz image &   0.046   \\
354 & total\_int\_flux\_115 & Jy & Total, integrated flux-density in 111--118\,MHz image &  6.557   \\
355 & err\_total\_int\_flux\_115 & Jy & Error on total, integrated flux-density in 111--118\,MHz image &  0.041  \\
356 & total\_int\_flux\_122 & Jy & Total, integrated flux-density in 118--126\,MHz image &   6.240  \\
357 & err\_total\_int\_flux\_122 & Jy & Error on total, integrated flux-density in 118--126\,MHz image &  0.035  \\
358 & total\_int\_flux\_130 & Jy & Total, integrated flux-density in 126--134\,MHz image &  5.682  \\
359 & err\_total\_int\_flux\_130 & Jy & Error on total, integrated flux-density in 126--134\,MHz image &  0.033   \\
360 & total\_int\_flux\_143 & Jy & Total, integrated flux-density in 139--147\,MHz image &   5.173  \\
361 & err\_total\_int\_flux\_143 & Jy & Error on total, integrated flux-density in 139--147\,MHz image & 0.032    \\
362 & total\_int\_flux\_151 & Jy & Total, integrated flux-density in 147--154\,MHz image &   4.920  \\
363 & err\_total\_int\_flux\_151 & Jy & Error on total, integrated flux-density in 147--154\,MHz image &  0.028   \\
364 & total\_int\_flux\_158 & Jy & Total, integrated flux-density in 154--162\,MHz image &  4.703   \\
365 & err\_total\_int\_flux\_158 & Jy & Error on total, integrated flux-density in 154--162\,MHz image &  0.028   \\
366 & total\_int\_flux\_166 & Jy & Total, integrated flux-density in 162--170\,MHz image &  4.514   \\
367 & err\_total\_int\_flux\_166 & Jy & Error on total, integrated flux-density in 162--170\,MHz image &  0.027   \\
368 & total\_int\_flux\_174 & Jy & Total, integrated flux-density in 170--177\,MHz image &  4.306   \\
369 & err\_total\_int\_flux\_174 & Jy & Error on total, integrated flux-density in 170--177\,MHz image &  0.025   \\
370 & total\_int\_flux\_181 & Jy & Total, integrated flux-density in 177--185\,MHz image &  4.064   \\
371 & err\_total\_int\_flux\_181 & Jy & Error on total, integrated flux-density in 177--185\,MHz image &   0.024  \\
372 & total\_int\_flux\_189 & Jy & Total, integrated flux-density in 185--193\,MHz image &   3.887  \\
373 & err\_total\_int\_flux\_189 & Jy & Error on total, integrated flux-density in 185--193\,MHz image &   0.024  \\
374 & total\_int\_flux\_197 & Jy & Total, integrated flux-density in 193--200\,MHz image &  3.732   \\
375 & err\_total\_int\_flux\_197 & Jy & Error on total, integrated flux-density in 193--200\,MHz image &  0.021   \\
376 & total\_int\_flux\_204 & Jy & Total, integrated flux-density in 200--208\,MHz image &   3.424  \\
377 & err\_total\_int\_flux\_204 & Jy & Error on total, integrated flux-density in 200--208\,MHz image &   0.024  \\
378 & total\_int\_flux\_212 & Jy & Total, integrated flux-density in 208--216\,MHz image &   3.289  \\
379 & err\_total\_int\_flux\_212 & Jy & Error on total, integrated flux-density in 208--216\,MHz image &  0.022   \\
380 & total\_int\_flux\_220 & Jy & Total, integrated flux-density in 216--223\,MHz image &  3.186   \\
381 & err\_total\_int\_flux\_220 & Jy & Error on total, integrated flux-density in 216--223\,MHz image &   0.022  \\
382 & total\_int\_flux\_227 & Jy & Total, integrated flux-density in 223--231\,MHz image &   2.992  \\
383 & err\_total\_int\_flux\_227 & Jy & Error on total, integrated flux-density in 223--231\,MHz image &  0.024   \\
\hline
\label{tab:catalogueexample}
\end{tabular}
\end{table}
\end{landscape}

\end{appendix}

\bibliographystyle{pasa-mnras}
\bibliography{SarahWhite_GLEAM_4Jy_DefinitionPaper}

\end{document}